\documentclass[11pt]{article}

\usepackage{amsmath,epsfig,amsfonts}
\usepackage{epstopdf}
\usepackage{subfigure}
\usepackage{amssymb,newlfont,setspace,rotating}
\usepackage{color}
\usepackage[normalem]{ulem}
\usepackage{graphicx,textcomp}%
\begin{document}

\title{Wave propagation and pattern formation in two-dimensional hexagonally-packed granular crystals under various configurations}

\author{Tianshu Hua and Robert A. Van Gorder \\  \small  Mathematical Institute, University of Oxford\\ \small Andrew Wiles Building, Radcliffe Observatory Quarter, Woodstock Road \\ \small Oxford, OX2 6GG, United Kingdom\\
\small Email: Robert.VanGorder@maths.ox.ac.uk}        
\date{\today}

\maketitle

\begin{center}
Abstract
\end{center}
We study wave propagation in two-dimensional granular crystals under the Hertzian contact law consisting of hexagonal packings of spheres under various basin geometries including hexagonal, triangular, and circular basins which can be tiled with hexagons. We find that the basin geometry will influence wave reflection at the boundaries, as expected, and also may result in bottlenecks forming. While exterior strikers the size of a single sphere have been considered in the literature, it is also possible to consider strikers which impact multiple spheres along a boundary, or to have multiple sides being struck simultaneously. It is also possible to consider obstructions or even strikers in the interior of the hexagonally packed granular crystal, as previously considered in the case of square packings, resulting in the basin geometry no longer forming a convex set. We consider various configurations of either boundary or interior strikers. We shall also consider the case where a granular crystal is composed of two separate crystals of differing material, with a single interface between the two distinct materials. Depending on the relative material properties of each type of sphere, this can result in a trapping of most of the wave energy within one of the two regions. While repeated reflections from the boundaries will cause the systems we study to fall into disorder for large time, there are a number of interesting wave structures and patters that emerge as transients at intermediate timescales. 

\section{Introduction}
Granular crystals consist of a closed packing of spherical particles having a periodic arrangement \cite{Burgoyne}. Extensive work has been done on one-dimensional granular chains or crystals, as the study of wave propagation in one-dimensional granular chains or crystals provides an experimentally realizable application of the motion of nonlinear coupled oscillators \cite{NES02,sen,M1,M2}. It is shown that under weak precompression or in the absence of precompression, the interactions between spheres are highly nonlinear, and highly localized waves can be supported, while the interactions will be approximately linear under higher precompression \cite{sen}. Compression waves can propagate as solitons or sound waves depending on the interaction potential between the beads and the external compression of the chain. The type of interaction that leads to non-linear solitary waves is the repulsive Hertz-like contact force that depends on the contact geometry and bead material \cite{JOH01}. Patterns of waves propagating in granular crystals can be controlled by the initial compression of the system, and observation can prove useful in applications, as granular crystals can be used to design shock and energy absorbing layers, actuating devices, acoustic lenses, acoustic diodes, and sound scramblers \cite{Leo14}. Even though solitary waves are stable in uniform granular chains \cite{NES02}, they can undergo drastic changes at interfaces between media with different properties \cite{sen,VER01,VER02,kekic2018wave}.

Theoretical investigations into two-dimensional granular crystals have begun in earnest more recently, and the patterns of waves in two-dimensional geometries are not as well understood as their one-dimensional counterparts. Still, several works have looked into individual two-dimensional cases. These studies can roughly be partitioned according to the geometries they consider.

In order to induce wave motion, strikers (particles or whole boundaries which are impacted by a force and then propagate into the neighboring particles) are often used. Wave propagation due to the location of boundary strikers for two-dimensional configurations of square packings have been studied \cite{Leo13}, and it was shown that the characteristics of wave propagation are similar to that of the one-dimensional case, with very little energy is lost to the adjacent chains. In addition to boundary strikers, in two-dimensional media it is also possible to have meaningful interior strikers or boundaries, and these can yield more interesting spatial dynamics. The wave propagation when several smaller intruders inserted into a square packing of spheres using the standard Hertzian model was studied \cite{Sze12}. If there is a strike at the end of one of the rows with a sphere next to the intruder, it has been shown that the intruder can redirect a small fraction of the wave energy into other directions. The fraction of energy diverted is relatively small, but the method can be employed to design materials with desired wave diversion effect by placing intruders at specific positions. This method was later extended \cite{Sze13} by considering intruders at every gap between the two rows that will be hit. Different patterns of wave fronts emerged when changing the combination of materials for spheres and intruders, and the main cause of variability of the experimental system appears to be the differences of the radius of individual spheres \cite{Leo12}. 

Sphere-intruder configurations with discontinuous lateral boundaries were imposed to give extra nonlinearity by \cite{Hasan}. Depending on the mass and stiffness ratios between the spheres and intruders, the system can be either in a propagation zone or an oscillatory zone. In the former case, solitary waves or travelling breathers will be supported as found in other studies. In the later case, passive wave arrest is observed. It was argued that such behaviour is a result of energy transfer between pulses of lower frequencies and higher frequencies, as well as the dynamics caused by the discontinuous boundaries. Configurations of square packings with intruders under strong precompression have also been studied \cite{Goncu}. Rather than considering an ODE at each sphere, \cite{Goncu} employ a finite element method  such that each particle is regarded as a node. They find that under various configurations, there will be different band gaps where waves with such frequencies will be tuned after propagating through the lattice.

Aside from two-dimensional granular crystals composed of square lattice packings, there are a variety of other geometric configurations which have been studied. Wattis et al. \cite{Wattis} investigated wave propagation on a honeycomb lattice within the context of a model is based on Kirchoff's laws of electrical charge on each node. The dispersion relation is shown to have two branches, acoustic and optical, and they will meet at the Dirac point. Yang et al.\cite{Yang} studied the wave propagation in a curved one-dimensional granular chain, resulting in a sort of quasi-two-dimensional problem with the force applied on each particle being the contact force with other particles and the boundaries. It was shown that the energy transmission is controlled by the curvature of the chains. A comparison was carried out between straight chains and curved chains, and it was observed that the solitary waves are sparser and carry less energy on curved chains. The work was later extended to include periodic arrangements of spherical and cylindrical particles in a curved elastic tube \cite{Yang2}. 

Instead of considering the spheres to be purely elastic, \cite{Burgoyne} considered elastic-plastic spheres; the contact law was still Hertzian under small deformations, but the contact law was different under large deformations. In the study, one-dimensional granular chains, two-dimensional hexagonally packed particles, and 3D packings with layers of hexagonal arrangements were compared, and it was shown that the dependence of wave velocity on the material properties are similar in all three cases. It was also shown that the wave can be directed away from the initial direction with an alternating arrangement of particles made by two materials in the elastic-plastic regime, which cannot happen in the purely elastic regime. Yi et al. \cite{Yi} studied wave propagation in a three-dimensional hexagonally packed geometry experimentally. Defects were added to the lattice, and it was shown that the existence of such defects can change the distribution of force in the lattice. Defects in main force chains will change the magnitude and direction of the wave, while defects off the main force chain have little influence on the system.

Wave propagation in a two-dimensional granular crystal composed of hexagonally packed spheres under a Hertzian contact model was studied in \cite{Leo14}. Using the symmetry of the packing pattern, two striking angles were investigated, and it was shown that there is a exponential decay of the maximum velocity of particles along the direction of propagation in both cases. (Additional results on angled strikes were discussed in \cite{manjunath2014plane}.) The propagation of conical waves in a hexagonal packing was considered in \cite{Chong}. Both a heuristic argument and an asymptotic analysis is employed to formulate the system for Dirac points, where Dirac points are defined as geometric point where the acoustic band and the optical band. Wallen and Boechler \cite{wallen2017shear} extended the Hertzian model to consider the influence of rotations and transverse displacement between particles, and then examined the weakly nonlinear case to obtain a theoretical prediction which gives an accurate result in the non-resonant case, and qualitatively describes the patterns in the resonant regime and anti-resonant regime. Interestingly, hexagonal packings result in waves which interact with all spheres on a large enough timescale, compared to square packings for which waves are localized to certain rows or columns \cite{Leo14}. Therefore, wave structures can be more complicated in hexagonal packings, and there are still many things to study in this setting. Shear to longitudinal mode conversion via second harmonic generation was very recently studied for plane waves in a two-dimensional, adhesive, hexagonally close-packed microscale granular media \cite{wallen2017shear}. Stress wave tuning in hexagonal lattices was further discussed at the nanoscale (rather than the more common macroscale) in \cite{xu2016stress}. Disorder in hexagonal lattices was shown to affect both wave transit time and peak force \cite{waymel2018propagation}. The interaction of certain solitary waves in hexagonally packed granular systems has been studied recently by \cite{li2018two}, with 1D approximations of the wave dynamics being obtained in appropriate regimes.

While \cite{Leo14} and others have demonstrated interesting behaviors of waves propagating within granular crystals composed of hexagonally packed spheres, there are still a number of interesting features to explore. In this paper, we will study wave propagation in two-dimensional granular crystals consisting of hexagonal packings of spheres under various basin geometries. While the hexagonal packing dictates the local dynamics along the granular crystal, assemblies of hexagonal packings can be tiled to produce other shapes, such as triangles and approximations to circles. We find that the basin geometry will influence wave reflection at the boundaries, as expected, and also may result in bottlenecks forming. While exterior strikers the size of a single sphere were considered in \cite{Leo14} for the hexagonal packing, it is also possible to consider strikers which impact many spheres along a boundary or even multiple sides being struck at once. As \cite{Sze12} considered in the case of spherical packings, it is also possible to consider obstructions or even strikers in the interior of the hexagonally packed granular crystal, resulting in the basin geometry no longer forming a convex set. Therefore, various configurations of either boundary or interior strikers will be studied in this work. We shall also consider the case where a granular crystal is composed of two separate crystals of differing material, with a single interface between the two distinct materials. Depending on the relative material properties of each type of sphere, this can result in a trapping of most of the wave energy within one of the two regions. 

As discussed in \cite{Leo14}, the Hertzian model is highly nonlinear, and this is only compounded by the contact geometry of the hexagonal packing. In addition to numerical simulations of the model, some analytical results were found in the heterogeneous crystal case studied by \cite{Leo14} in certain special limits and under some assumptions (for instance, under a ternary collision approximation). However, for the heterogeneous situations we are more interested in for the present paper, such simplifying assumptions will fail to capture the interesting dynamics. Therefore, we shall resort to full numerical simulations of the Hertzian contact model under hexagonal packing in order to draw conclusions about the dynamics. While not as robust as analytical results, we have sampled a wide assortment of configurations, and feel that many of our findings from numerical experiments are sufficiently robust so as to suggest the emergence of interesting dynamics and patterning. The remainder of the paper is organized as follows. In Section 2 we will discuss the Hertzian contact model and the numerical solution method employed in our work. In Section 3, we study wave propagation in various configurations when one or more spheres are struck at a boundary, while in Section 4 we consider the corresponding results for interior strikers. In Section 5, we shall study the effects of interfaces between two distinct media on wave propagation due to a boundary striker. We then discuss the results in Section 6.

\section{Two-dimensional Hertzian contact model for hexagonal sphere packing}
In order to model wave propagation within hexagonal packings of spheres, we shall employ the Hertzian contact model, which is also the focus of much of current work in this field. For a spherical particle $i$ surrounded by a set of spherical particles, denoted by $S_i$, the contact law in two-dimensional planar geometry can be described as \cite{Leo13}
\begin{equation}\label{deGen}
M_i \ddot{\mathbf{u}}_i = -\sum_{j\in S_i} A_{i,j} (R_i+R_j-|\mathbf{l}_{ij}|)^{3/2}_{{+}}\frac{\mathbf{l}_{ij}}{| \mathbf{l}_{ij}|},
\end{equation}
where $\mathbf{l}_{ij}$ is a vector from the center of particle $i$ to particle $j$ defined as
\begin{equation}
\mathbf{l}_{ij} = (R_i+R_j-\delta_0)\mathbf{e}_{ij}-\mathbf{u}_i+\mathbf{u}_j,
\end{equation}
$M_i$ is the mass of the particle, $\mathbf{u}_i$ is the displacement of particle $i$, and $\mathbf{e}_{ij}$ is a unit vector pointing from the initial position of the center of particle $i$ to the initial position of the center of particle $j$. The coefficient $A_{ij}$ is defined as
\begin{equation}
A_{i,j}= \frac{4}{3}\left(\frac{1-\nu^2_i}{E_i}+\frac{1-\nu^2_j}{E_j}\right)^{-1}\left(\frac{R_iR_j}{R_i+R_j}\right)^{1/2},
\end{equation} 
where $R_i$ is the radius of the particle, $\nu_i$ is the Poisson ratio, and $E$ stands for Young's modulus. The Heaviside step function $(a)_{{+}}$ will return $a$ if $a>0$, and return 0 otherwise, which corresponds to the fact that there will be no elastic force between particles if there is no compression between them. This law is valid in the elastic regime of particles \cite{Burgoyne}, and will need to be altered to account for large deformations. In this study, we control the initial velocity of the striker so that the displacements of particles in the configuration are sufficiently small to allow the Hertzian law to remain valid. We therefore assume that all the interactions will follow the Hertzian contact law throughout the process.

In order to impose physically relevant boundary conditions, we shall also consider the contact between particles and a wall which contains all of the particles within a basin or domain. For a sphere $i$ in contact with a set of walls $W_i$, the contact mechanics are described by \cite{hanaor}
\begin{equation}\begin{aligned}
M_i \ddot{\mathbf{u}}_i & = -\sum_{j\in S_i} A_{i,j} (R_i+R_j-|\mathbf{l}_{ij}|)^{3/2}_{{+}}\frac{\mathbf{l}_{ij}}{| \mathbf{l}_{ij}|} \\
& \quad +\sum_{l\in W_i} B_{i,l} (R_i-d_{il})^{3/2}_{{+}} \mathbf{e}_{il},
\end{aligned}\end{equation}
where $d_{il}$ is the distance from the center of the particle $i$ to the wall $l$, $\mathbf{e}_i$ is a unit vector pointing from the wall to the particle, and 
\begin{equation}
B_{i,l} = \frac{4}{3}\left(\frac{1-\nu^2_i}{E_i}+\frac{1-\nu^2_l}{E_l}\right)^{-1}\sqrt{R_i}.
\end{equation} 
These equations will be the governing equations in this study. The movement of the spheres within the basin due to a strike at the initial time can then be fully described given initial conditions.

We now specialize the general two-dimensional equations to the case of hexagonal packings of spheres, where each interior sphere in the packing will be surrounded by six spheres and the centers of those spheres will form a hexagon. Writing $\mathbf{u}_{(m,n)}=(u_{(m,n)},v_{(m,n)})$, equation \eqref{deGen} then becomes \cite{Leo14}
\begin{equation}\begin{aligned}
M_{(m,n)} \ddot{u}_{(m,n)} & =  A_{(m,n),(m-1,n-1)}(2R-\rho_1)^{3/2}_{{+}} \frac{\rho_1^x}{\rho_1}\\
& \quad -A_{(m,n),(m-1,n)}(2R-\rho_2)^{3/2}_{{+}} \frac{\rho_2^x}{\rho_2}\\
&\quad +A_{(m,n),(m,n-1)}(2R-\rho_3)^{3/2}_{{+}} \frac{\rho_3^x}{\rho_3}\\
& \quad -A_{(m,n),(m,n+1)}(2R-\rho_4)^{3/2}_{{+}} \frac{\rho_4^x}{\rho_4}\\
& \quad +A_{(m,n),(m+1,n)}(2R-\rho_5)^{3/2}_{{+}} \frac{\rho_5^x}{\rho_5}\\
&  -A_{(m,n),(m+1,n+1)}(2R-\rho_6)^{3/2}_{{+}} \frac{\rho_6^x}{\rho_6},
\end{aligned}\end{equation}
\begin{equation}\begin{aligned}
M_{(m,n)} \ddot{v}_{(m,n)} & = -A_{(m,n),(m-1,n-1)}(2R-\rho_1)^{3/2}_{{+}} \frac{\rho_1^y}{\rho_1}\\
& \quad +A_{(m,n),(m-1,n)}(2R-\rho_2)^{3/2}_{{+}} \frac{\rho_2^y}{\rho_2}\\
& \quad +A_{(m,n),(m,n-1)}(2R-\rho_3)^{3/2}_{{+}} \frac{\rho_3^y}{\rho_3}\\
& \quad -A_{(m,n),(m,n+1)}(2R-\rho_4)^{3/2}_{{+}} \frac{\rho_4^y}{\rho_4}\\
& \quad +A_{(m,n),(m+1,n)}(2R-\rho_5)^{3/2}_{{+}} \frac{\rho_5^y}{\rho_5}\\
& \quad -A_{(m,n),(m+1,n+1)}(2R-\rho_6)^{3/2}_{{+}} \frac{\rho_6^y}{\rho_6},
\end{aligned}\end{equation}
where $\rho_i =\sqrt{(\rho_i^x)^2+(\rho_i^y)^2}$ and
$$
\rho_1^x = R-\frac{1}{2}\delta_0 -u_{(m-1,n-1)}+ u_{(m,n)},
$$
$$
\rho_1^y = \sqrt{3}R - \frac{\sqrt{3}}{2}\delta_0 +v_{(m-1,n-1)}- v_{(m,n)},
$$
$$
\rho_2^x = R-\frac{1}{2}\delta_0 -u_{(m,n)}+u_{(m-1,n)},
$$
$$
\rho_2^y = \sqrt{3}R - \frac{\sqrt{3}}{2}\delta_0 -v_{(m,n)}+v_{(m+1,n)},
$$
$$
\rho_3^x = 2R -\delta_0 +u_{(m,n)}-u_{(m,n-1)},
$$
$$
\rho_3^y = v_{(m,n)}-v_{(m,n-1)},
$$
$$
\rho_4^x = 2R -\delta_0 +u_{(m,n+1)}-u_{(m,n)},
$$
$$
\rho_4^y = v_{(m,n+1)}-v_{(m,n)},
$$
$$
\rho_5^x = R-\frac{1}{2}\delta_0 -u_{(m+1,n)}+ u_{(m,n)},
$$
$$
\rho_5^y = \sqrt{3}R- \frac{\sqrt{3}}{2}\delta_0 - v_{(m+1,n)} + v_{(m,n)},
$$
$$
\rho_6^x = R-\frac{1}{2}\delta_0 -u_{(m,n)}+u_{(m+1,n+1)},
$$
$$
\rho_6^y = \sqrt{3}R - \frac{\sqrt{3}}{2}\delta_0 -v_{(m,n)}+v_{(m+1,n+1)}.
$$

For the initial condition, it is usually supposed that the system is at rest in the beginning. There would either be a precompression among the spheres or the spheres are packed so that they will be in contact with each other but no compression is applied. At time $t=0$, a striker will impact the system. In most works, the strikers will hit a particle at the boundary of a basin. Then the striker will be deflected away and will not come back to the system. 
A wave generated by the strike will propagate within the basin. Sen et al.\cite{sen} claim that without precompression, the interactions between particles will be highly nonlinear, and solitary trains will emerge. 
The setting of the model will allow us to consider spheres of different materials and the interaction between different types of particles. It is possible to consider interfaces within the geometries, appropriately modifying the parameters in the model given above. The interface will have spheres of different physical properties on each of the two sides.

The radius of spheres and material properties used in the simulations will follow \cite{Leo13}. The radius of the spheres is set to 9.025 mm, and the properties of materials used in this paper are listed in Table \ref{table}. Unless specified otherwise, the simulations will be conducted assuming all the particles and walls are made of Stainless steel (type 316). The results of the system will be simulated with a standard fourth order Runge-Kutta scheme using MATLAB, which shall allow adaptive control of the integration timestep so that the relative error is to within $10^{-5}$.

\begin{table}
\begin{tabular}{|c|c|c|}
\hline
Material & Stainless steel & Polycarbonate\\
& (type 316) & \\
\hline
Mass density & 8000 $\text{kg}/\text{m}^3$ & 1200 $\text{kg}/\text{m}^3$\\
\hline
Young's modulus & 193 GPa & 2.3 GPa\\
\hline
Poisson's ratio & 0.3 & 0.37\\
\hline
\end{tabular}
\caption{Properties of materials we use in our simulations, as reported in \cite{Leo13}. \label{table}}
\end{table}

The results are presented in several ways. An important representation of the wave evolution in this paper is by giving colored graphs similar to \cite{Leo14}. The graphs will have colors at each disc which indicate the norms of velocities of the spheres at corresponding positions at a particular time, hence in such cases we shall plot $\left|\dot{\mathbf{u}}_{(m,n)}\right|$ for all $(m,n)$ in the relevant basin geometry. Other quantities of interest will include the velocity magnitude ($|\dot{\mathbf{u}}_{(m,n)}|$) and acceleration relative to the direction of the strike ($\ddot{\mathbf{u}}_{(m,n)}\cdot \mathbf{\eta}$ where $\mathbf{\eta}$ is the inward normal in the direction of the strike) for individual beads $(m,n)$ within a given basin geometry. We also consider the wave front speed, which is measured from the time taken for the maximal amplitude of the compression wave to travel from the boundary to the tenth row within the geometry. We consider the evolution of the total kinetic energy for the system, where the total kinetic energy is defined by
\begin{equation}\begin{aligned}
\text{Kinetic Energy} & = \frac{1}{2}\sum_{(m,n)\in \mathcal{D}} M_{(m,n)}\left|\dot{\mathbf{u}}_{(m,n)}\right|^2 \\
& = \frac{1}{2}\sum_{(m,n)\in \mathcal{D}} M_{(m,n)}\left( u_{(m,n)}^2 + v_{(m,n)}^2\right)\,,
\end{aligned}\end{equation}
where $\mathcal{D}$ represents the basin geometry, and $(m,n)\in \mathcal{D}$ only if there is a sphere at location $(m,n)$ within a fixed geometry $\mathcal{D}$. A large number of the beads will be activated almost simultaneously after the strike, though many of the beads will only have displacements below machine precision.

\section{Wave propagation due to boundary strikers}
We first consider wave propagation in a hexagonally packed granular crystal due to boundary strikers. We only consider the highly nonlinear case, where no precompression is applied to the geometry. The hexagonal packing allows the waves to scatter, so that the waves are expected to go through all the beads in the packing, in contrast to square packings (as discussed in Section 1). From the results in \cite{Leo14}, the system will be expected to go into disorder eventually as a result of collisions of multiple reflected waves, but there are a variety of interesting transient dynamics observed. 

In addition to the hexagonal packing, there is another geometric consideration: the shape of the basin or domain within which the spheres are held. While it certainly makes sense to consider hexagon-shaped basins, note that triangular and near-circular basins are also a possibility. We shall therefore perform numerical experiments for each case, under a variety of boundary striker configurations.

\subsection{Hexagonal basin}
We start with the configuration where the striker will hit the midpoint of one of the edges and the direction of the strike will be perpendicular to the edge. In this case, the geometry consists of stainless steel spheres and has 11 particles on each edge. This setting is identical to the setting of Leonard et al.\cite{Leo14}, which can help to verify the accuracy of our extensions to these results. 

\begin{figure*}
\centering
\subfigure[]{\includegraphics[width=0.48\linewidth]{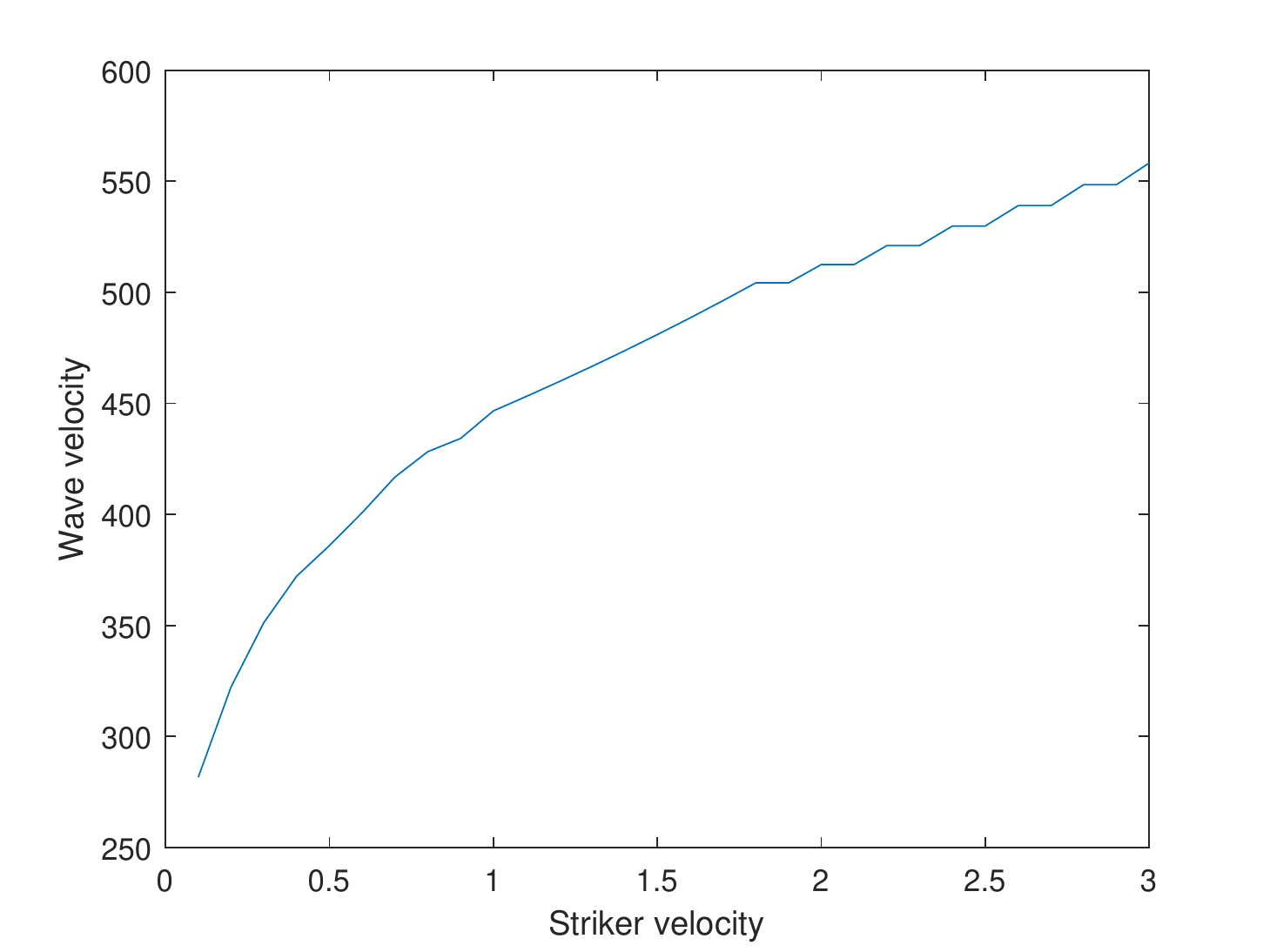}}
\subfigure[]{\includegraphics[width=0.48\linewidth]{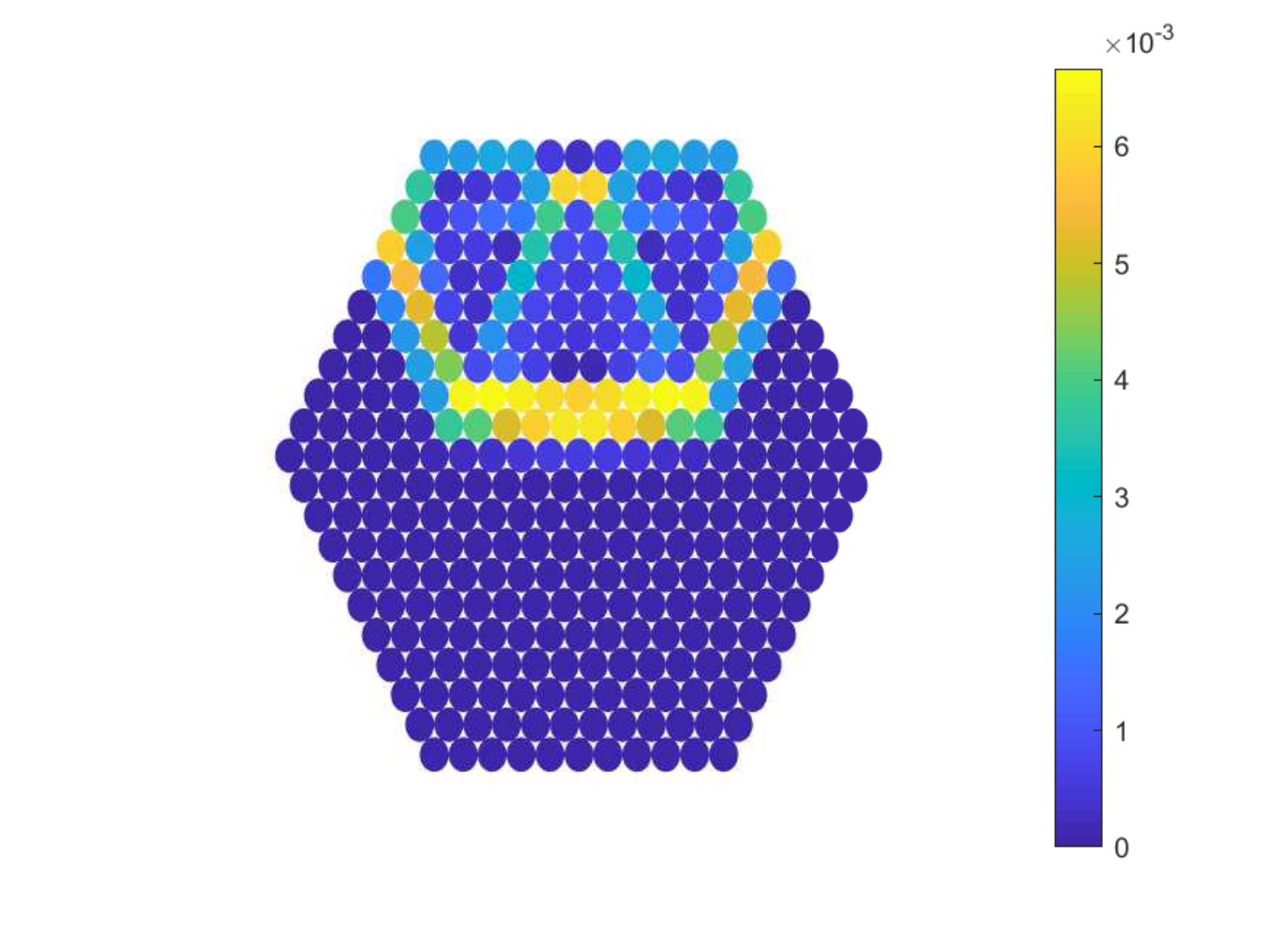}}
\caption{\label{wavespeed} (a) The wave front speed (m/s) as a function of striker velocity (m/s) for a hexagonal packing of 11 spheres on each edge which is struck perpendicularly on one of the edges at $t=0$. The curve is somewhat jagged due to the discrete geometry. The wave front speed is measured from the time taken for the maximal amplitude of the compression wave to travel from the boundary to the tenth row within the geometry. (b) An example of a case where the compression wave has reached the tenth row within a hexagonal geometry. Here the velocity magnitudes are plotted.}
\end{figure*}

\begin{figure*}
\centering
\subfigure[]{\includegraphics[width=0.48\linewidth]{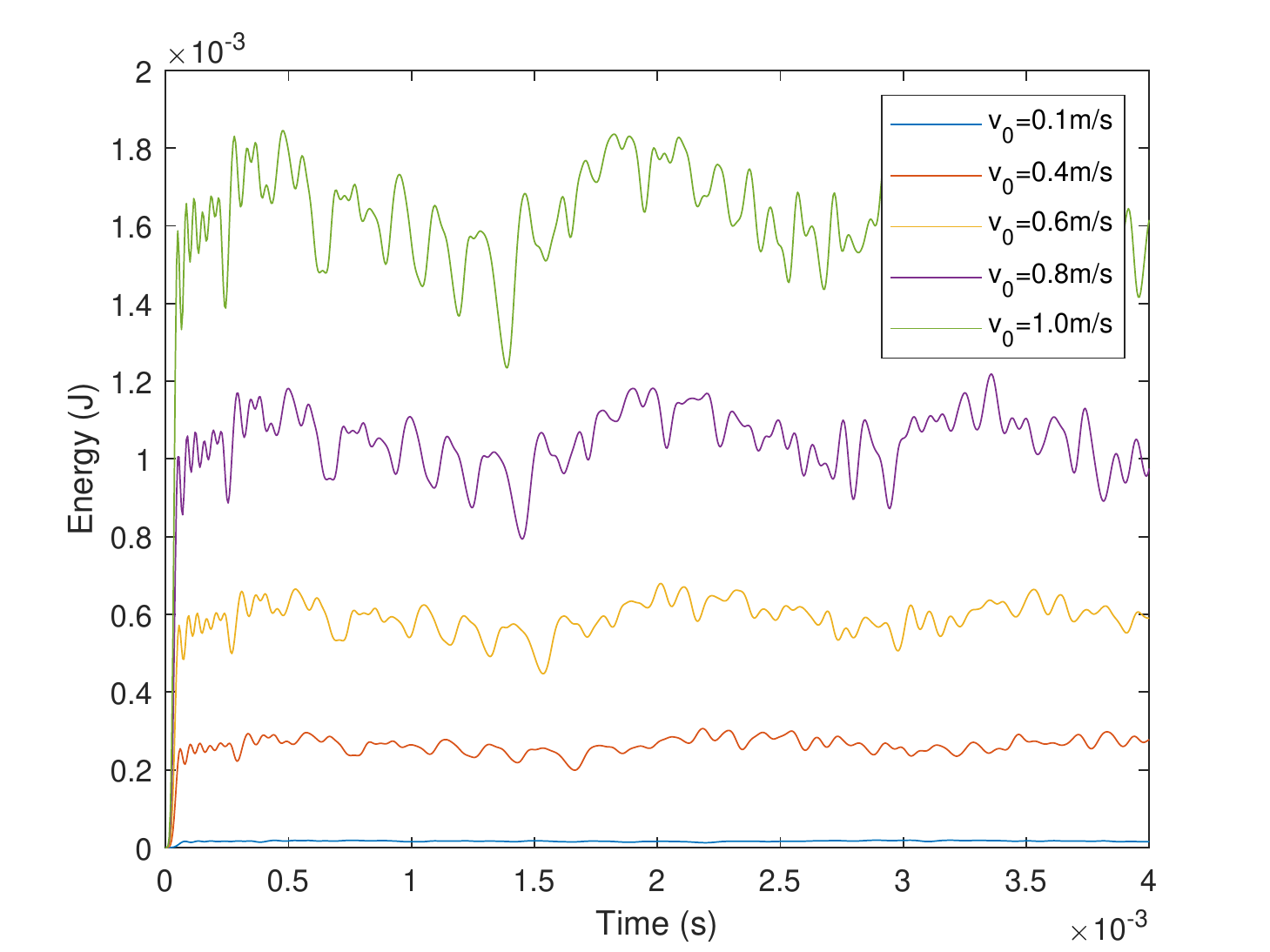}}
\subfigure[]{\includegraphics[width=0.48\linewidth]{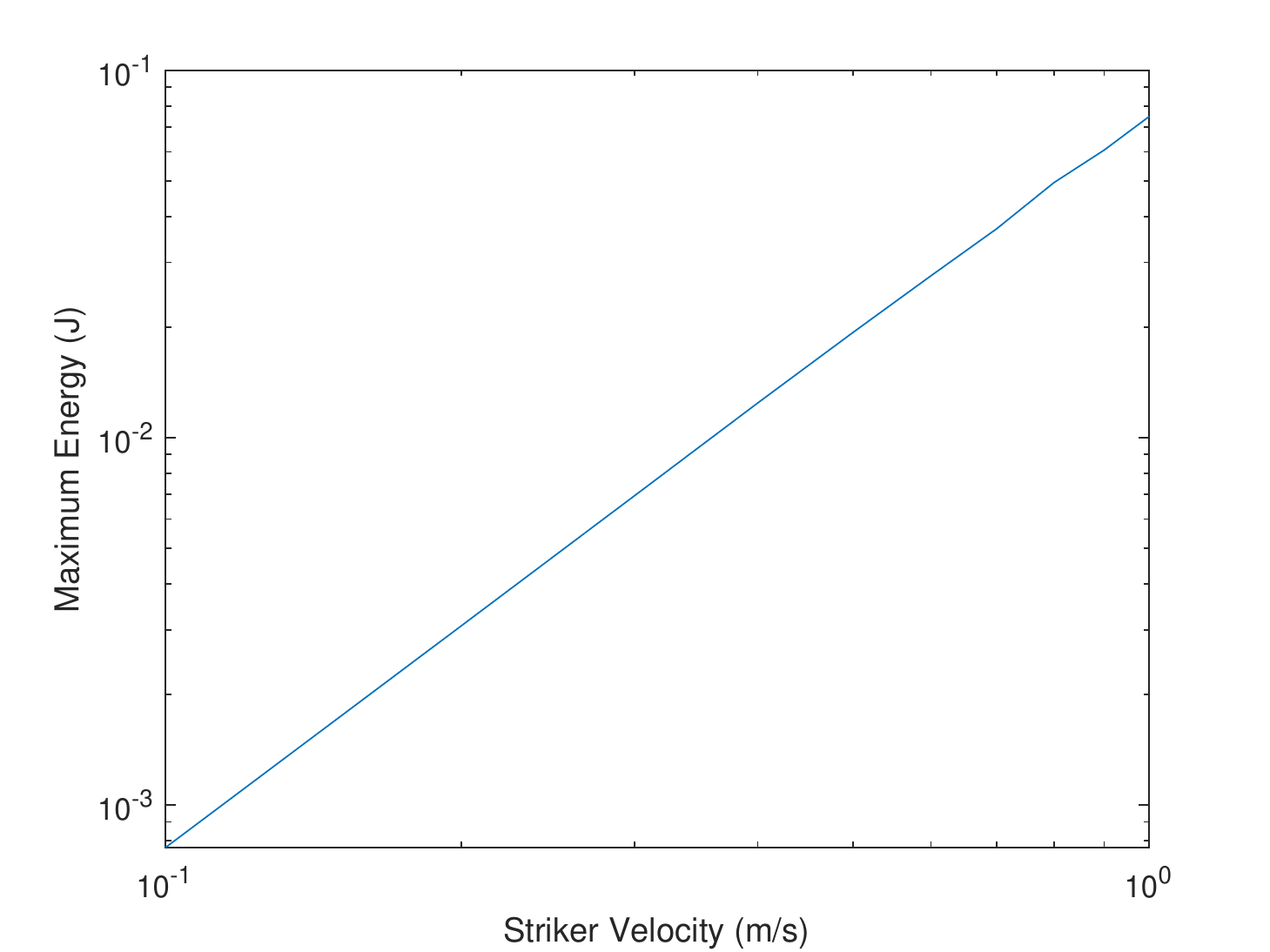}}
\caption{\label{energy} The kinetic energy evolution for a hexagonal packing of 11 spheres on each edge which is struck perpendicularly on one of the edges at $t=0$. (a) We plot total kinetic energy of the system over time for various striker velocities, $v_0$. (b) We plot the maximum total kinetic energy over the striker velocity. As expected, $\text{kinetic energy} \sim v_0^2$.}
\end{figure*}

\begin{figure}
\centering
\includegraphics[width=1\linewidth]{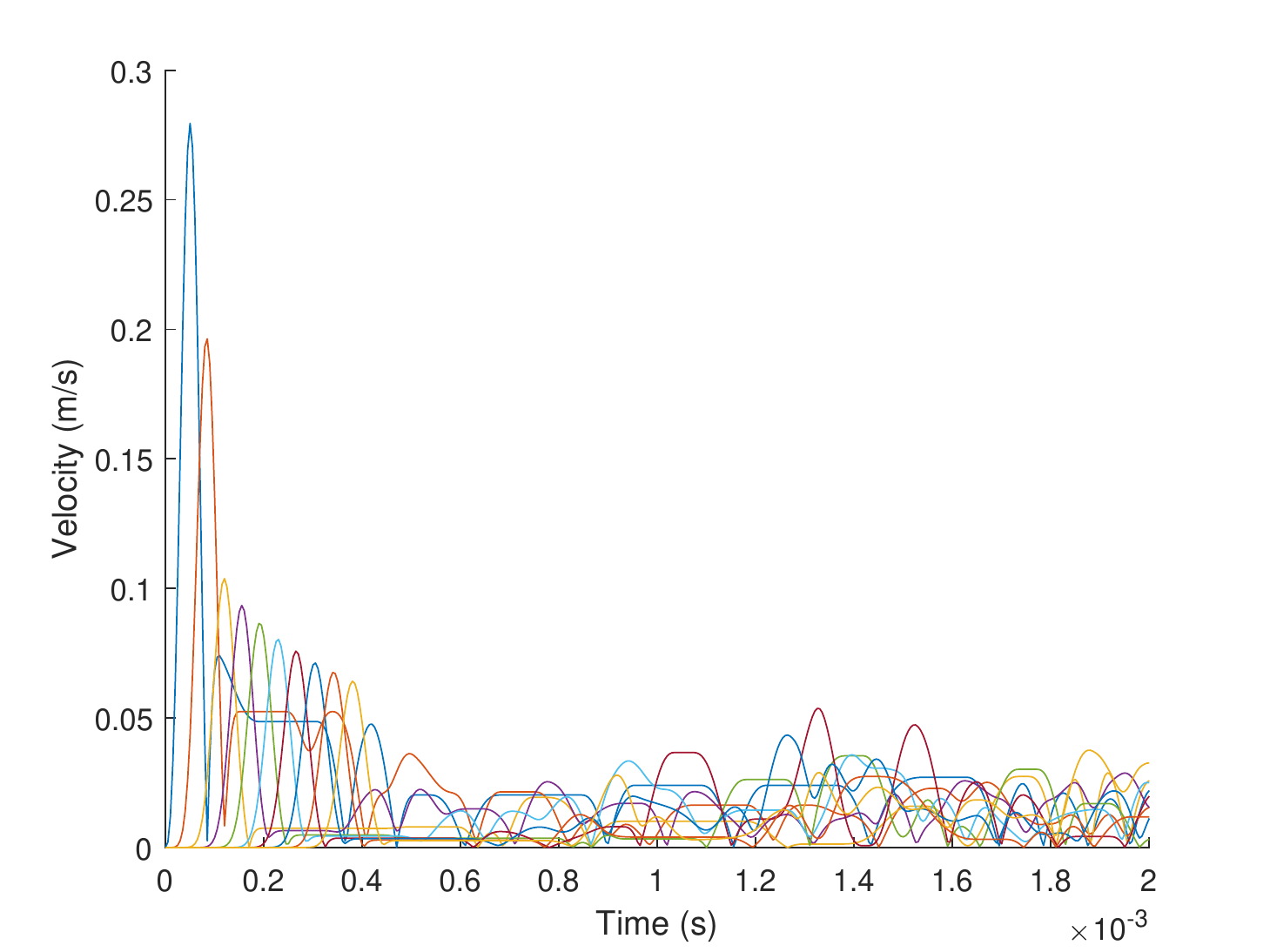}
\caption{\label{velocity} The velocity magnitude of individual spheres over time given a geometry consisting of a hexagonal packing with 11 spheres on each edge which is struck perpendicularly on one of the edges at time $t=0$. The velocity of the striker is taken as 0.4m/s. }
\end{figure}

\begin{figure}
\centering
\includegraphics[width=1\linewidth]{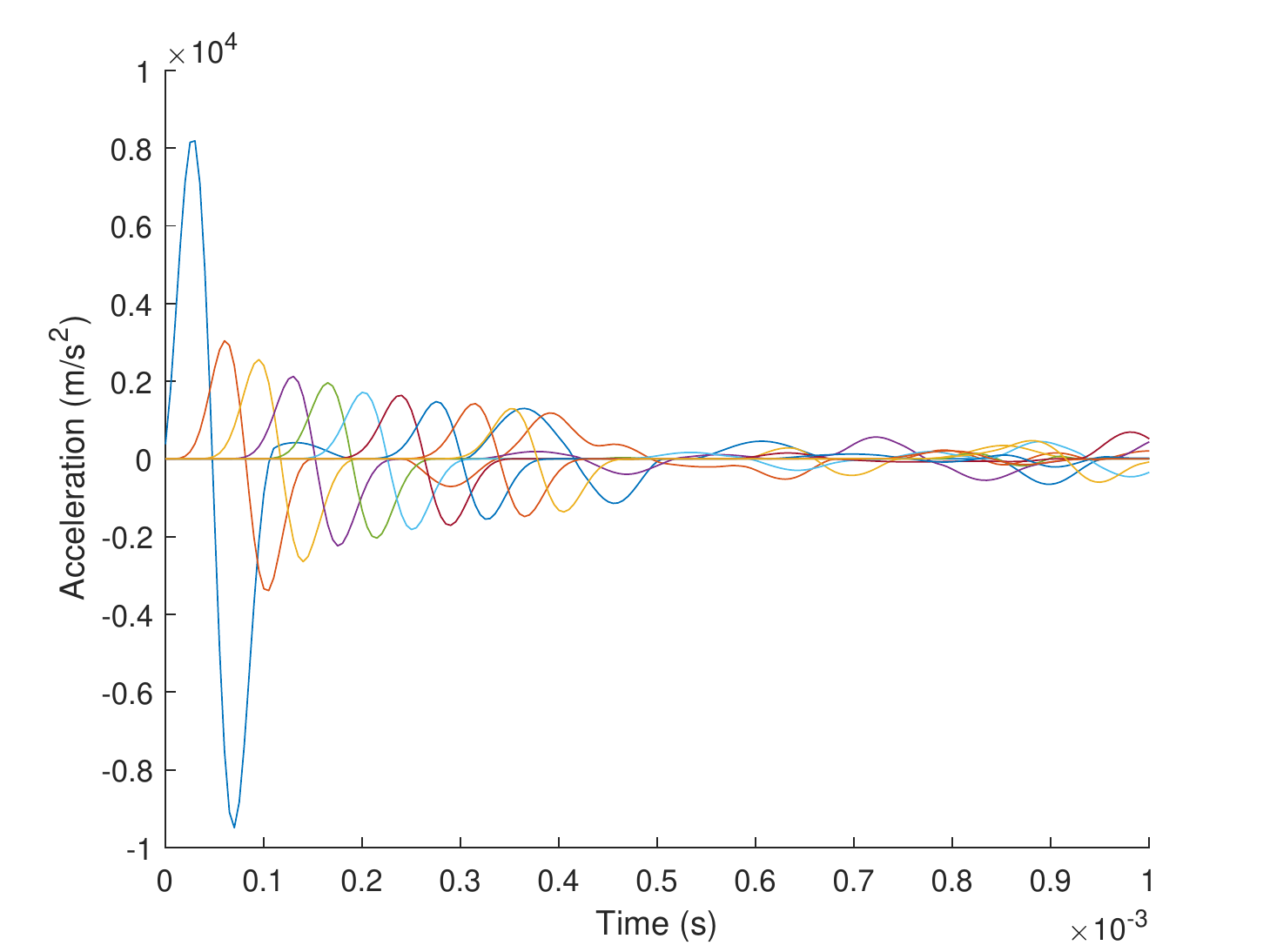}
\caption{\label{accel} The acceleration relative to the strike direction of individual spheres over time given a geometry consisting of a hexagonal packing with 11 spheres on each edge which is struck perpendicularly on one of the edges at time $t=0$.  The velocity of the striker is taken as 0.4m/s.}
\end{figure}

In Figs. \ref{wavespeed} - \ref{accel}, summary results of a perpendicular strike on a hexagonal packing is presented. The wave front speed shown in Fig. \ref{wavespeed} again shows a log-like increase in striker velocity, but the speed is slower compared to a perpendicular strike on a square packing (we omit the numerical simulations for that case). This is the result of lost energy in other directions. The kinetic energy over time is shown in Fig. \ref{energy} as a function of initial striker velocity. From Figs. \ref{velocity} - \ref{accel}, which show individual bead velocities and accelerations, we can see the patterns of motions of spheres on the direction of the strike. The spheres will first be activated with a solitary wave, and the velocities will drop to almost zero when the solitary wave has passed. There are other waves following the wave front, and many of the waves are solitary, which can be confirmed by the pulses of velocities after $t=0.5$ms.

\begin{figure*}
\centering
\subfigure[]{\includegraphics[width=0.32\linewidth]{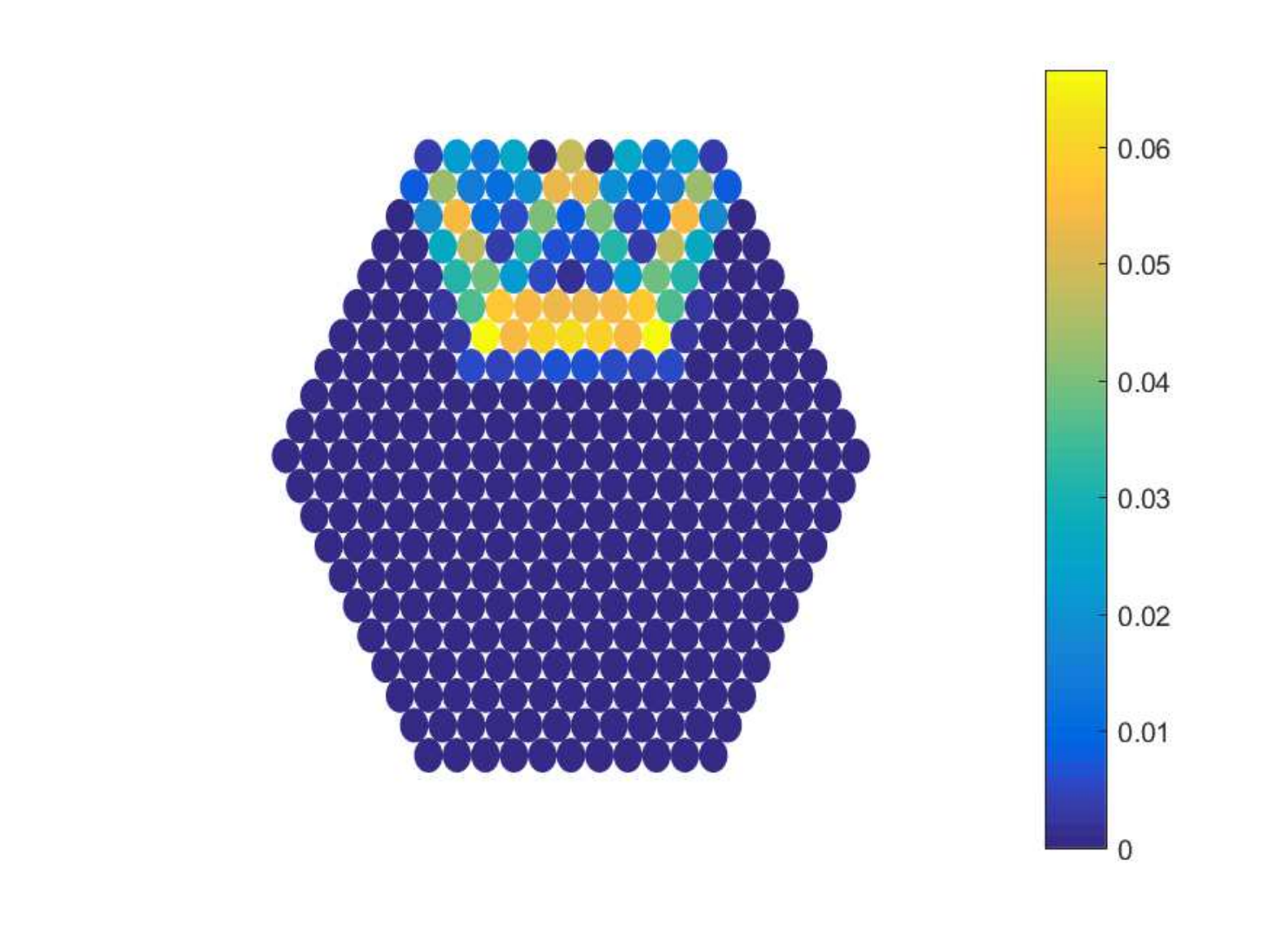}}
\subfigure[]{\includegraphics[width=0.32\linewidth]{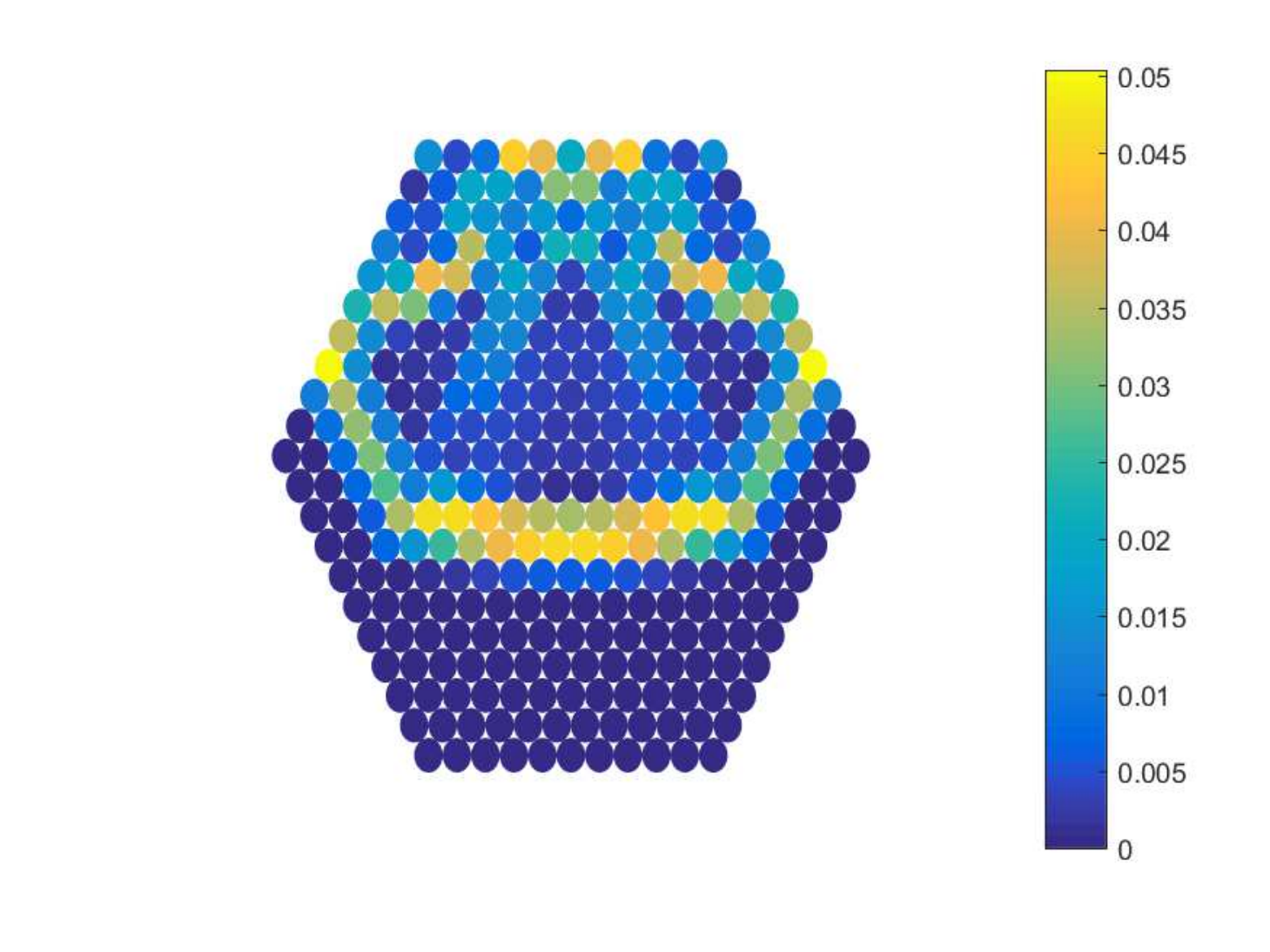}}
\subfigure[]{\includegraphics[width=0.32\linewidth]{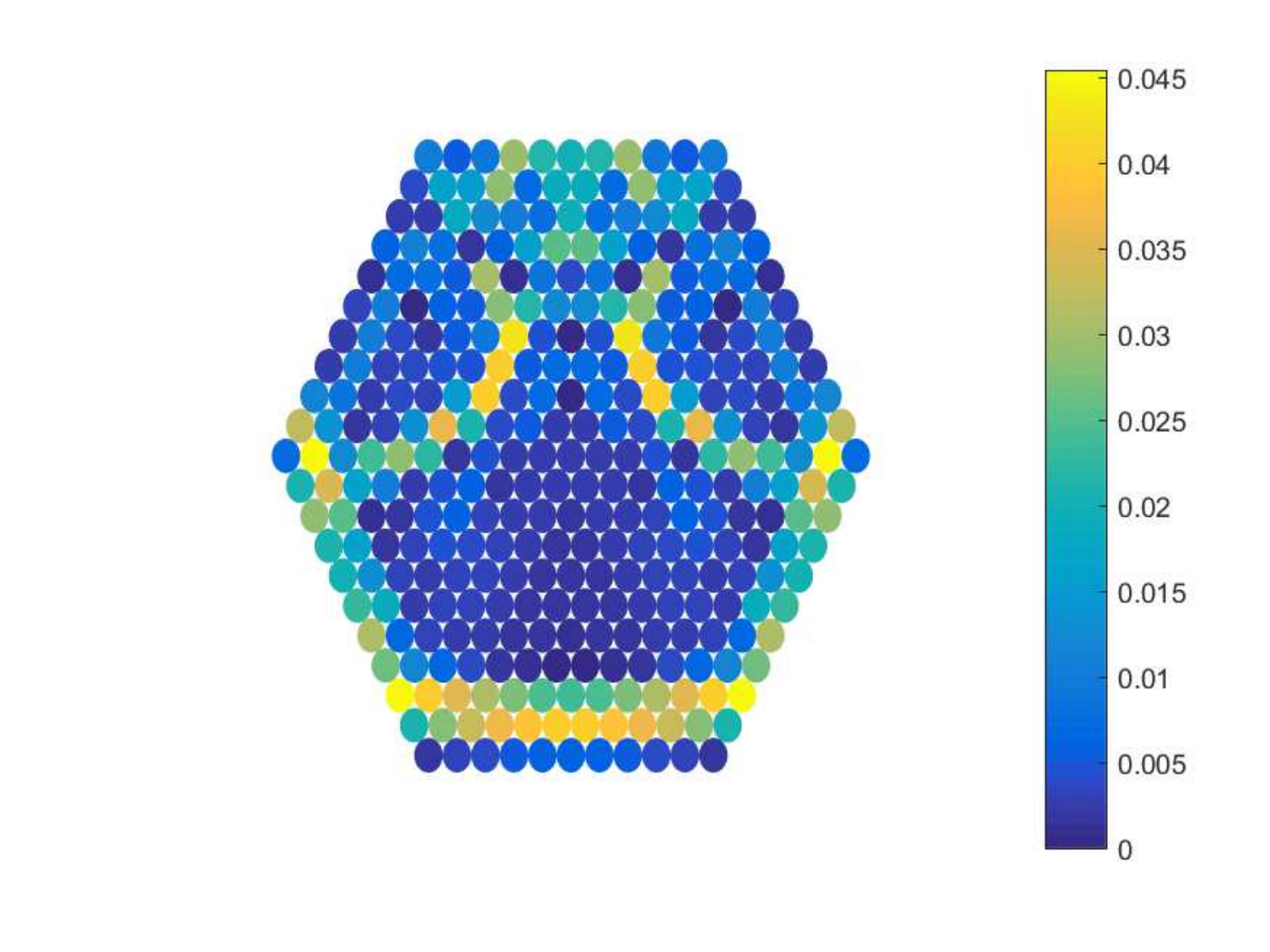}}
\subfigure[]{\includegraphics[width=0.32\linewidth]{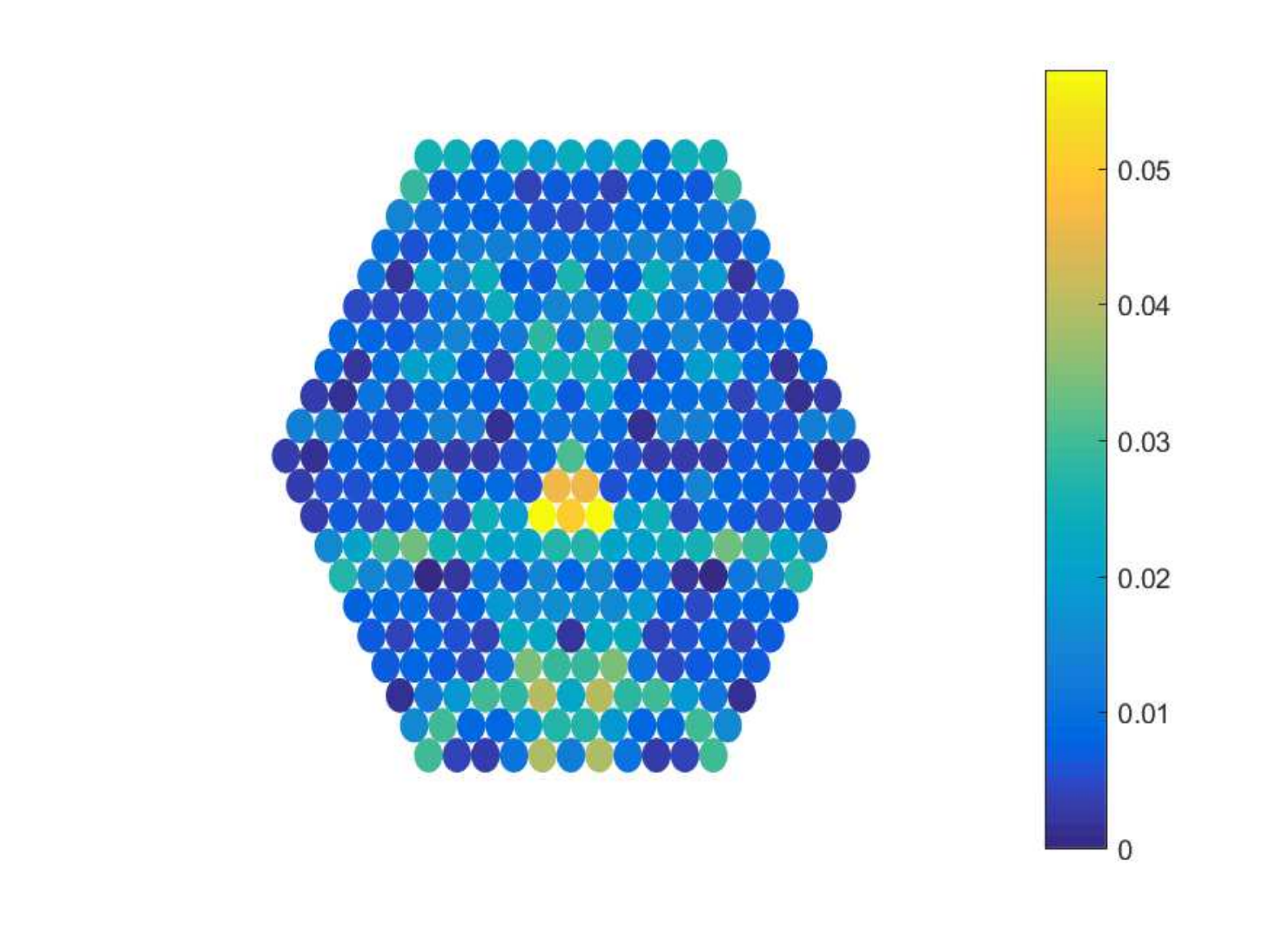}}
\subfigure[]{\includegraphics[width=0.32\linewidth]{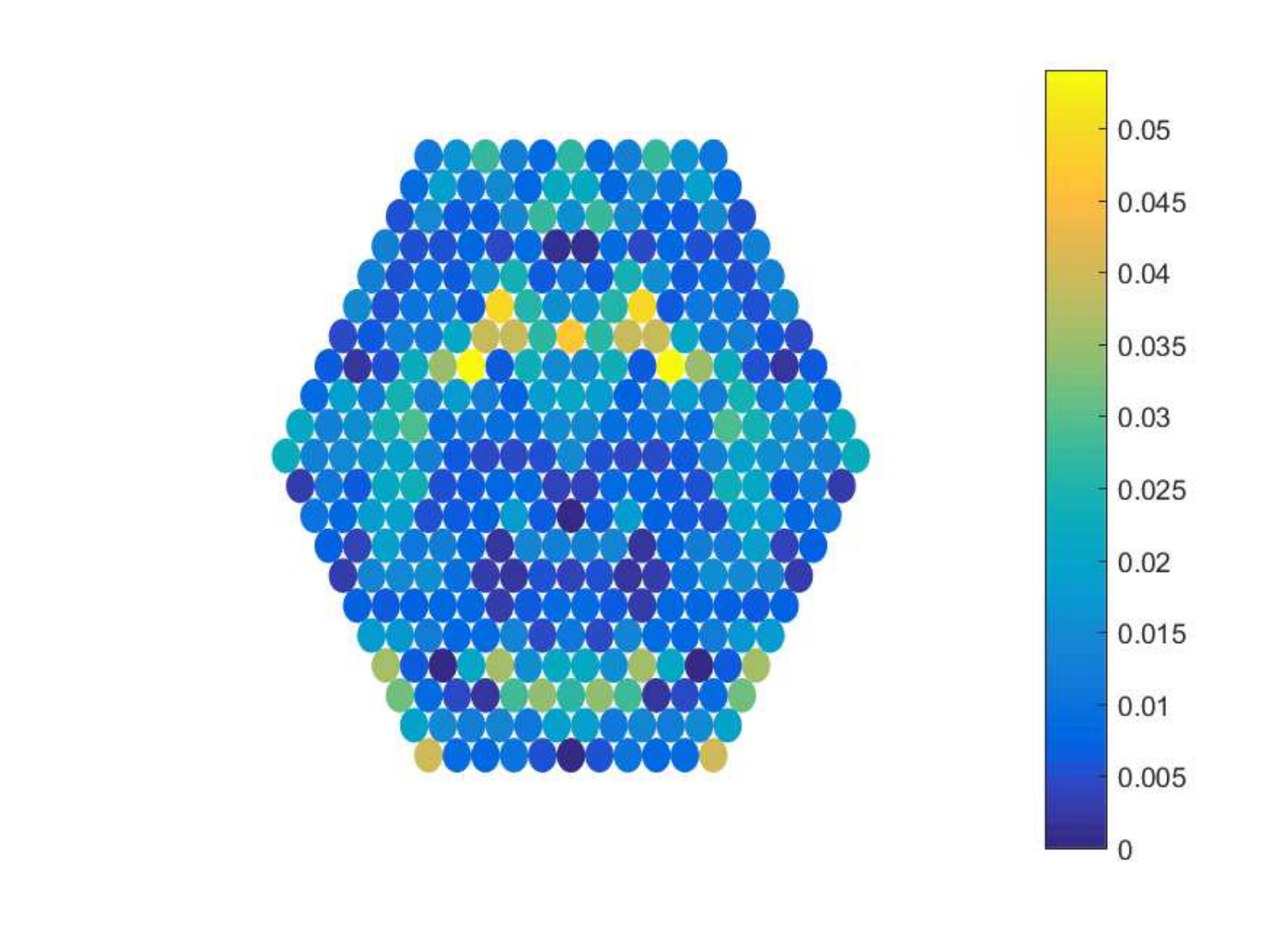}}
\subfigure[]{\includegraphics[width=0.32\linewidth]{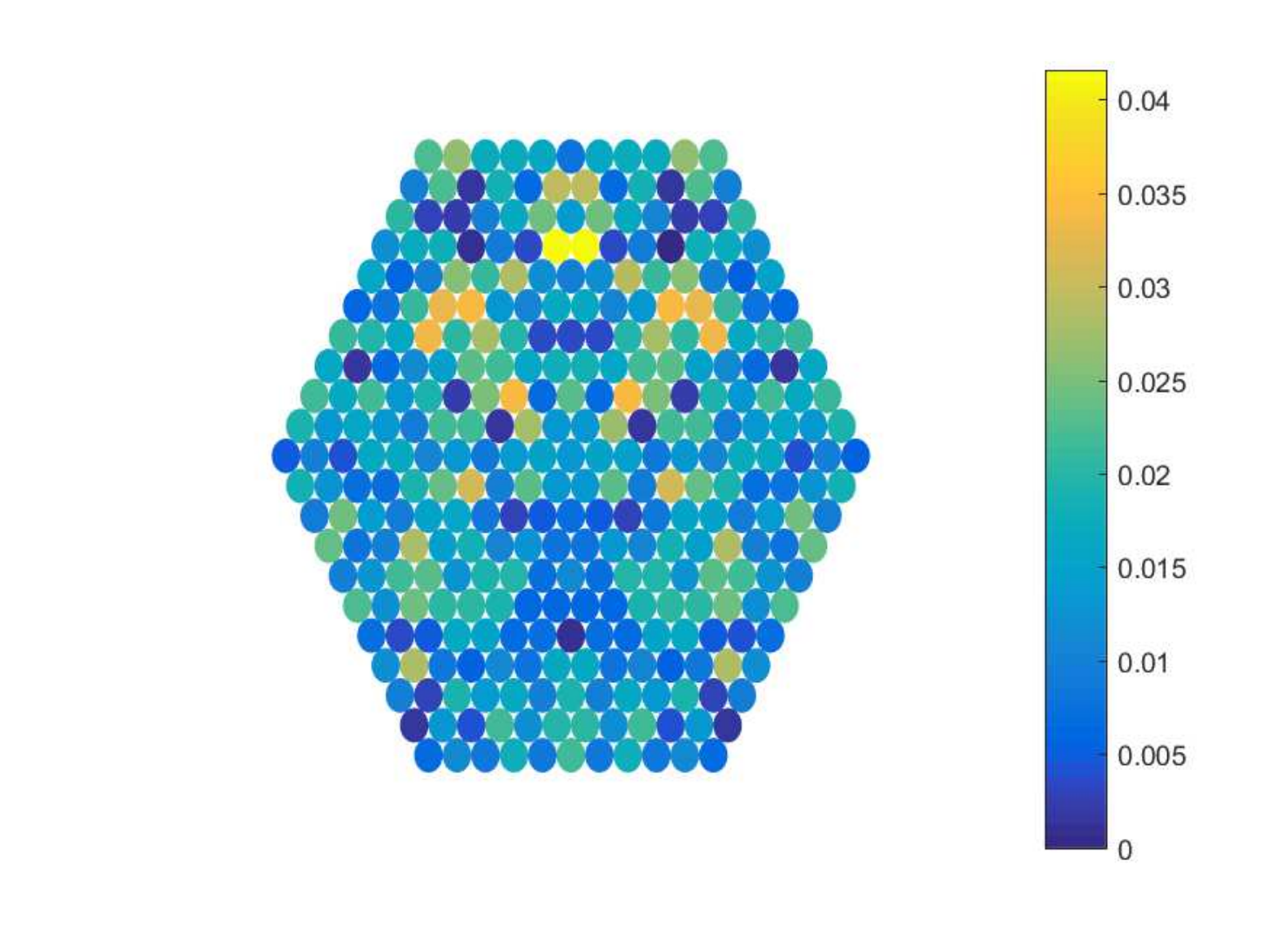}}
\vspace{-0.2in}
\caption{\label{WaveHexUpOnestriker} Velocity magnitudes at each sphere as a function of time for a hexagonal basin filled with a hexagonal packing of spheres. The packing has 11 spheres on each edge and is struck perpendicularly on one of the edge. The initial velocity of the striker is $0.4m/s$. Here (a) $t=0.275$ms, (b) $t=0.525$ms, (c) $t=0.775$ms, (d) $t=1.20$ms, (e) $t=1.50$ms, (f) $t=10.0$ms. Panels (a) and (b) coincide with results from \cite{Leo14}. Note that (a)-(c) are before the main wave reflection, while (d)-(f) are after the main wave reflection.}
\end{figure*}

In Fig. \ref{WaveHexUpOnestriker}, we can see that the solitary wave caused by the striker is propagating in three directions. The shape of the wave front looks similar to part of a hexagon. One of the wave fronts is traveling toward the lower boundary, and the other two are propagating toward the left and right boundaries. The latter two wave fronts will first hit the boundaries, and are reflected. After the reflection of the wave front originally heading downward, the three reflected wave fronts will approach the same location. For larger time, the system will fall into disorder. These results are in complete agreement with the behaviors observed in \cite{Leo14}.

\begin{figure*}
\centering
\subfigure[]{\includegraphics[width=0.32\linewidth]{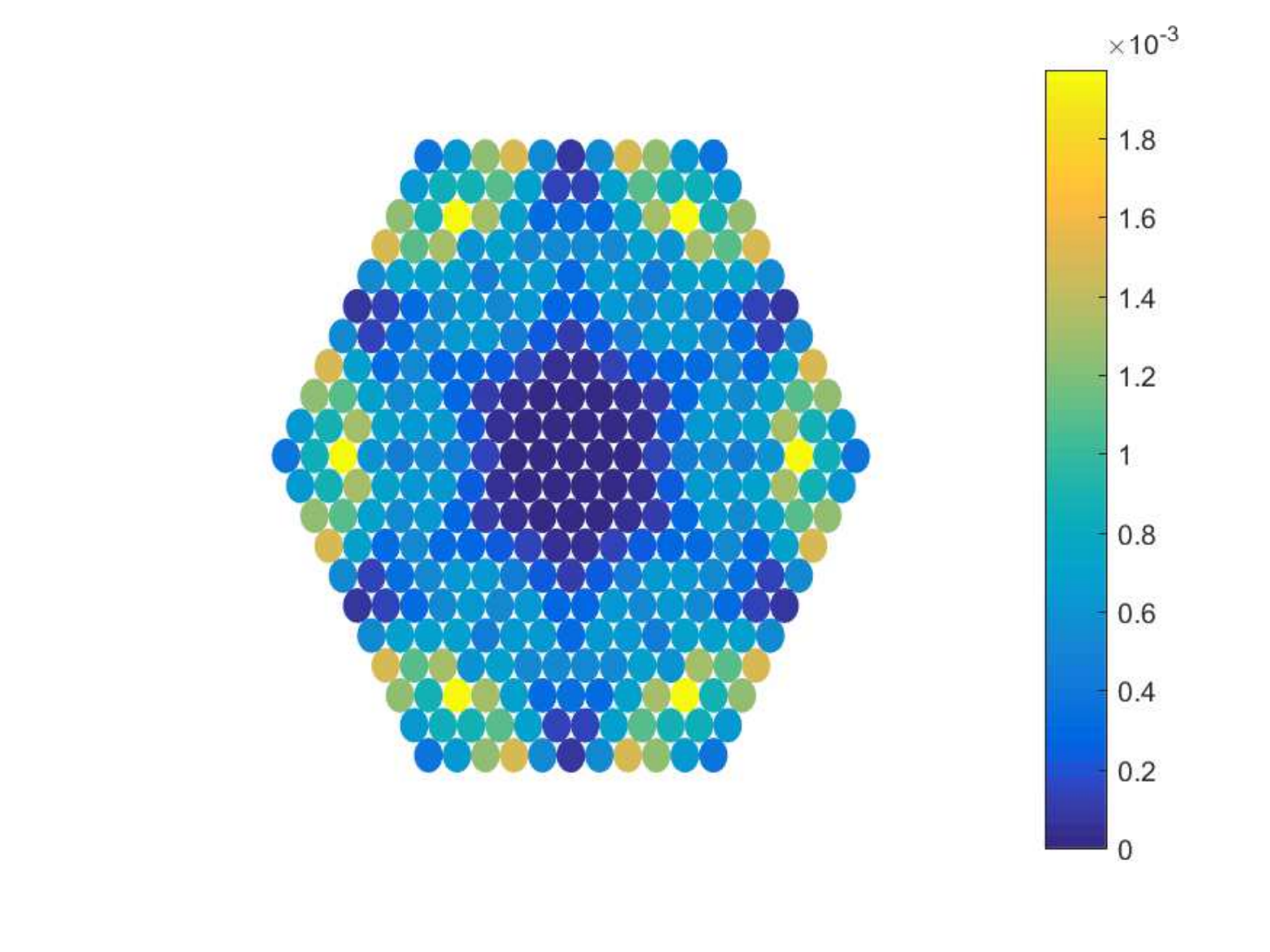}}
\subfigure[]{\includegraphics[width=0.32\linewidth]{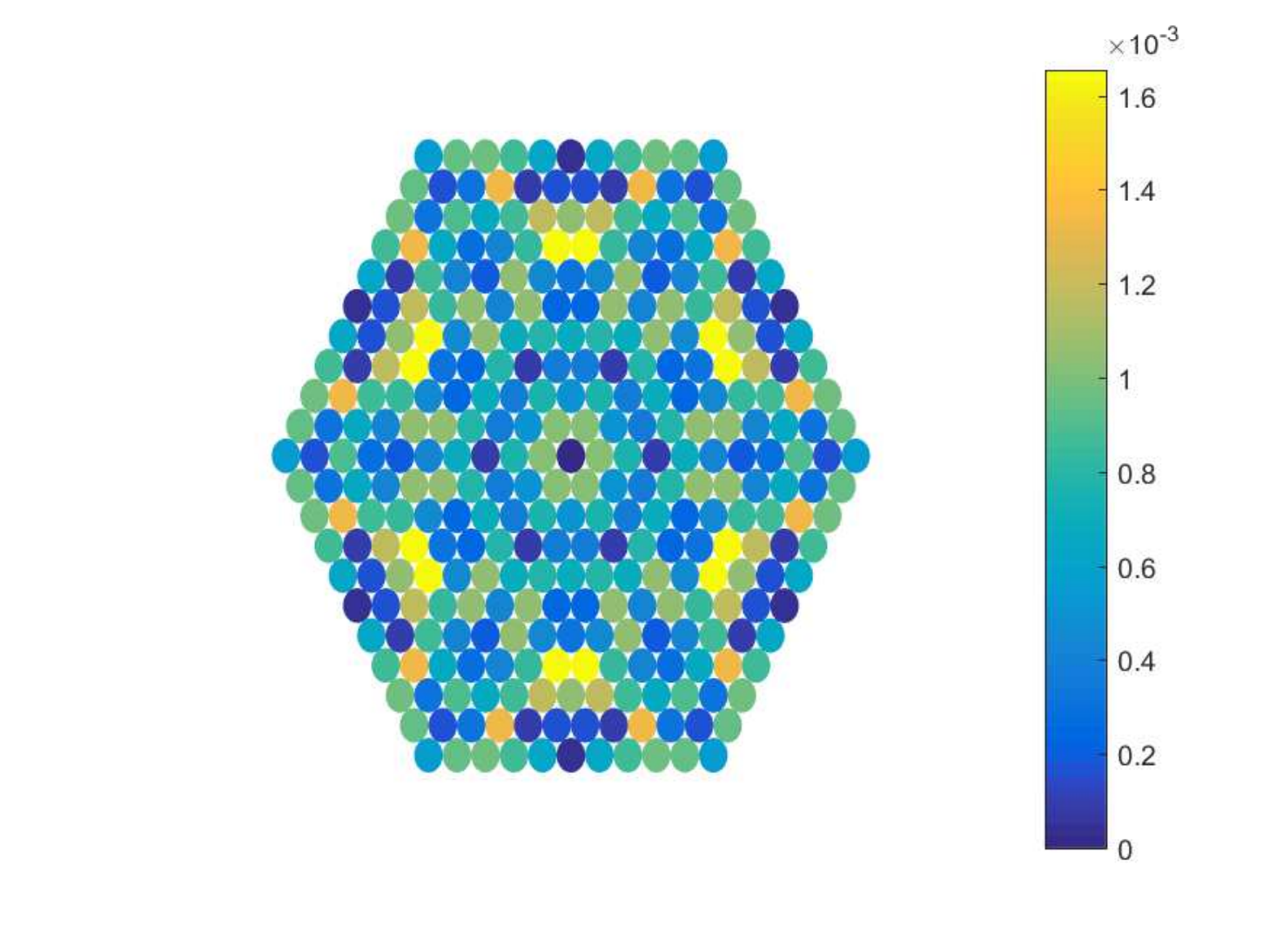}}
\subfigure[]{\includegraphics[width=0.32\linewidth]{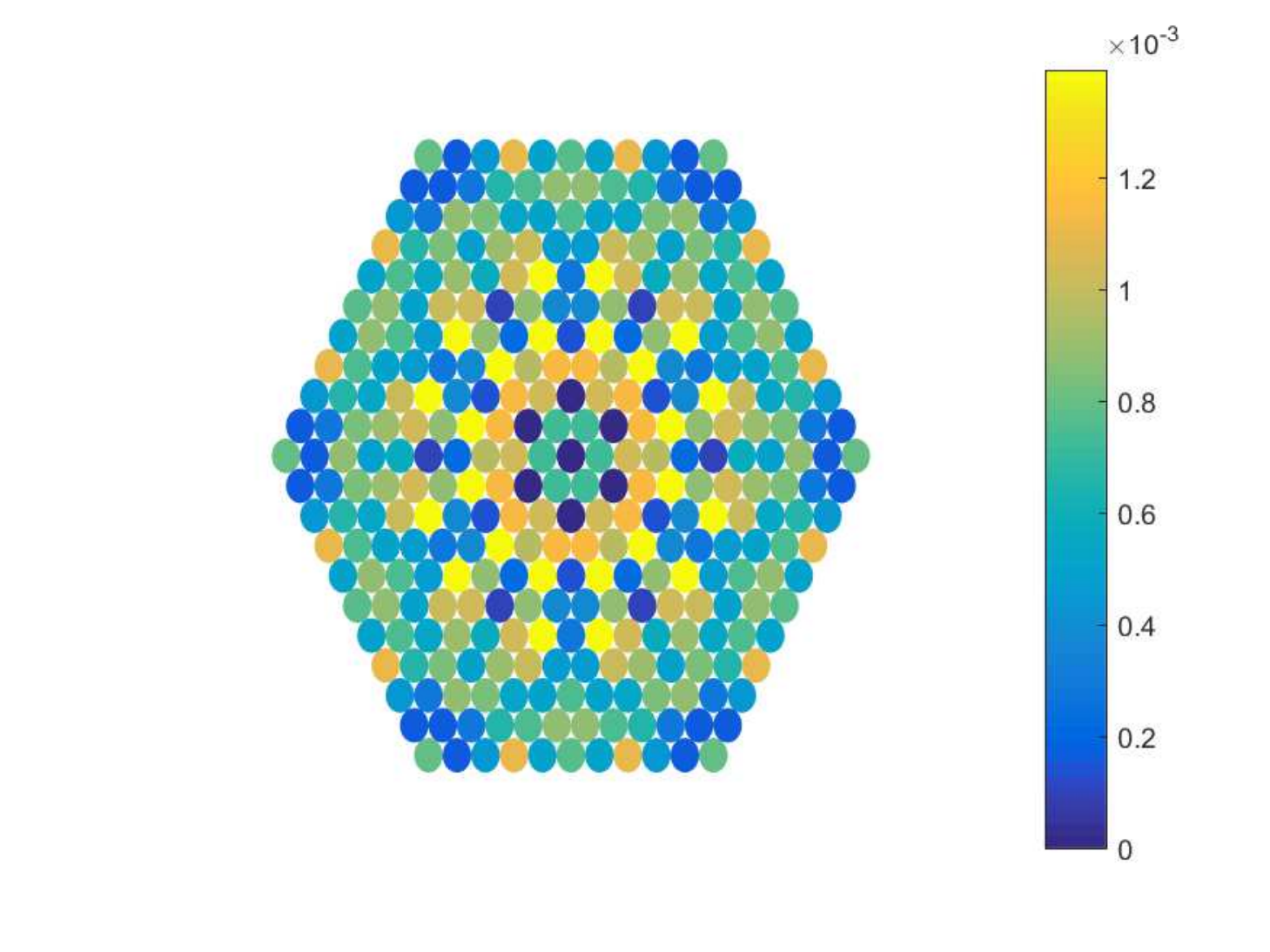}}
\subfigure[]{\includegraphics[width=0.32\linewidth]{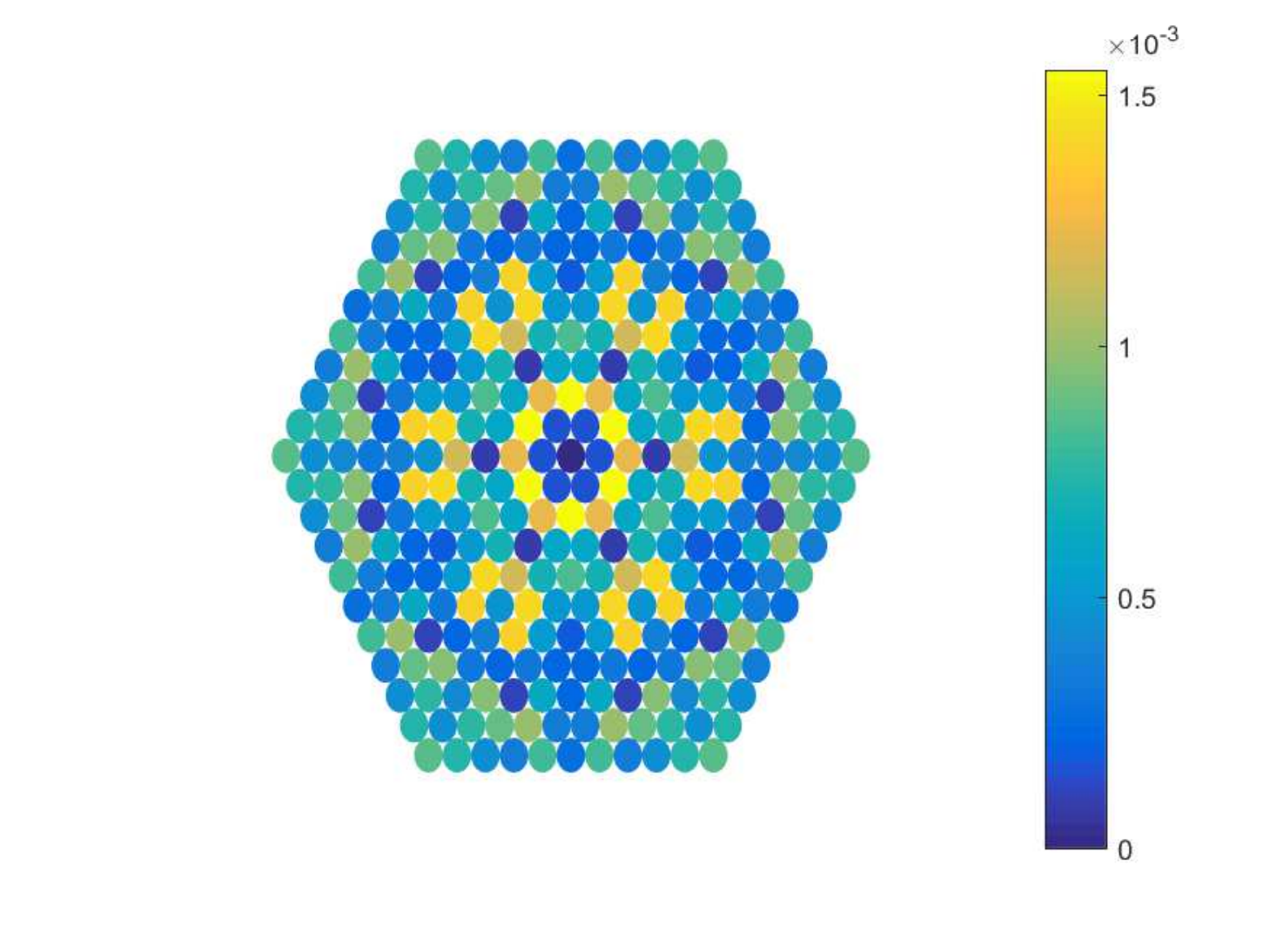}}
\subfigure[]{\includegraphics[width=0.32\linewidth]{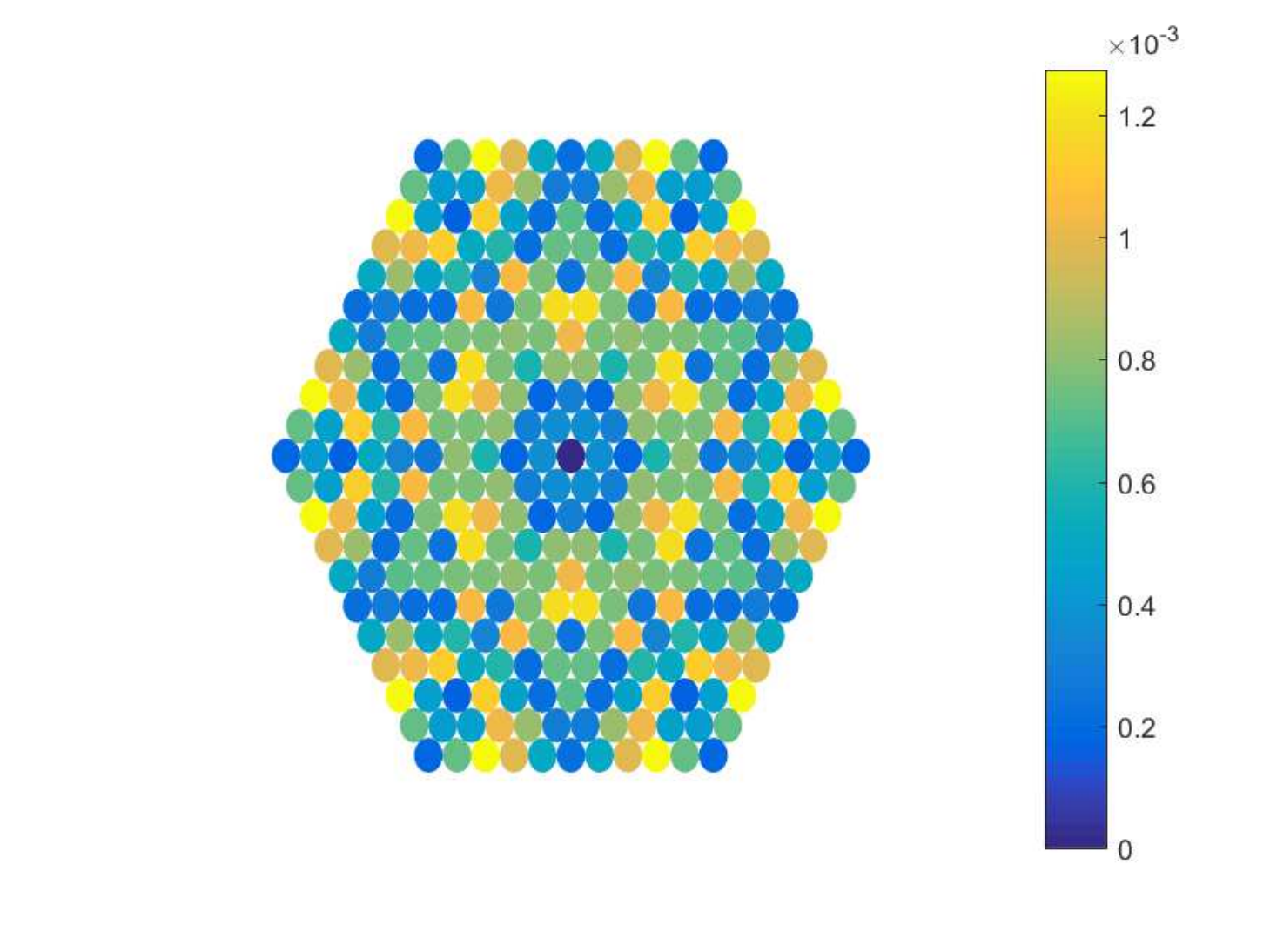}}
\subfigure[]{\includegraphics[width=0.32\linewidth]{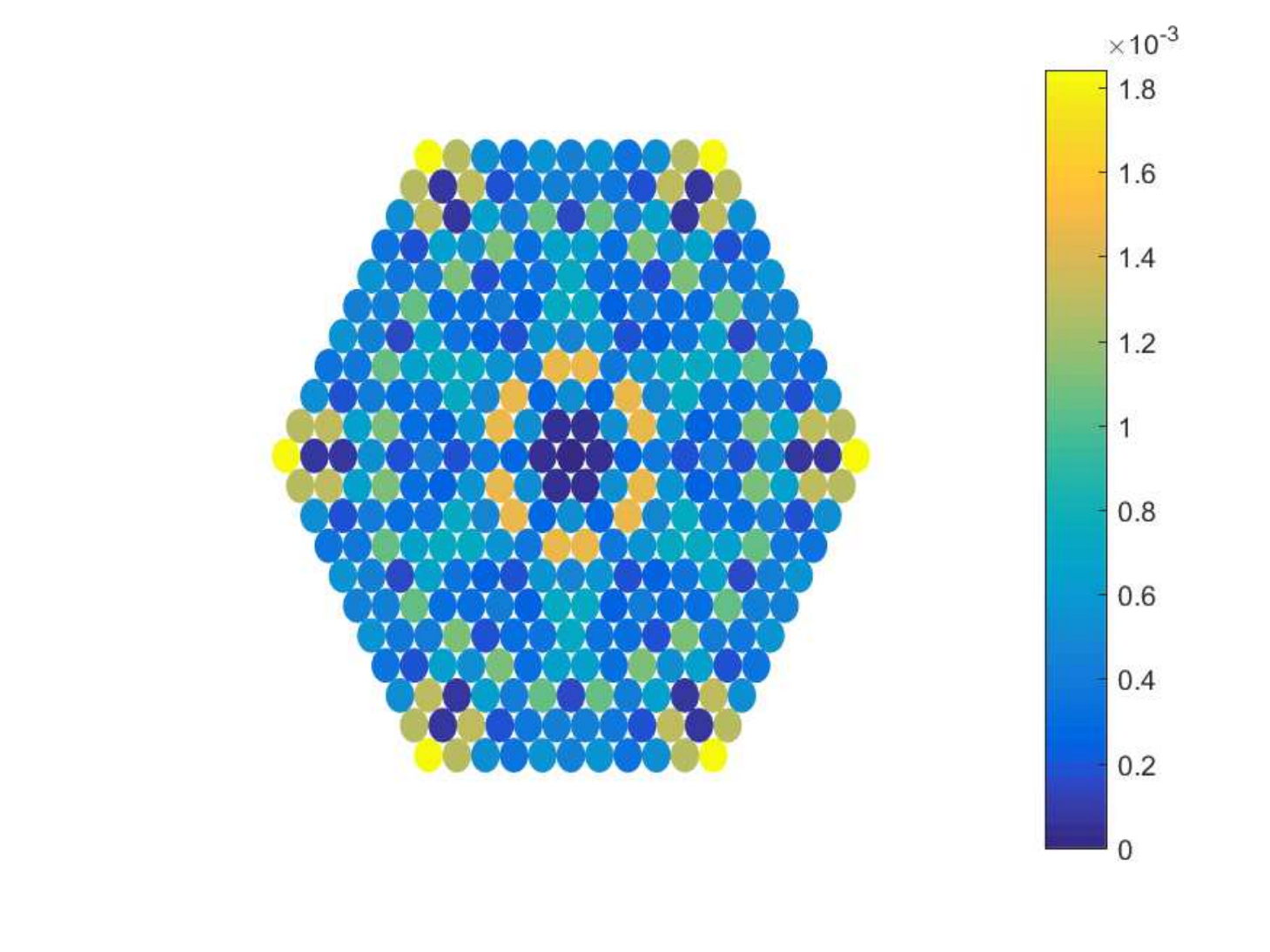}}
\vspace{-0.2in}
\caption{\label{WaveHexNoStriker} Velocity magnitudes at each sphere as a function of time for a hexagonal basin filled with a hexagonal packing of spheres. The packing is compressed so that the deformation of an interior sphere will be $10^{-4}$ of its radius. Even if the initial deformation set all the spheres in the interior and on the edges balanced, the spheres at the corners will be unbalanced and generate oscillating waves. Here (a) $t=0.275$ms, (b) $t=0.600$ms, (c) $t=1.00$ms, (d) $t=1.20$ms, (e) $t=4.00$ms, (f) $t=10.0$ms. }
\end{figure*}

\begin{figure*}
\centering
\subfigure[]{\includegraphics[width=0.32\linewidth]{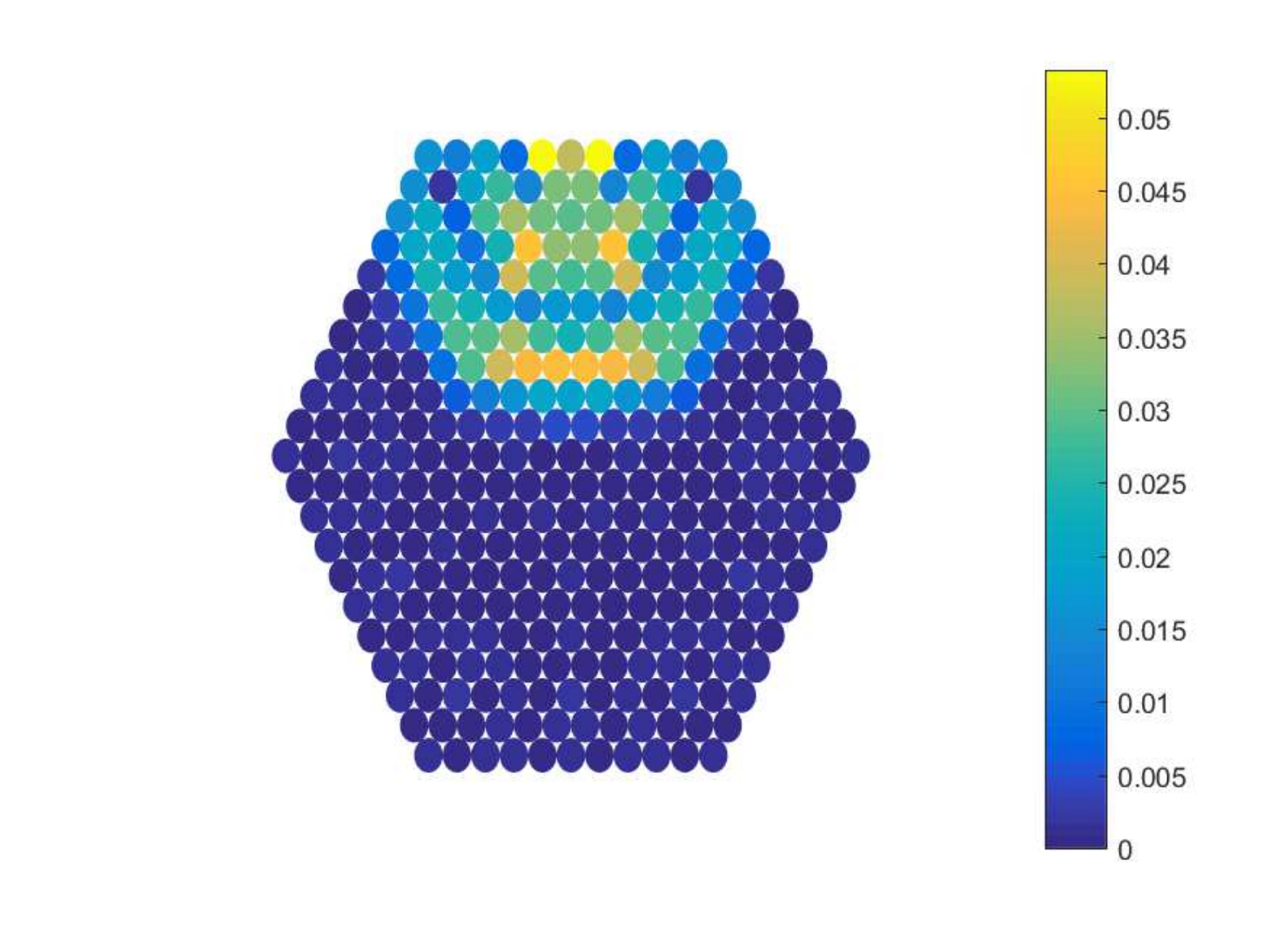}}
\subfigure[]{\includegraphics[width=0.32\linewidth]{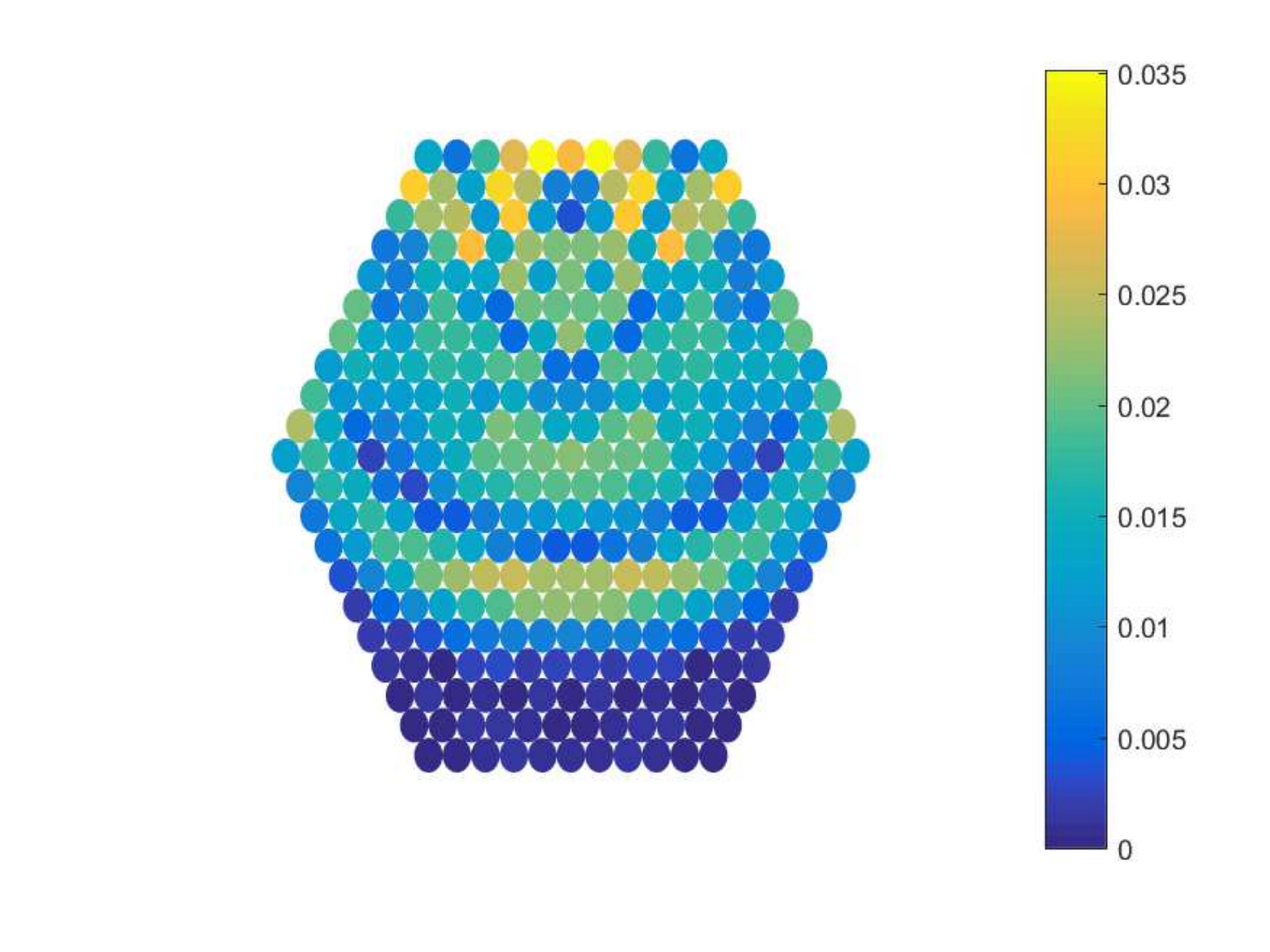}}
\subfigure[]{\includegraphics[width=0.32\linewidth]{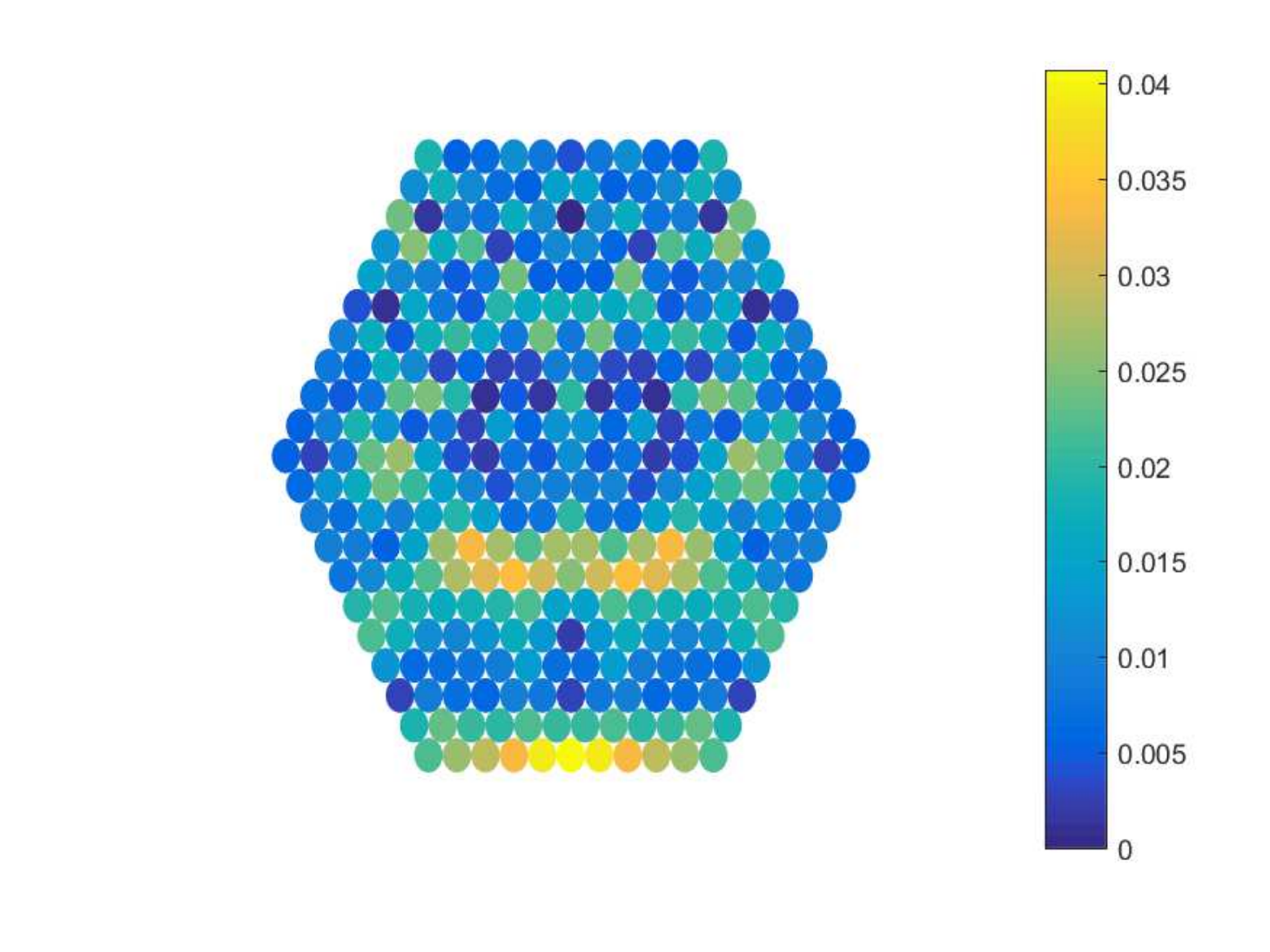}}
\subfigure[]{\includegraphics[width=0.32\linewidth]{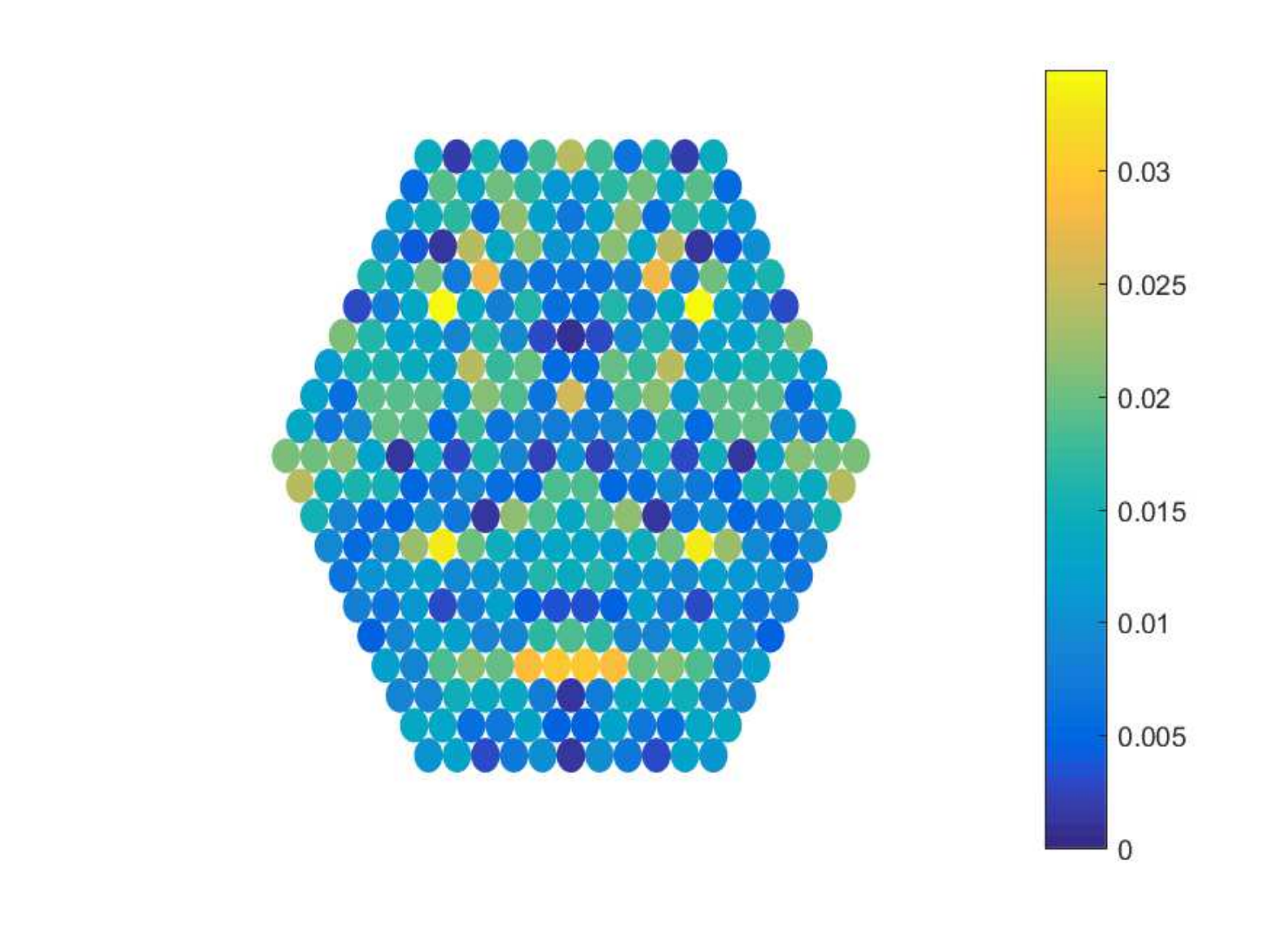}}
\subfigure[]{\includegraphics[width=0.32\linewidth]{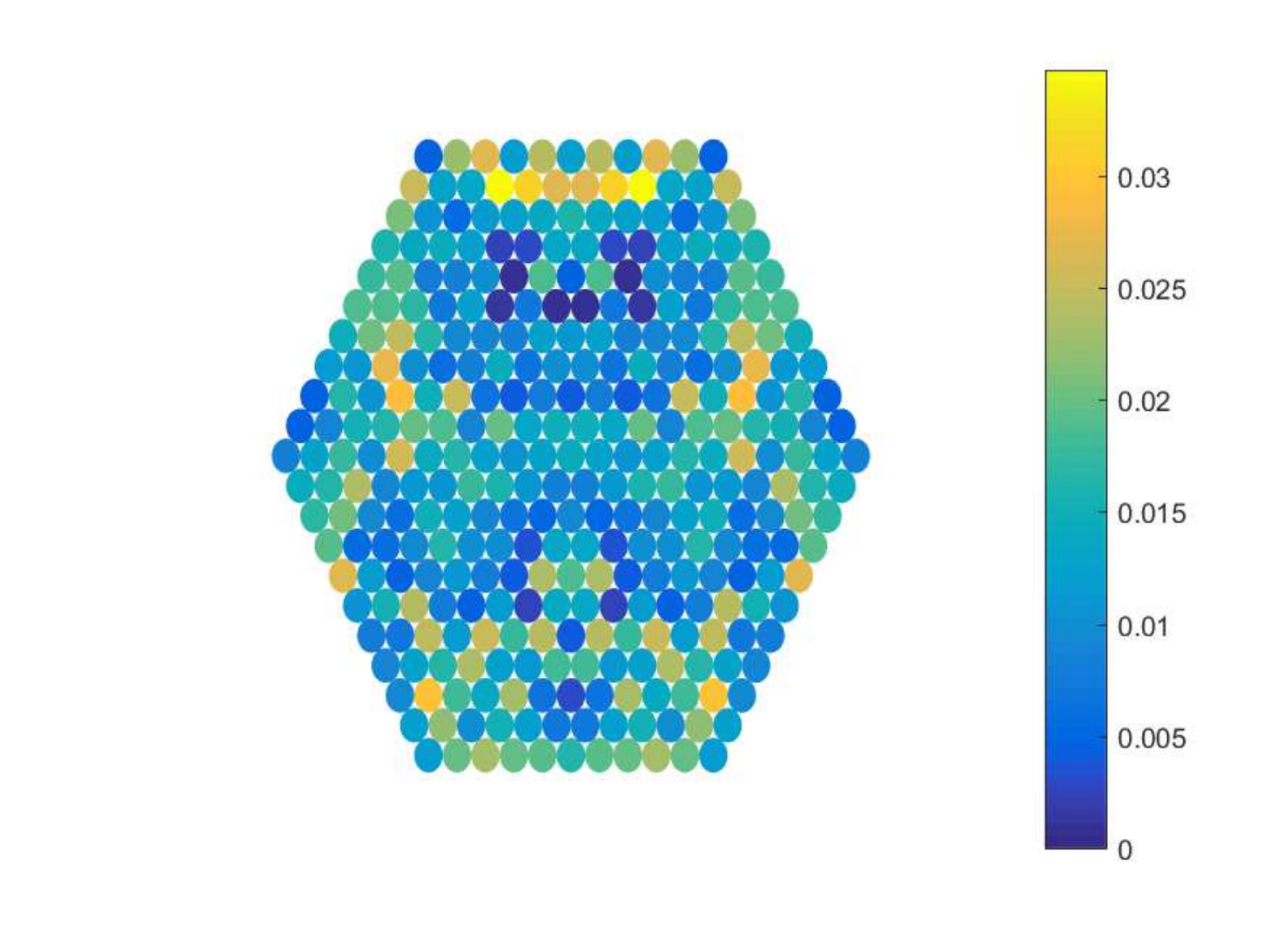}}
\subfigure[]{\includegraphics[width=0.32\linewidth]{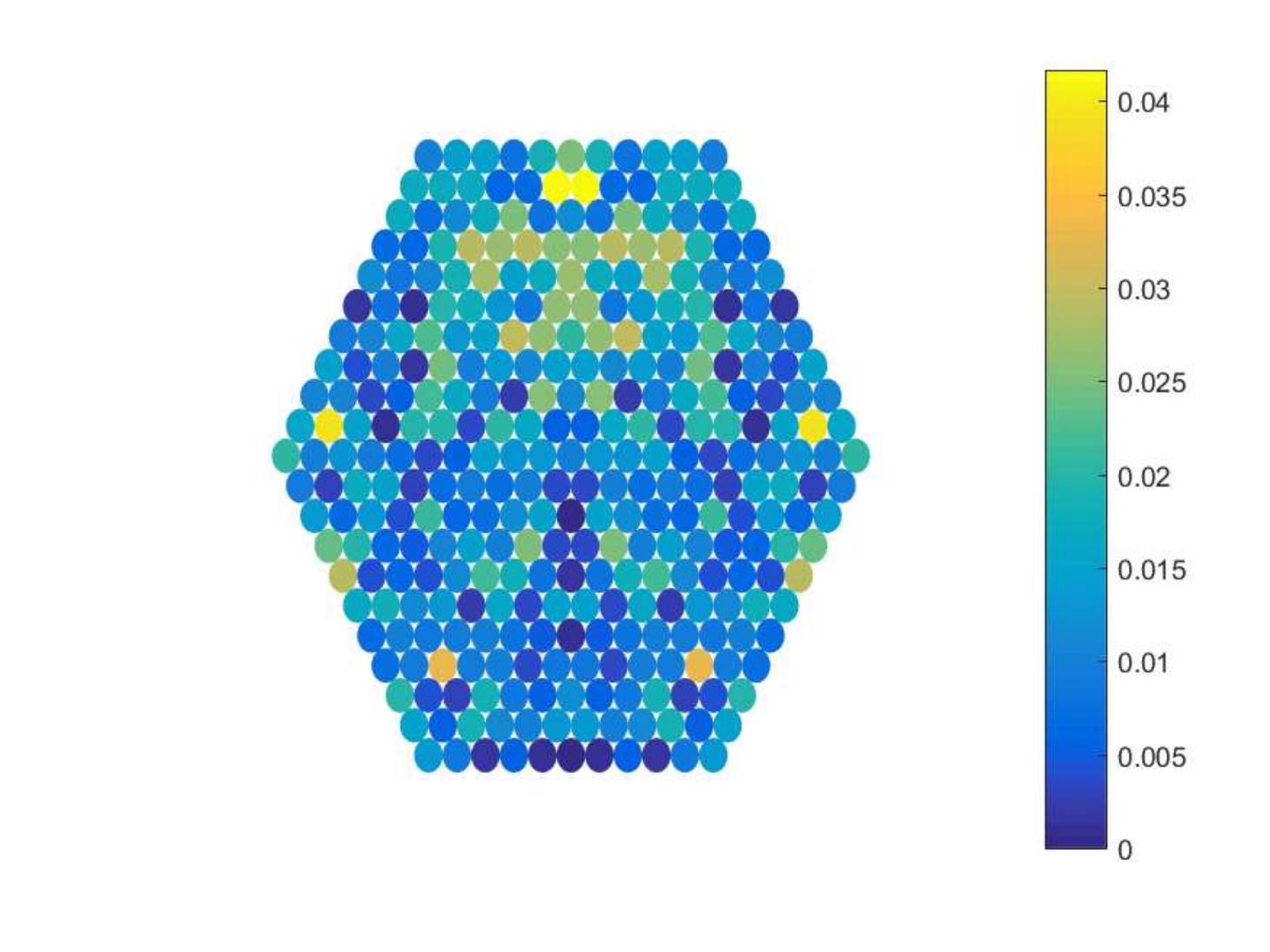}}
\vspace{-0.2in}
\caption{\label{WaveHexCompressed} Velocity magnitudes at each sphere as a function of time for a hexagonal basin filled with a hexagonal packing of spheres. The packing is compressed so that the deformation of an interior sphere will be $10^{-4}$ of its radius. The system is let alone for 10ms and then hit by a vertical striker with velocity $0.4m/s$. Here (a) $t=0.275$ms, (b) $t=0.525$ms, (c) $t=1.00$ms, (d) $t=1.20$ms, (e) $t=1.88$ms, (f) $t=4.00$ms. Note that (a)-(c) are before the main wave reflection, while (d)-(f) are after the main wave reflection.}
\end{figure*}

We compare the results in Fig. \ref{WaveHexUpOnestriker} with a case where the domain is slightly compressed rather than impacted by a striker at the boundary. For square packing, it is possible to find an initial state where a uniform deformation is applied to the contacts between spheres, and work out a corresponding deformation between a boundary sphere and the wall, so that the granular crystal is of balance. However, if we use the same approach for hexagonal packings, the six spheres at the corners will be unbalanced, and results in waves propagating in the domain. In Fig. \ref{WaveHexNoStriker} we present results for where there is a deformation of $10^{-4}$ of the radius between two interior spheres. We can see the waves propagating from each corner to the centre in a symmetric manner. We see that after 10ms, there are still waves travelling in the domain, and the maximum velocity is still similar to the maximum velocity in the beginning. The nature of the waves is different from the case where no precompression is applied, as no qualitative change takes place in the wave propagation. Now, if we apply a boundary striker after some time, we can see the propagation of waves due to the striker superimposed on the waves due to precompression; see Fig. \ref{WaveHexCompressed}. This results in a smoothing of the waves, compared with Fig. \ref{WaveHexUpOnestriker}. However, as the velocity of waves generated from the unbalance will be at least the same order as waves caused by the striker under larger precompression, we do not look into the precompressed case further in this paper, as it will not be useful to study waves due to striker impact superimposed on waves due to precompression. 

\begin{figure*}
\centering
\subfigure[]{\includegraphics[width=0.32\linewidth]{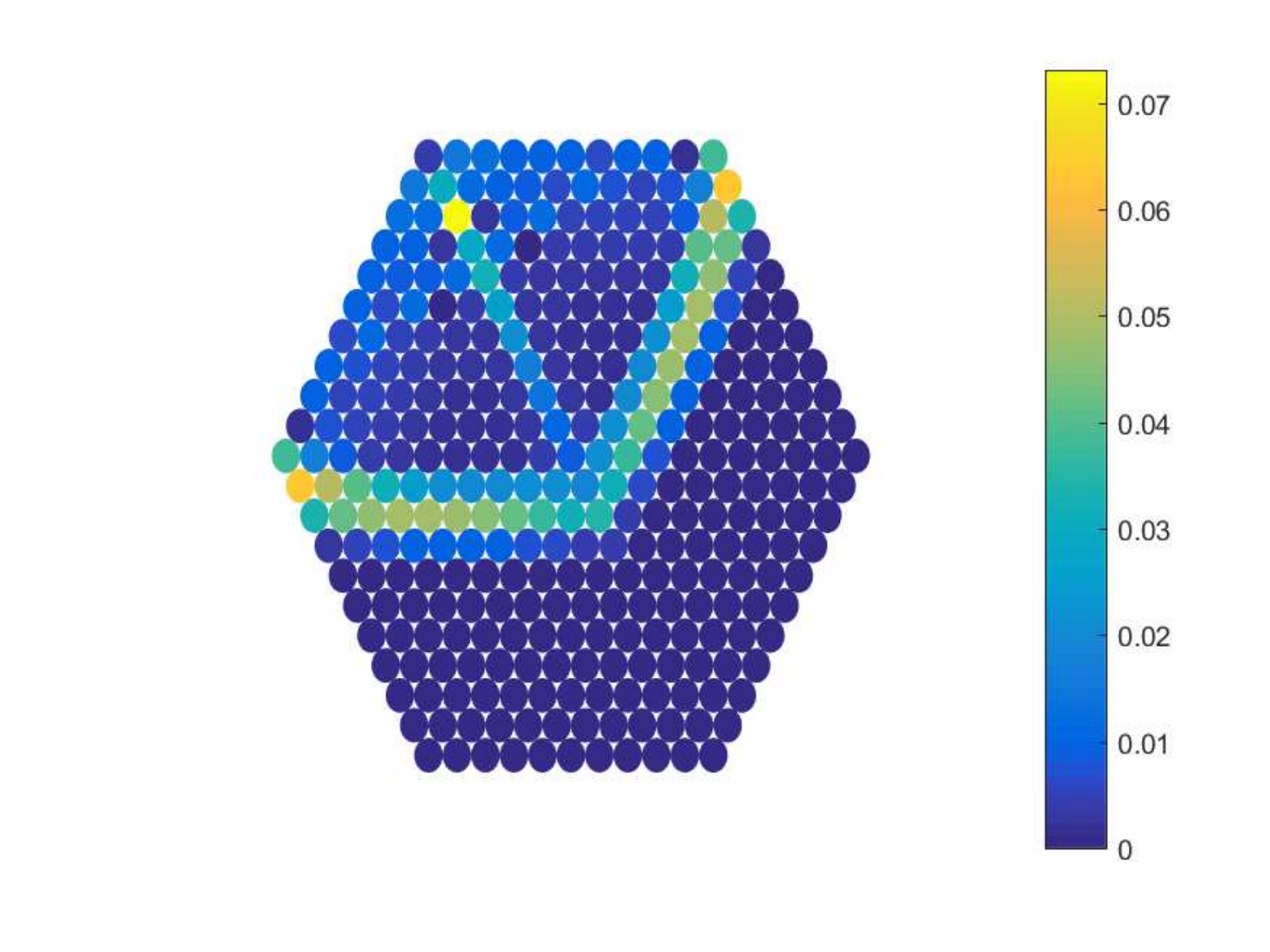}}
\subfigure[]{\includegraphics[width=0.32\linewidth]{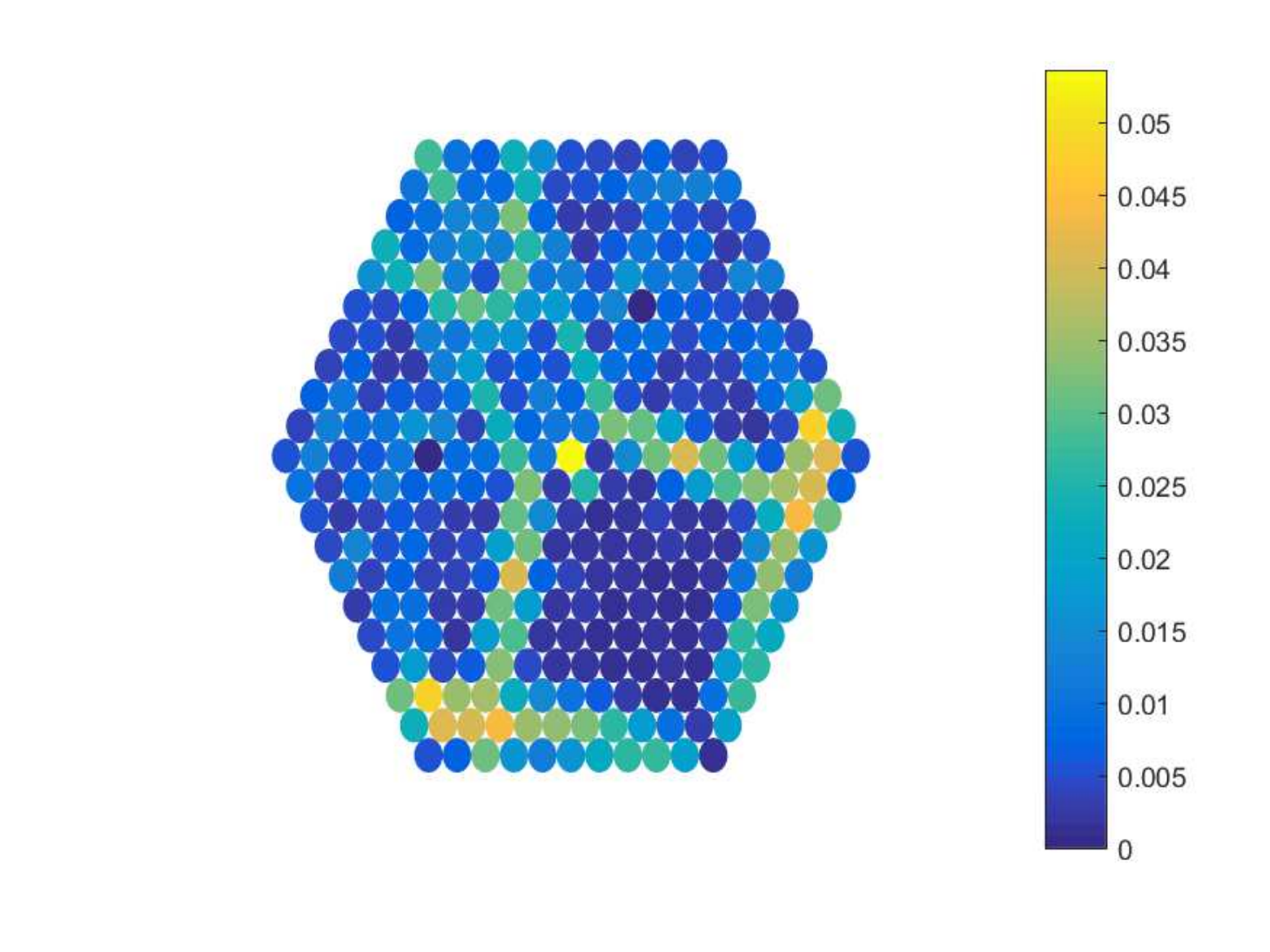}}
\subfigure[]{\includegraphics[width=0.32\linewidth]{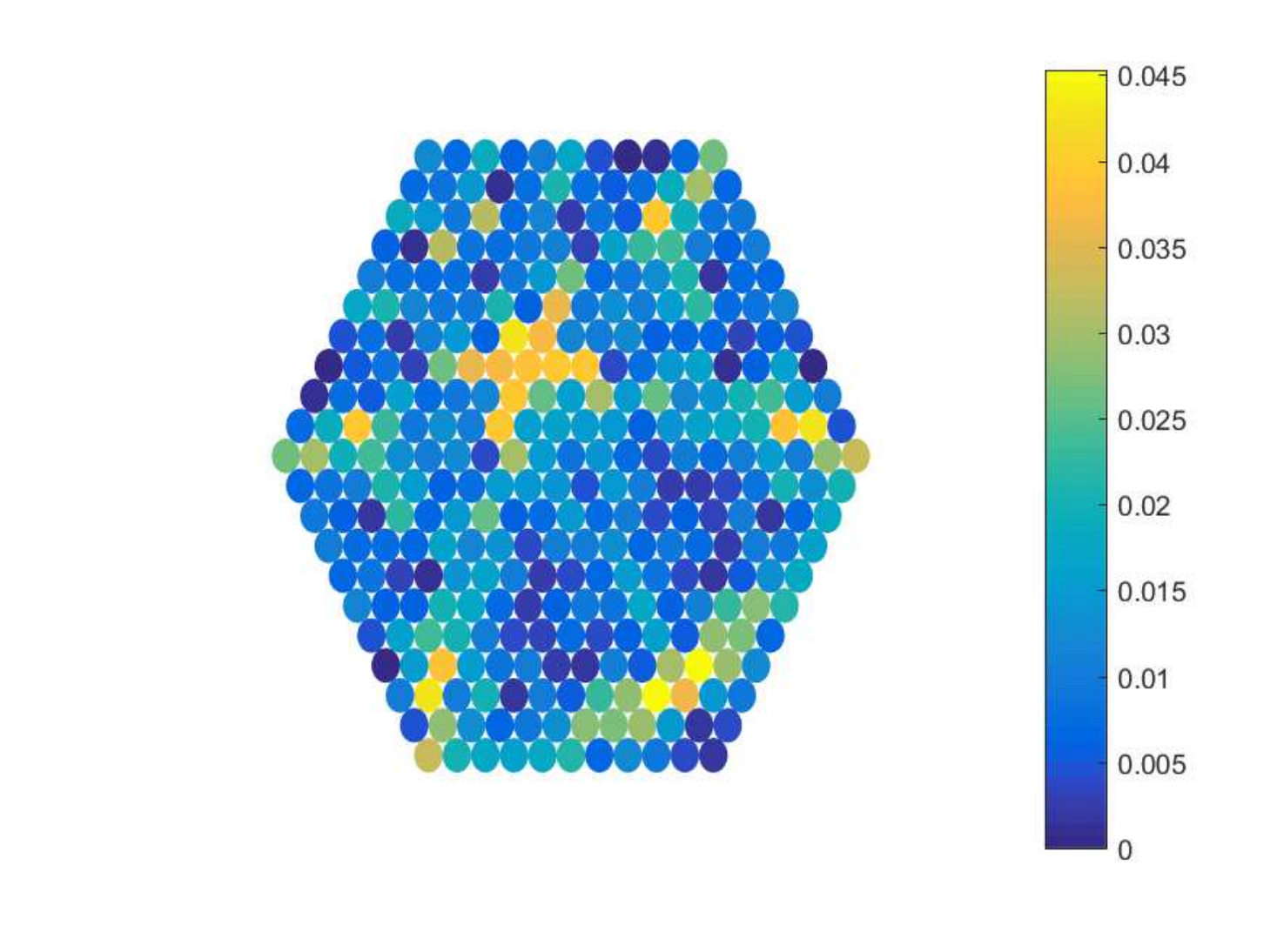}}
\vspace{-0.2in}
\caption{\label{Hex:StainlessSteel_OneCornerstriker} Velocity magnitudes at each sphere as a function of time for a hexagonal basin filled with a hexagonal packing of spheres. The packing has 11 spheres on each edge. A striker will hit at one of the corners. The initial velocity of the striker is $0.4m/s$. Here (a) $t=0.525$ms, (b) $t=1.00$ms, (c) $t=1.55$ms. Note that (a)-(b) are before the main wave reflection, while (c) is after the main wave reflection.}
\end{figure*}

\begin{figure*}
\centering
\subfigure[]{\includegraphics[width=0.32\linewidth]{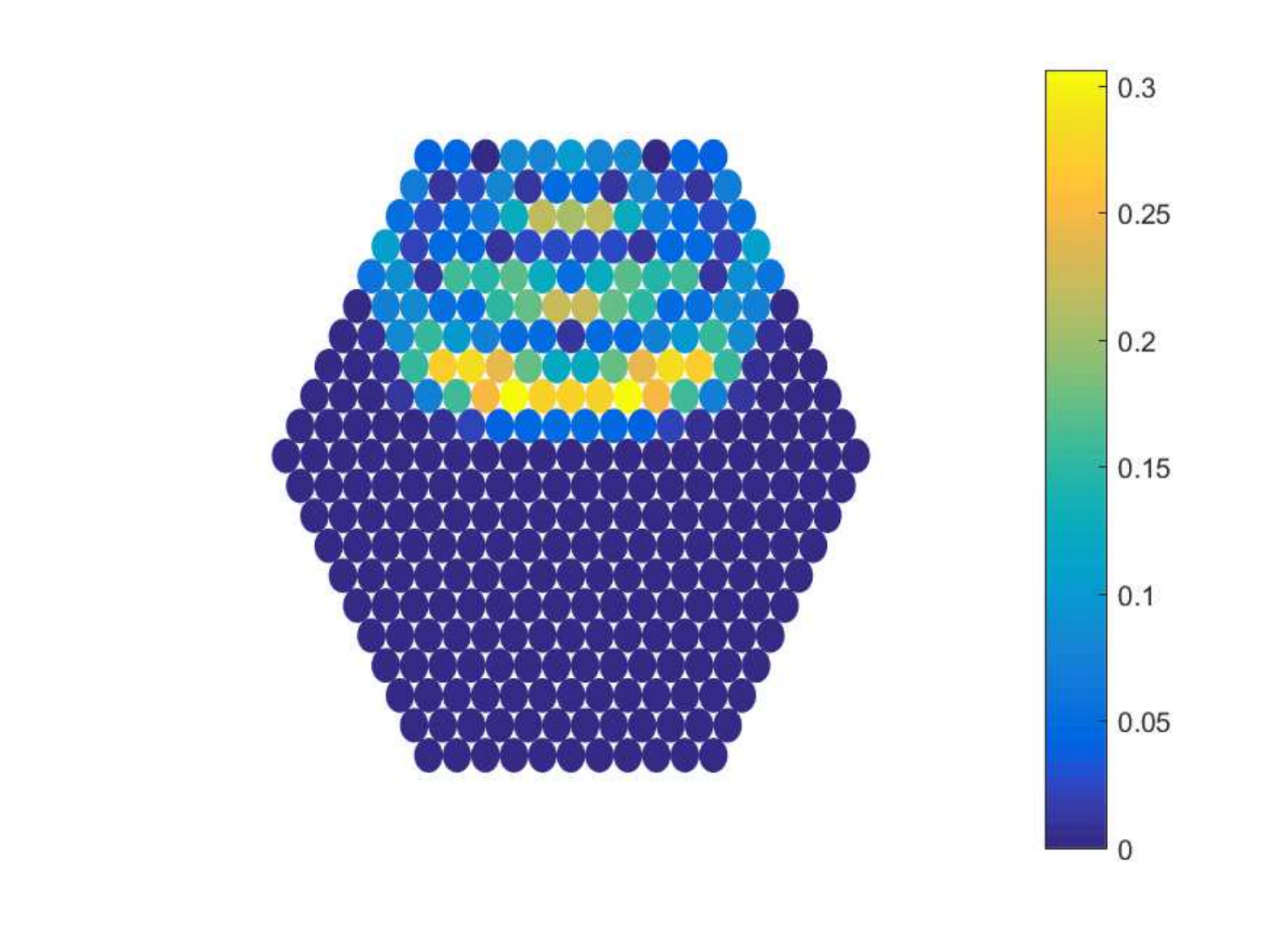}}
\subfigure[]{\includegraphics[width=0.32\linewidth]{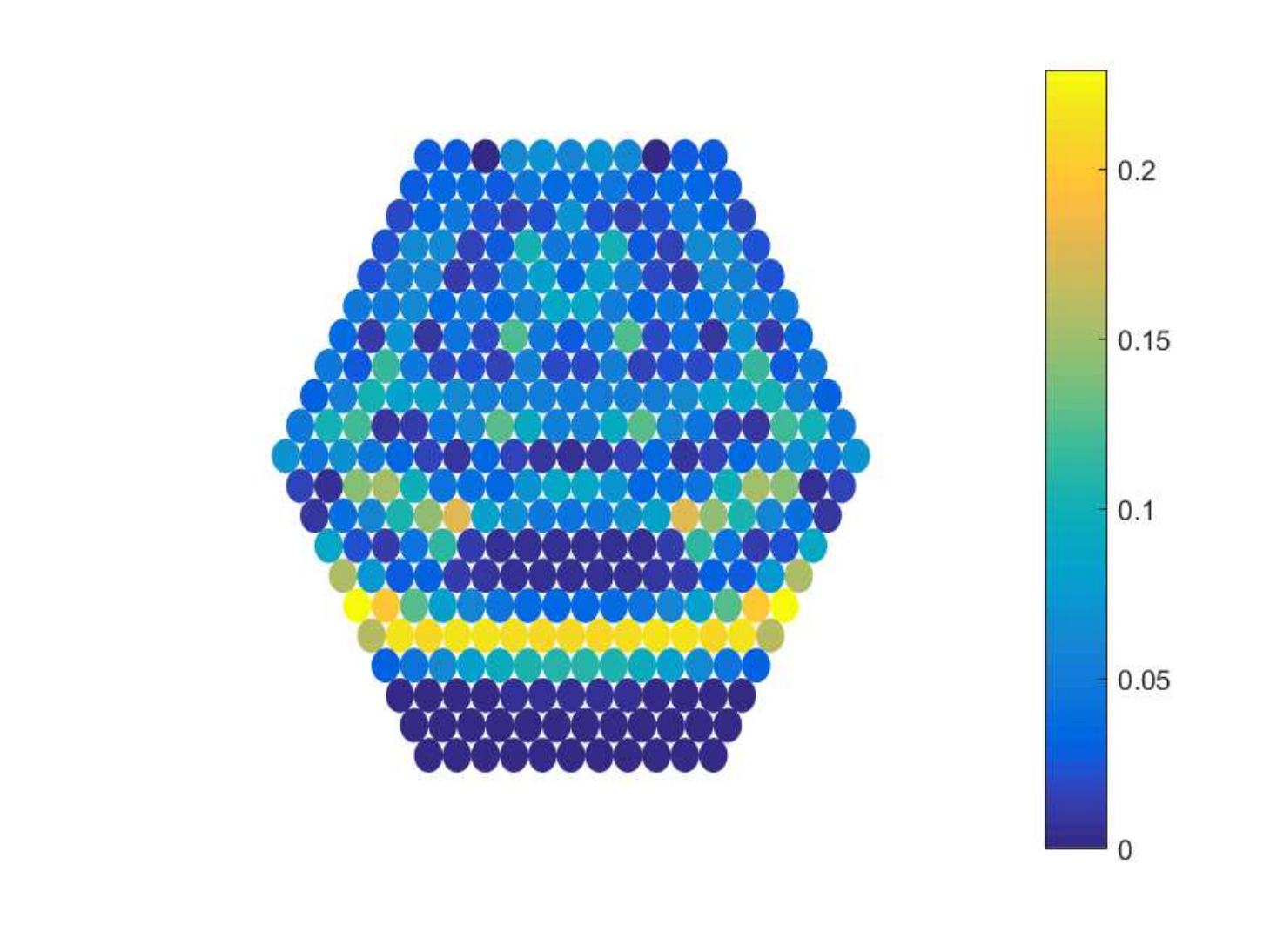}}
\subfigure[]{\includegraphics[width=0.32\linewidth]{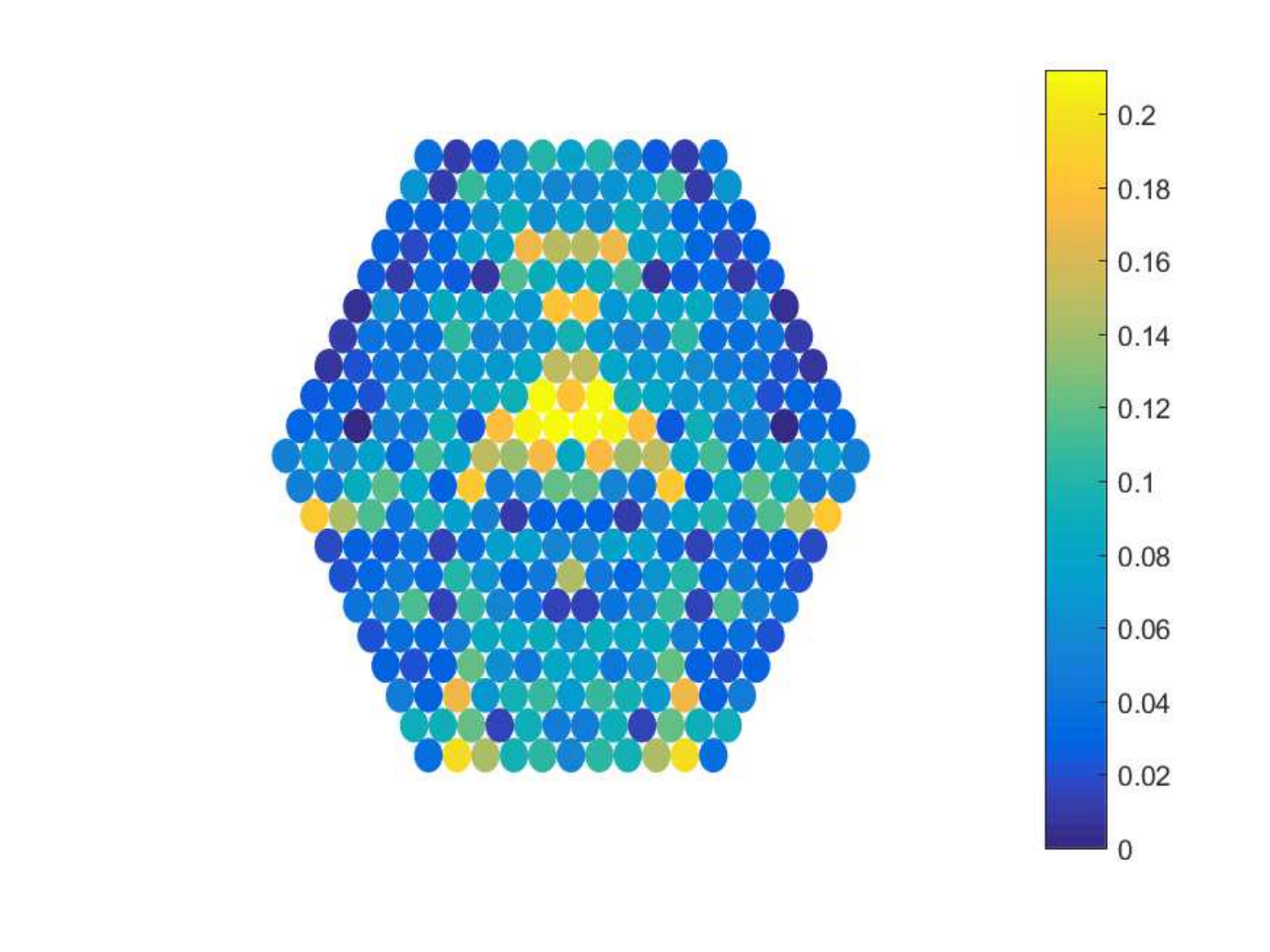}}
\vspace{-0.2in}
\caption{\label{Hex:StainlessSteel_ArrayStiker} Velocity magnitudes at each sphere as a function of time for a hexagonal basin filled with a hexagonal packing of spheres. The packing has 11 spheres on each edge. An array of strikers will hit vertically at one of the edges. The initial velocity of the strikers are $0.4m/s$. Here (a) $t=0.275$ms, (b) $t=0.525$ms, (c) $t=1.00$ms. Note that (a)-(b) are before the main wave reflection, while (c) is after the main wave reflection.}
\end{figure*}

Another configuration would be to place the striker at a corner. The striker will then hit a corner with a direction so that the angle between the direction of the striker and two edges will be the same. There are only two wave fronts in this case, and the shape will look like an arrow as shown in Fig. \ref{Hex:StainlessSteel_OneCornerstriker}. The wave is strongest along the boundaries. The reflected waves will collide and strike opposite to the direction of the initial strike. The wave will be concentrated near corners and also in a larger region close to the center which is gradually reflected back to the corner which was struck, after which the waves scatter and the dynamics become less regular in structure.

Instead of using a single striker to strike one sphere at the boundary, one may use multiple strikers to impact multiple spheres simultaneously. To this end, in Fig. \ref{Hex:StainlessSteel_ArrayStiker} we consider the case where 11 strikers impact an entire edge at $t=0$. We see that there are several waves behind the wave front (even though there is a single strike), in contrast to the case where the system is hit by one striker. The leading wave also propagates faster and carries more energy compared to the one-striker case. After a reflection, the wave train will interact at the middle of the basin, these localized interactions resulting in greater velocities near the center of the basin. 

\begin{figure}
\centering
\includegraphics[width=0.9\linewidth]{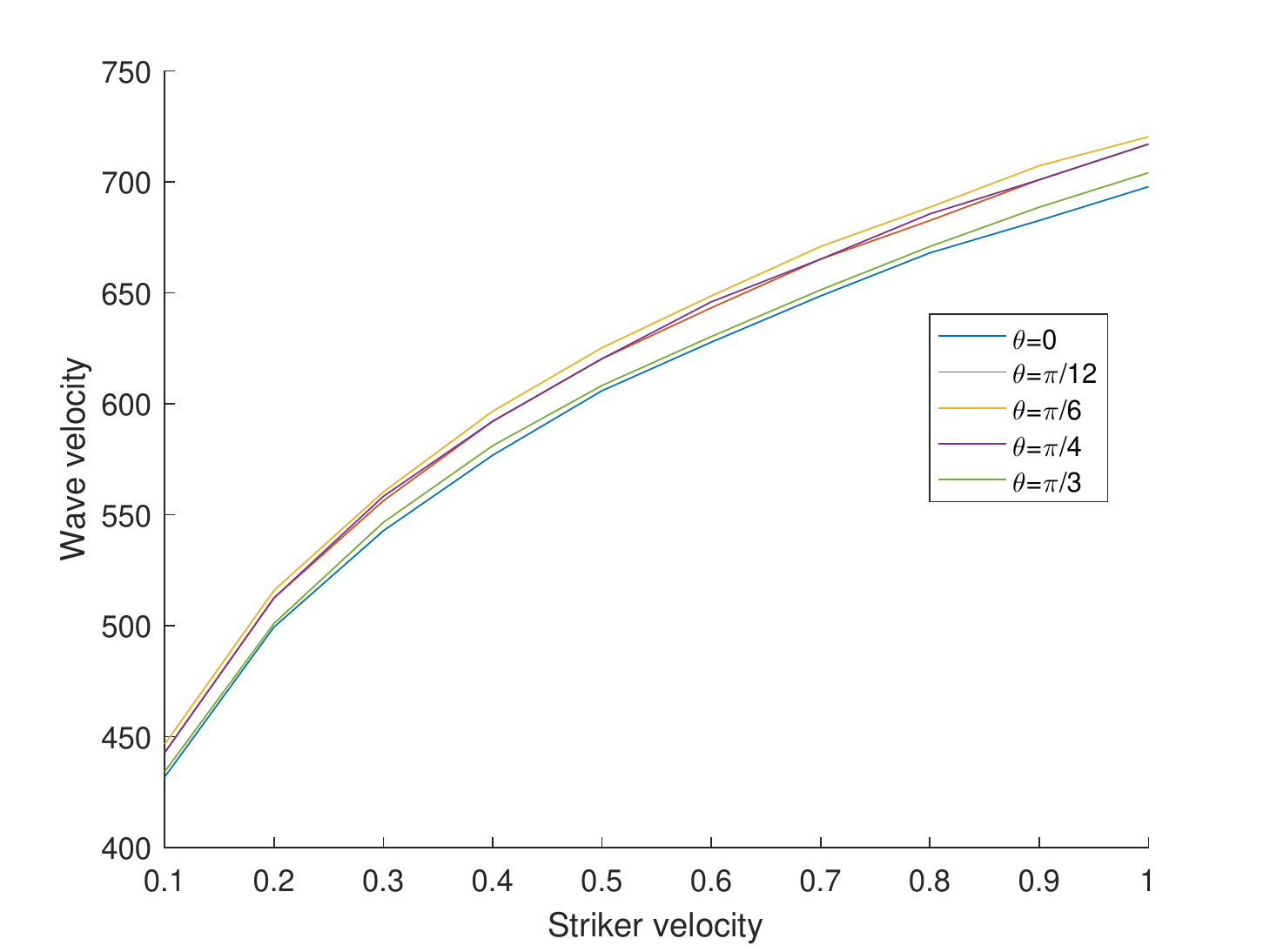}
\vspace{-0.05in}
\caption{The wave front speed (m/s) as a function of striker velocity (m/s) for a hexagonal packing of 11 spheres on each edge which is struck from different angles $\theta$ out of the perpendicular. The angle $\theta$ is measured by the angle between the striking direction and the normal direction of the top edge. The curve is somewhat jagged due to the discrete geometry. The method of calculating the wave front speed is the same as that used in Figure \ref{wavespeed}.}
\label{VaryingAngle}
\end{figure}

\begin{figure*}
\centering
\subfigure[]{\includegraphics[width=0.32\linewidth]{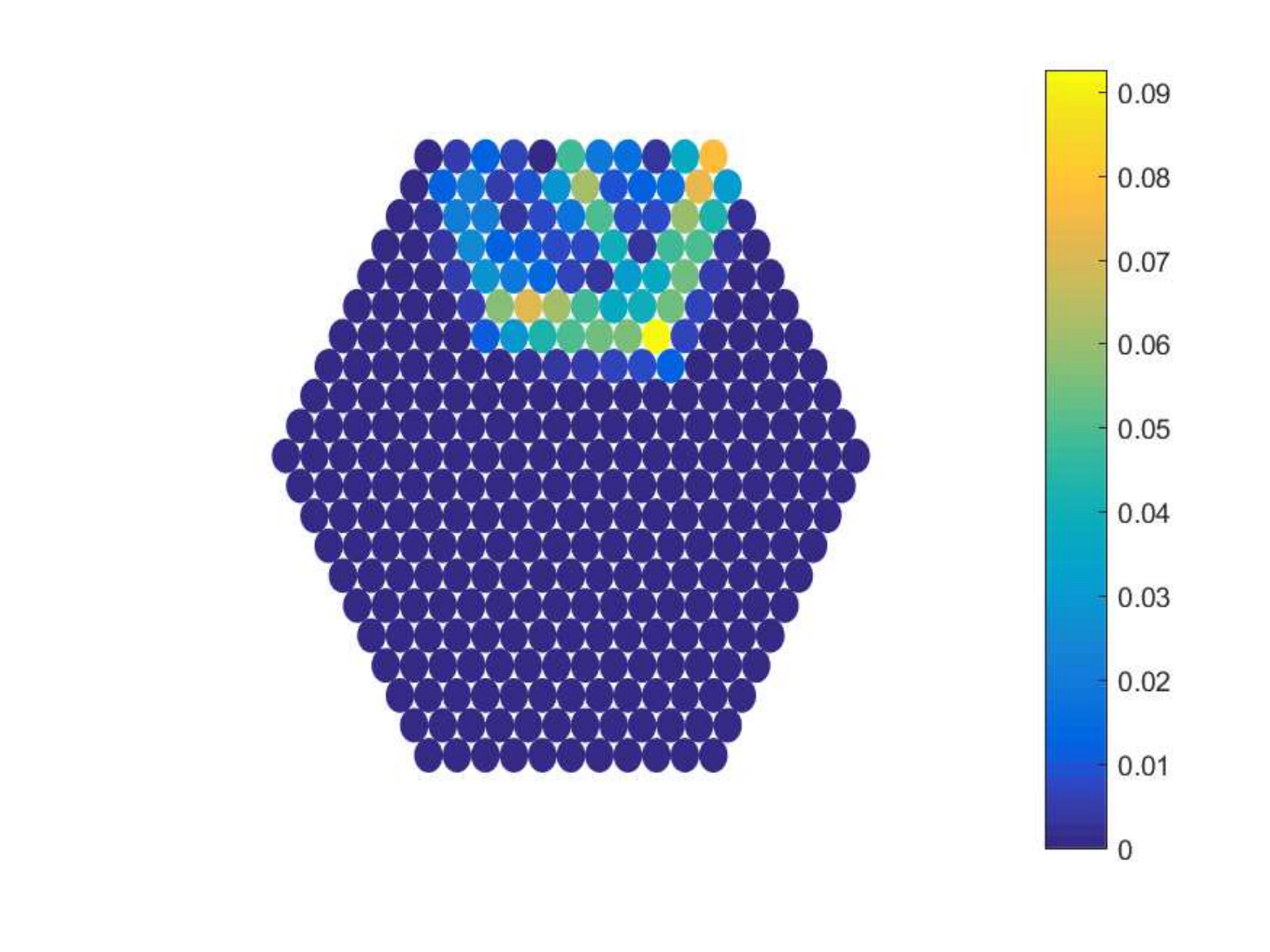}}
\subfigure[]{\includegraphics[width=0.32\linewidth]{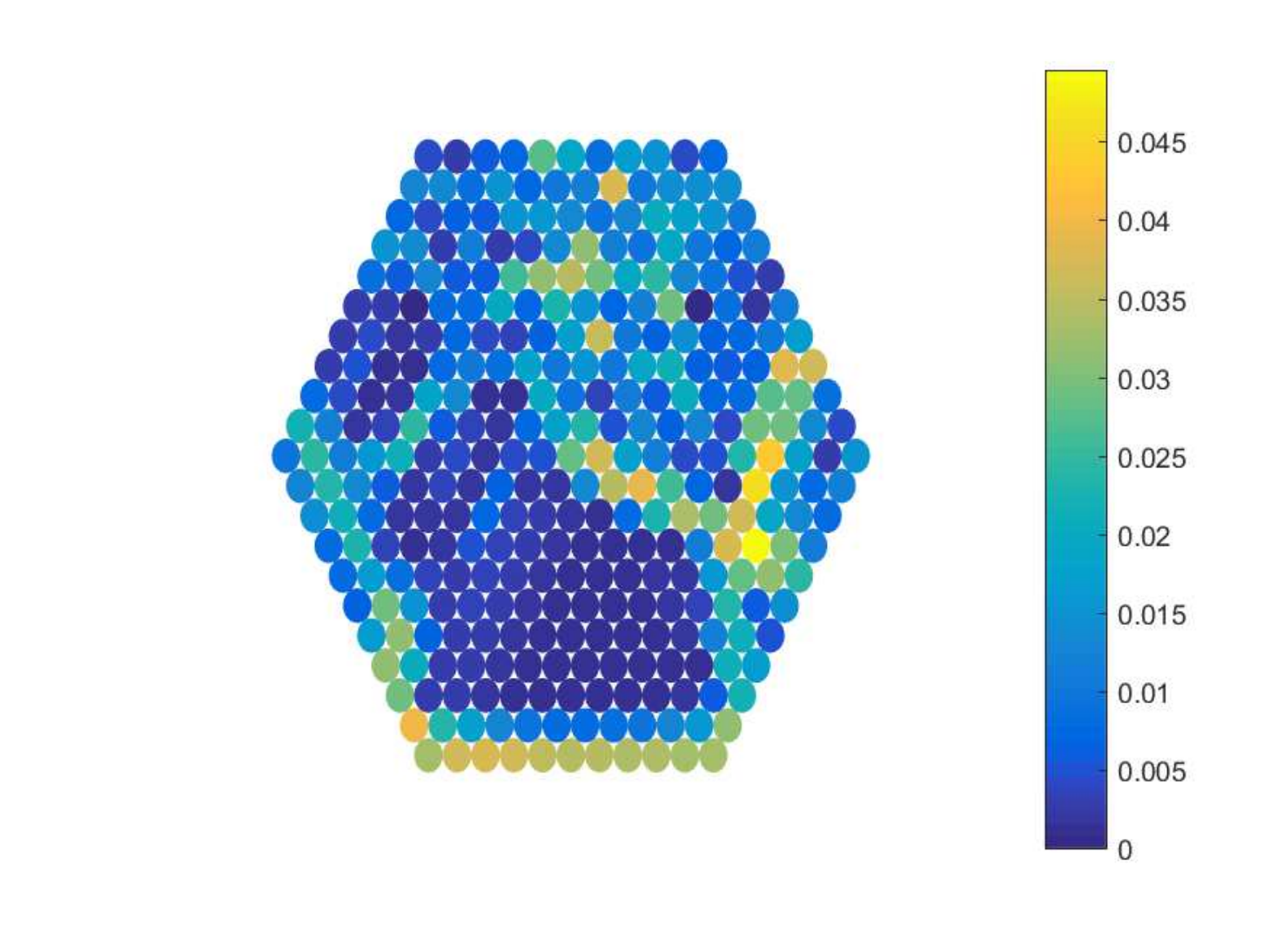}}
\subfigure[]{\includegraphics[width=0.32\linewidth]{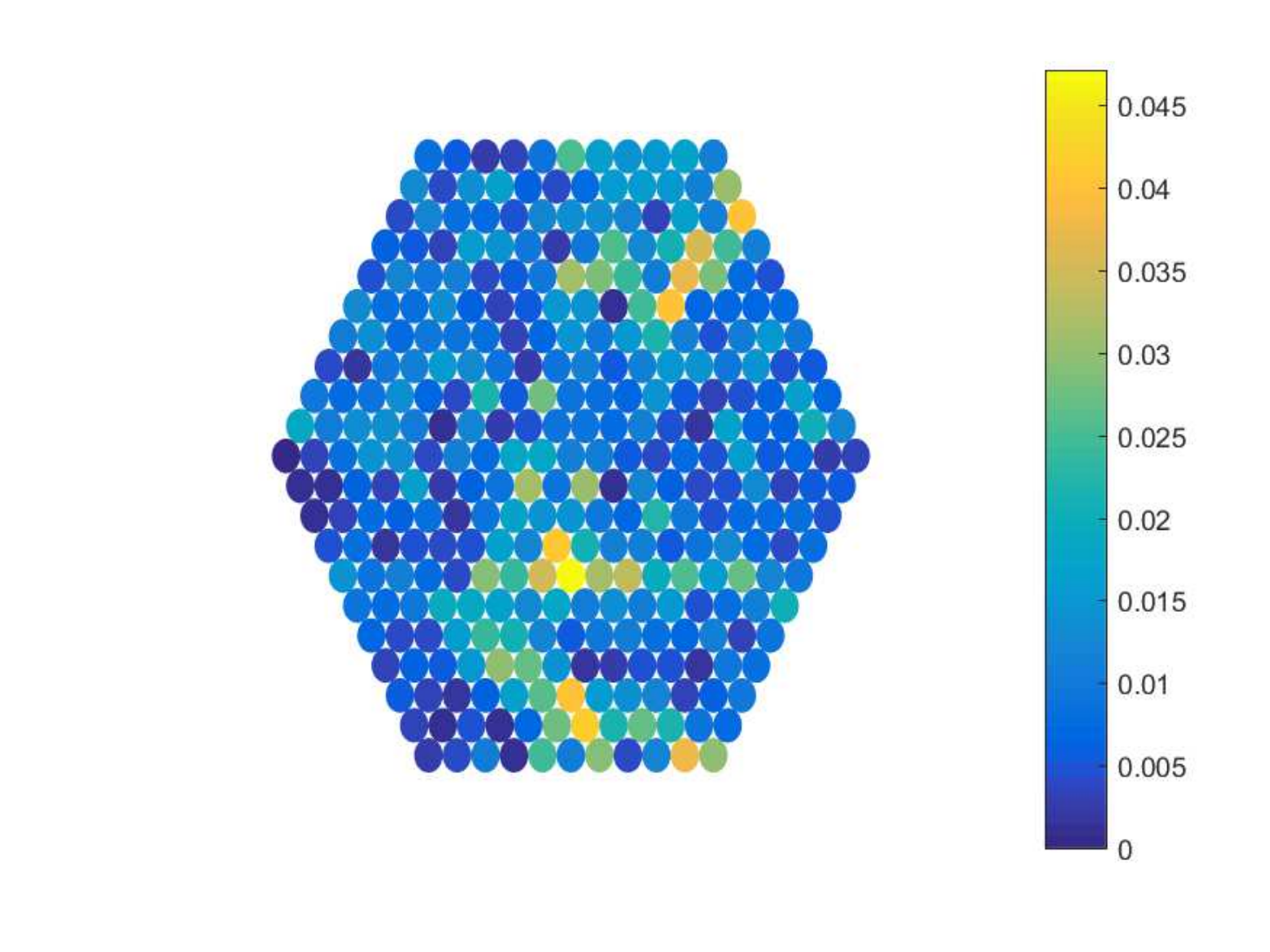}}
\subfigure[]{\includegraphics[width=0.32\linewidth]{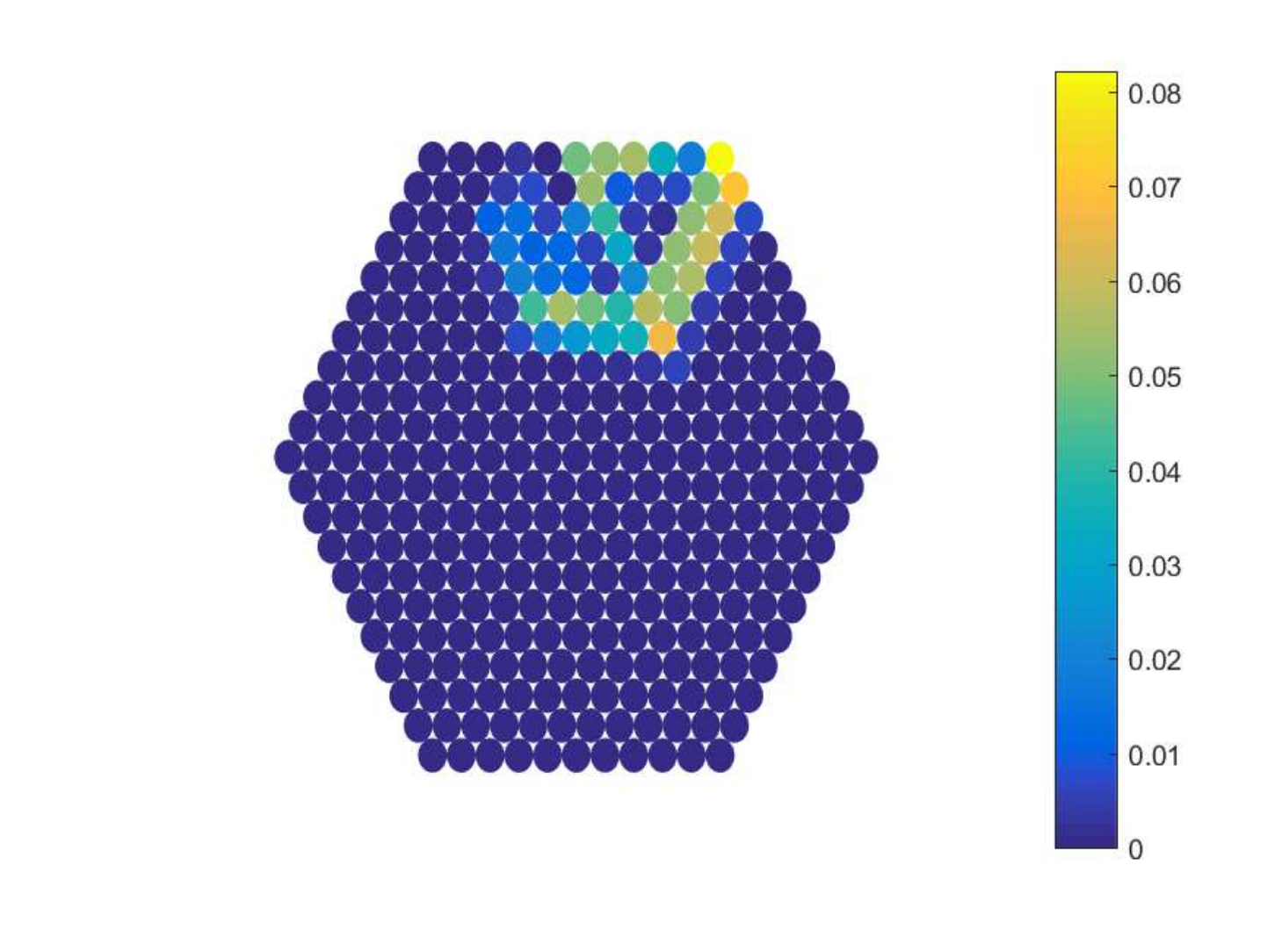}}
\subfigure[]{\includegraphics[width=0.32\linewidth]{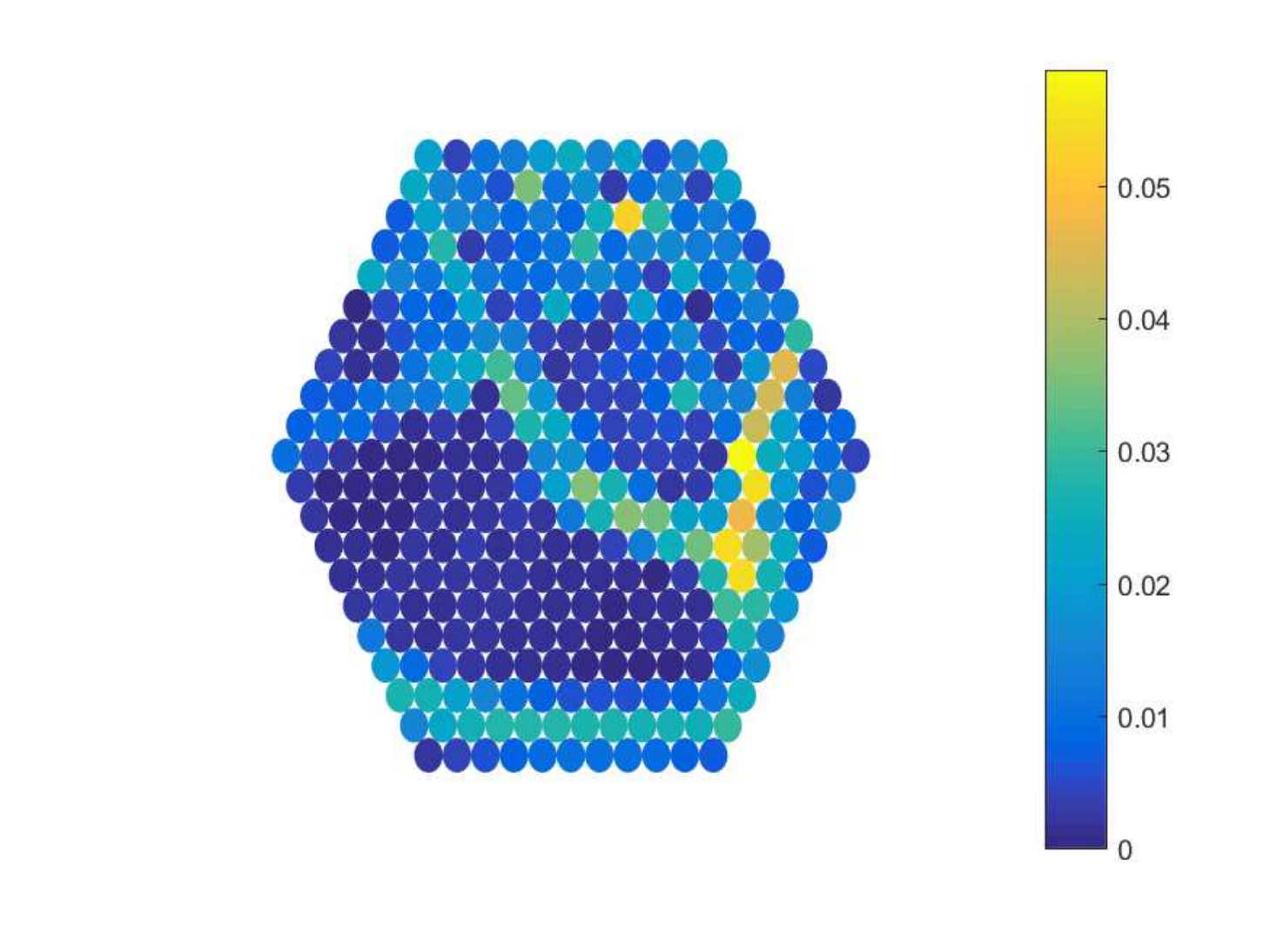}}
\subfigure[]{\includegraphics[width=0.32\linewidth]{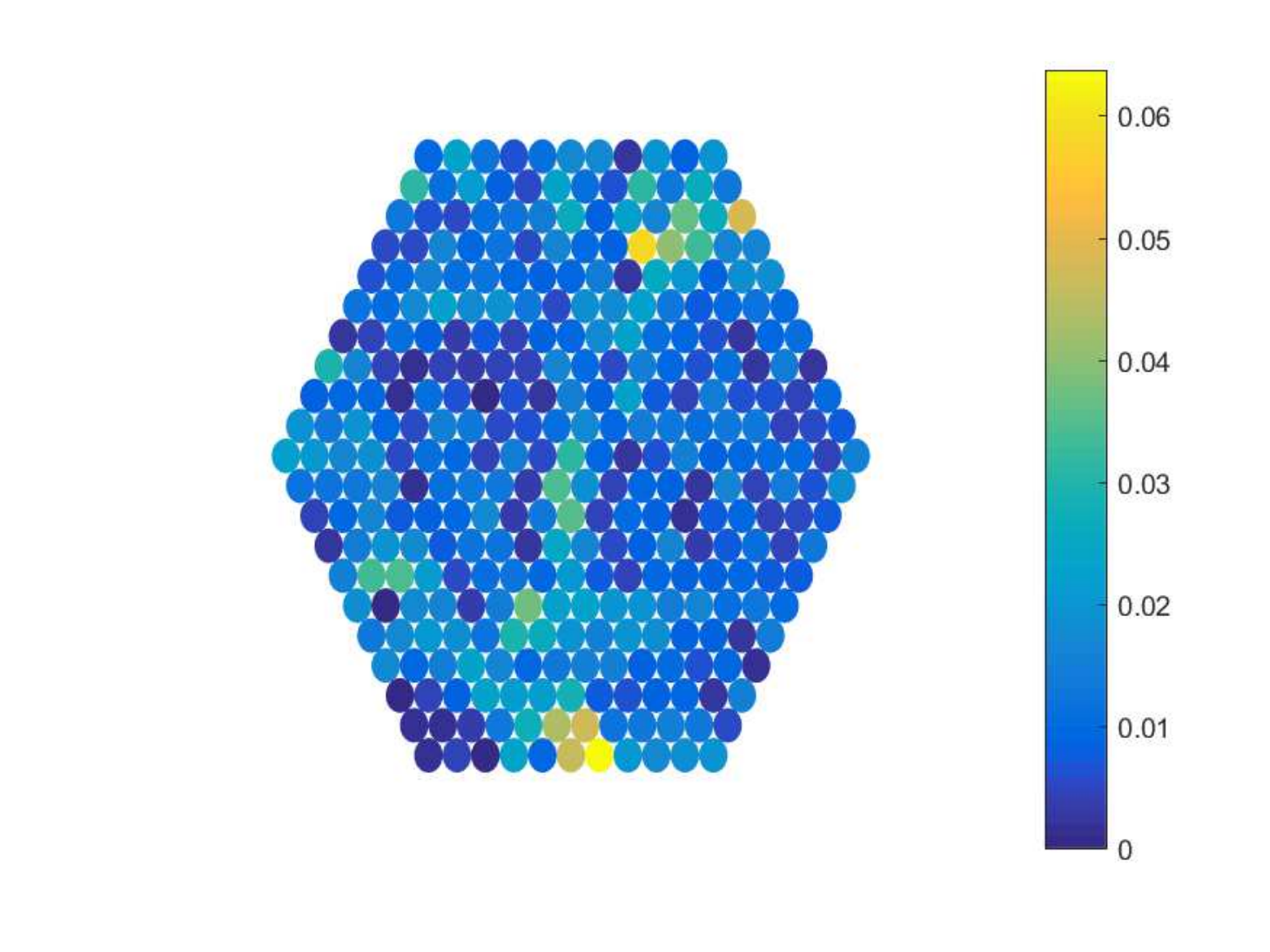}}
\vspace{-0.2in}
\caption{\label{WaveHexAngle} Velocity magnitudes at each sphere as a function of time for a hexagonal basin filled with a hexagonal packing of spheres. The packing has 11 spheres on each edge. Results from a strike with a $30^\circ$ degree with the normal direction of the top edge are shown at (a) $t=0.275$ms, (b) $t=0.875$ms, (c) $t=1.20$ms, and results from a strike with a $60^\circ$ degree with the normal direction of the top edge are shown at (d) $t=0.275$ms, (e) $t=0.875$ms, (f) $t=1.20$ms. The initial velocity of the strikers are both $0.4m/s$. Note that (a)-(b) are before the main wave reflection, while (c)-(f) are after the main wave reflection.}
\end{figure*}

As discussed in \cite{Leo14,manjunath2014plane}, it is possible to strike boundary spheres at an angle, rather than perpendicular to the boundary. Doing so, we break the symmetry of the configurations considered above, and the waves appear to scatter sooner.  The velocity of the wave front can be modified by striking at an angle, and we show this for one configuration in Fig. \ref{VaryingAngle}. When the impact is perpendicular or close to parallel, the wave front speed will be similar, while then the angle is in between these two extremes, the wave front speed will be somewhat faster. This is because the energy is focused more narrowly along part of the granular crystal, so the leading section of the wave front will propagate more quickly. We demonstrate this for two cases in Fig. \ref{WaveHexAngle}. Due to the asymmetry, the wave patterns more quickly fall into disorder. 

\begin{figure*}
\centering
\subfigure[]{\includegraphics[width=0.32\linewidth]{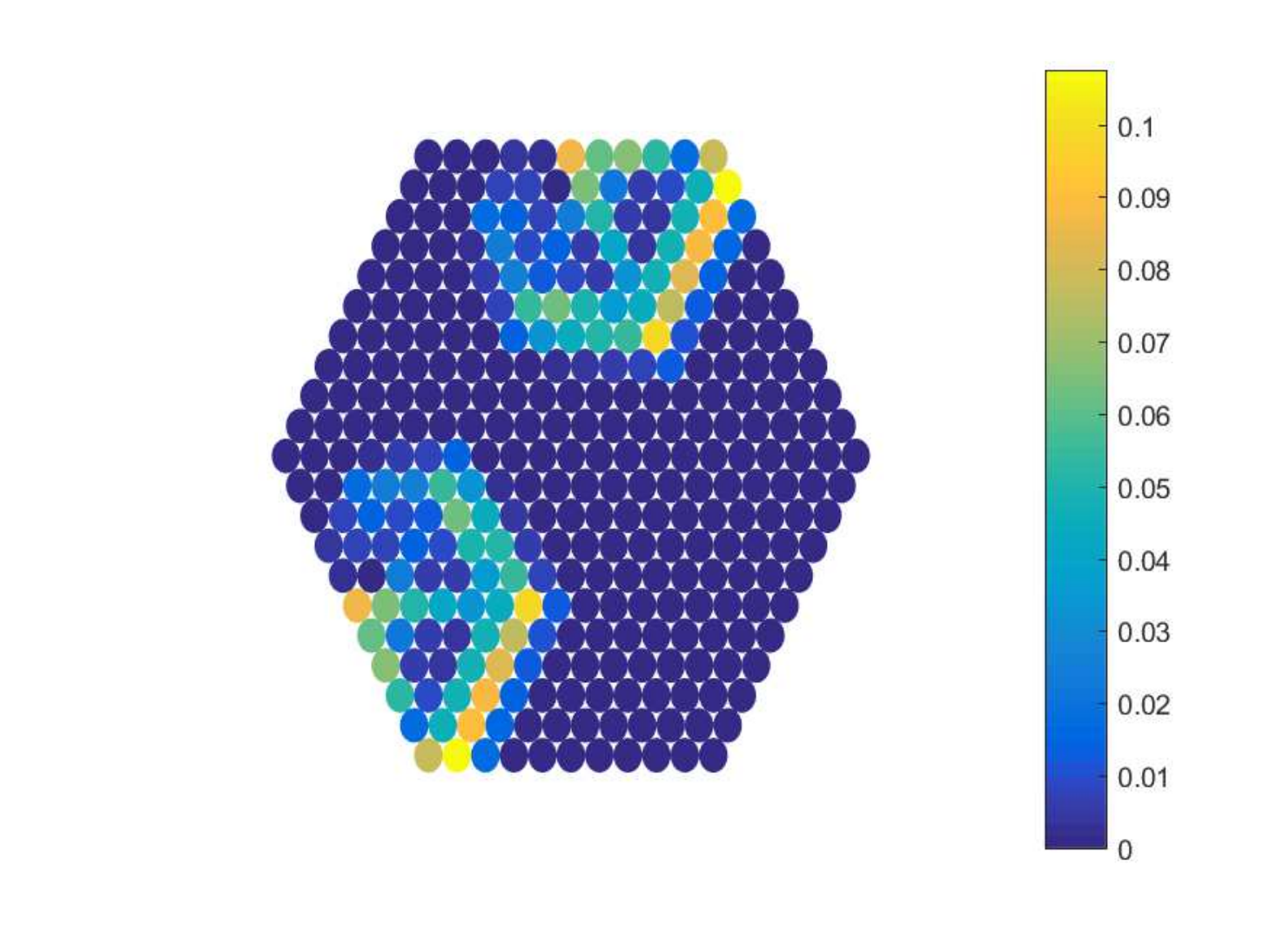}}
\subfigure[]{\includegraphics[width=0.32\linewidth]{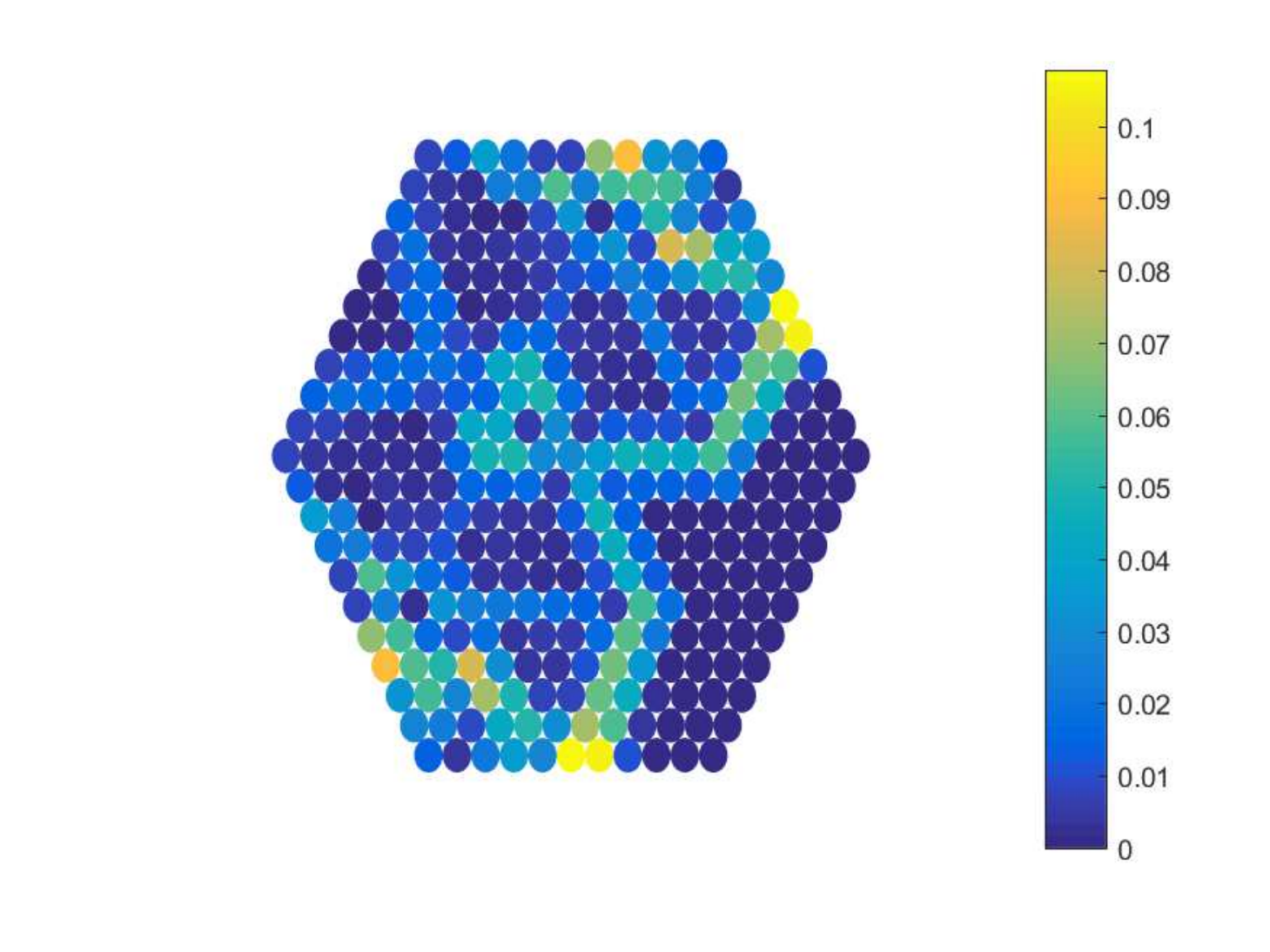}}
\subfigure[]{\includegraphics[width=0.32\linewidth]{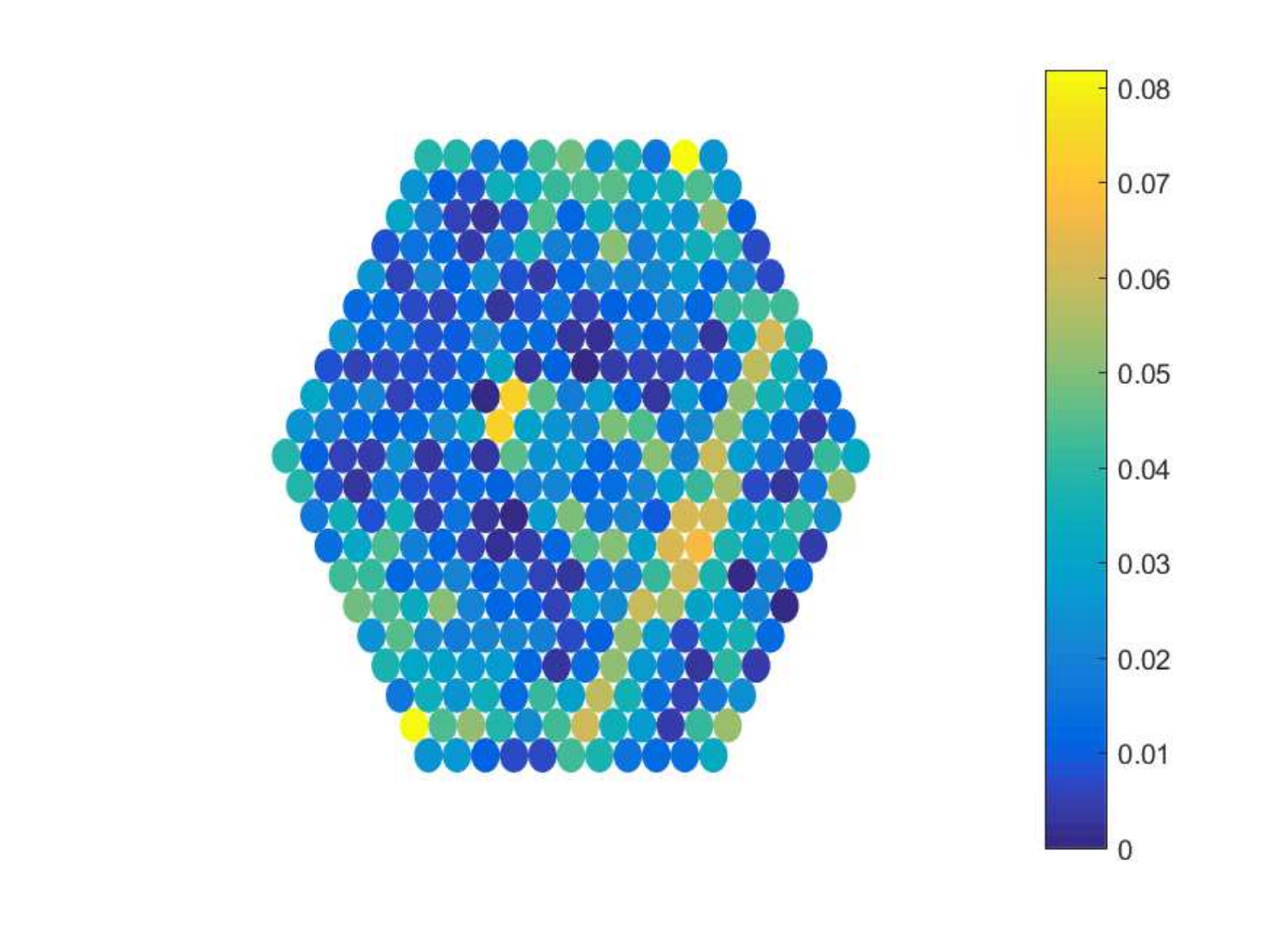}}
\vspace{-0.2in}
\caption{\label{WaveHexTwostriker} Velocity magnitudes at each sphere as a function of time for a hexagonal basin filled with a hexagonal packing of spheres. The packing has 11 spheres on each edge. Two strikers will hit at two edges. The initial velocity of the striker is $0.4m/s$. Here (a) $t=0.275$ms, (b) $t=0.450$ms, (c) $t=0.875$ms. Note that (a)-(b) are before the main wave reflection, while (c) is after the main wave reflection.}
\end{figure*}

It is also possible to impose two strikers on different edges, instead of a single striker, and we give one example in Fig. \ref{WaveHexTwostriker}. For this configuration, the wave velocity is suppressed toward the middle, while the most distant sides of each wave propagate with a larger velocity. This is seemingly result of the instantaneous yet subtle compression caused by the strike before the arrival of wave fronts. A wave in the direction of the composition of the two waves will emerge shortly after, but will soon break.

\subsection{Triangular basin}
After considering hexagonal basins consisting of hexagonal packings of spheres, it is interesting to look into different basin geometries which are assembled using hexagonal packings. One such possible basin is triangular, with three (as opposed to six) boundaries. Again, no precompression will be applied initially.

\begin{figure*}
\centering
\subfigure[]{\includegraphics[width=0.32\linewidth]{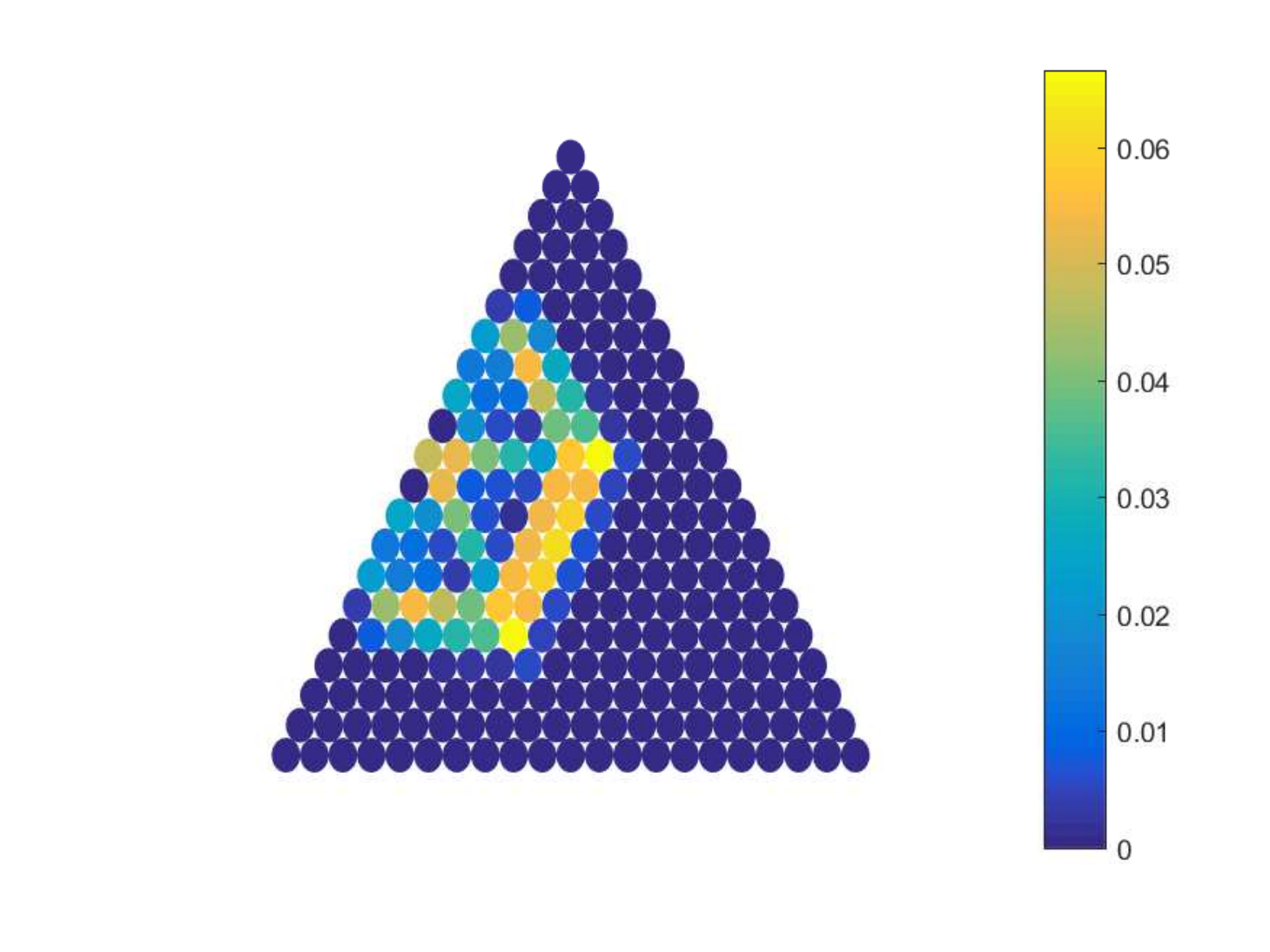}}
\subfigure[]{\includegraphics[width=0.32\linewidth]{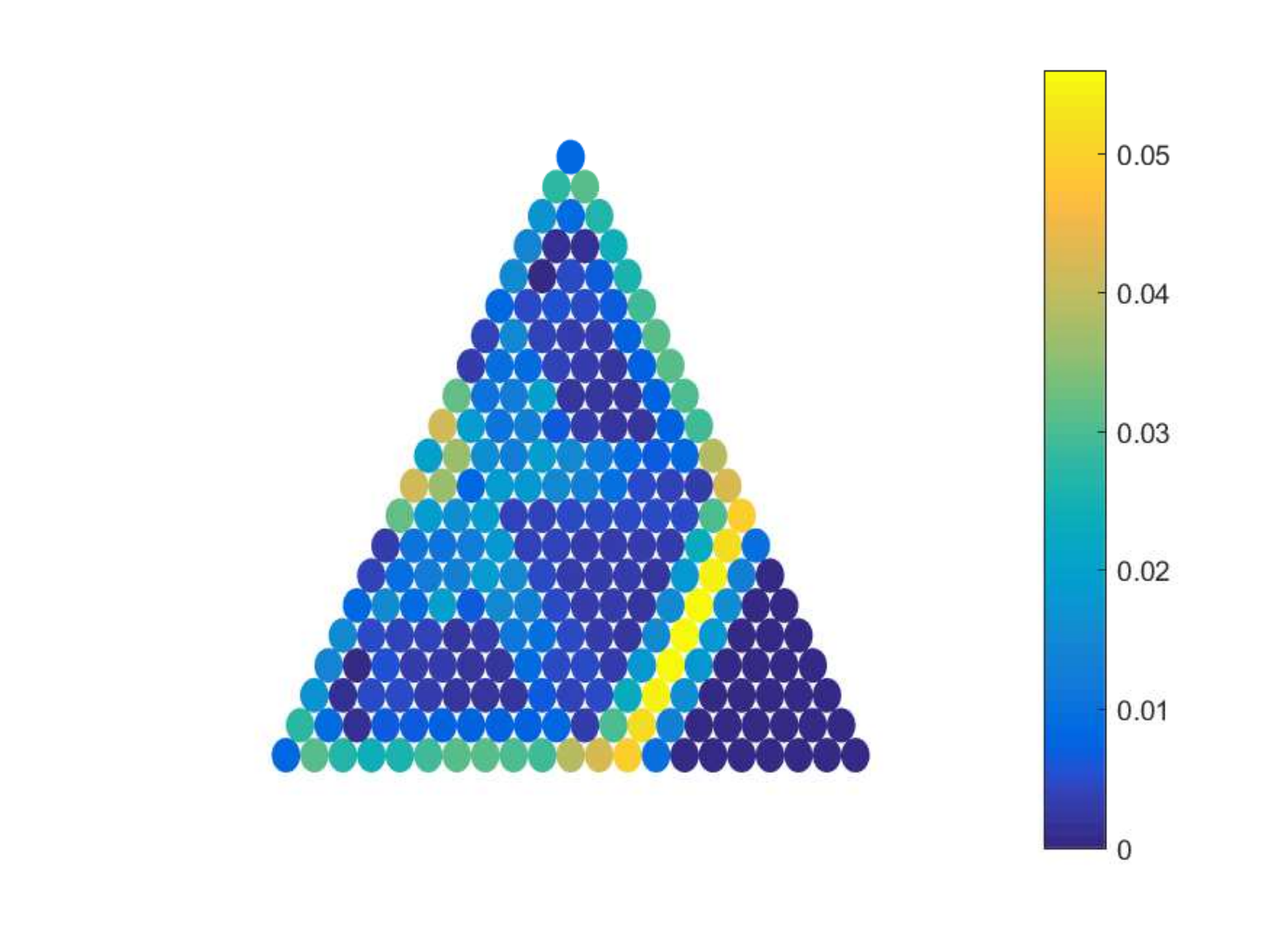}}
\subfigure[]{\includegraphics[width=0.32\linewidth]{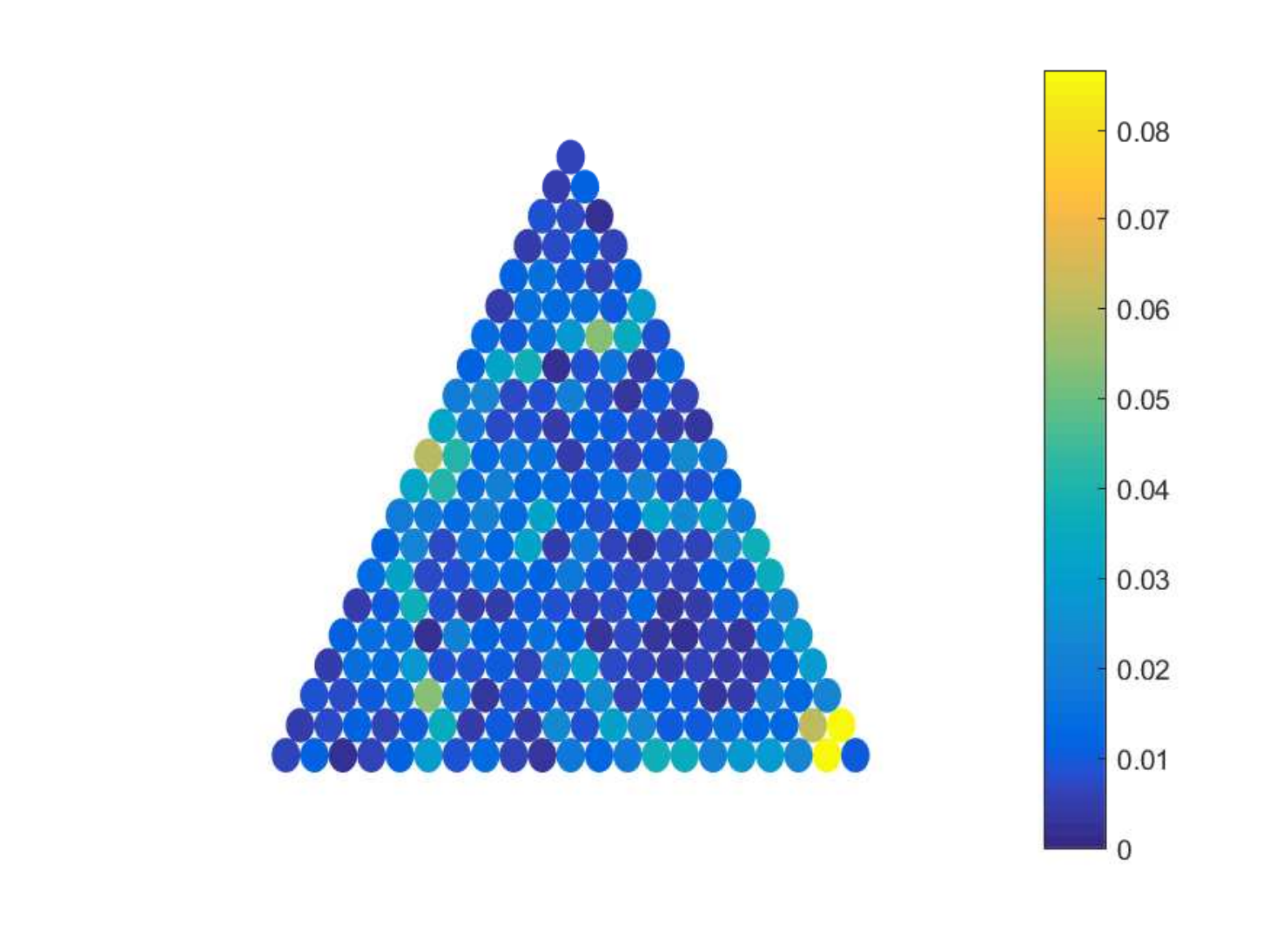}}
\vspace{-0.2in}
\caption{Velocity magnitudes at each sphere as a function of time for a triangular basin filled with a hexagonal packing of spheres. The packing has 21 spheres on each edge. A striker will hit perpendicularly at the centre of an edge. The initial velocity of the striker is $0.4m/s$. Here (a) $t=0.275$ms, (b) $t=0.525$ms, (c) $t=1.00$ms. Note the increase in maximal amplitude as the wave is focused into a corner, in panel (c). Note that (a)-(c) are before the main wave reflection, which occurs shortly after (c).}
\label{WaveTriPer}
\end{figure*}

\begin{figure*}
\centering
\subfigure[]{\includegraphics[width=0.32\linewidth]{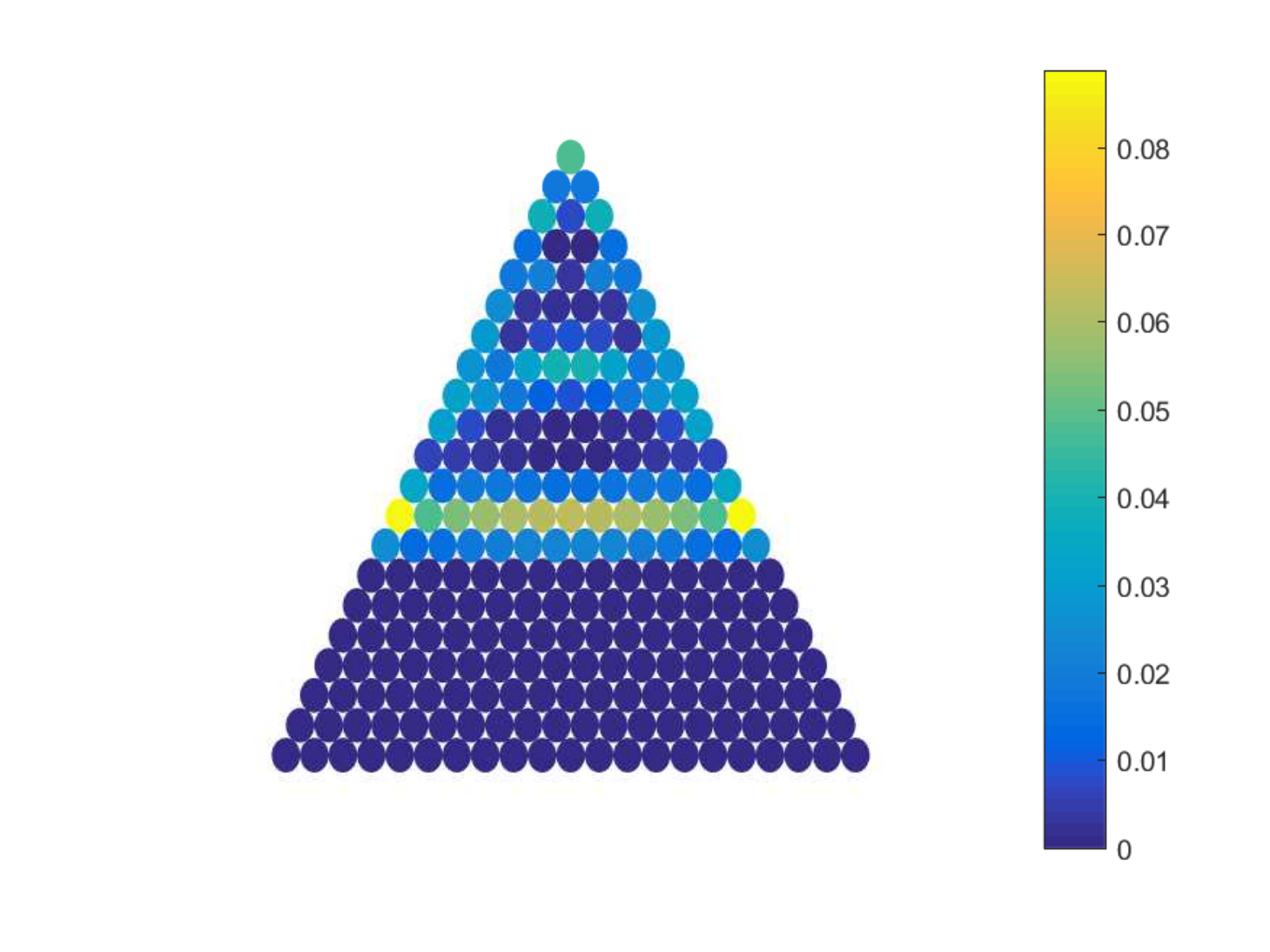}}
\subfigure[]{\includegraphics[width=0.32\linewidth]{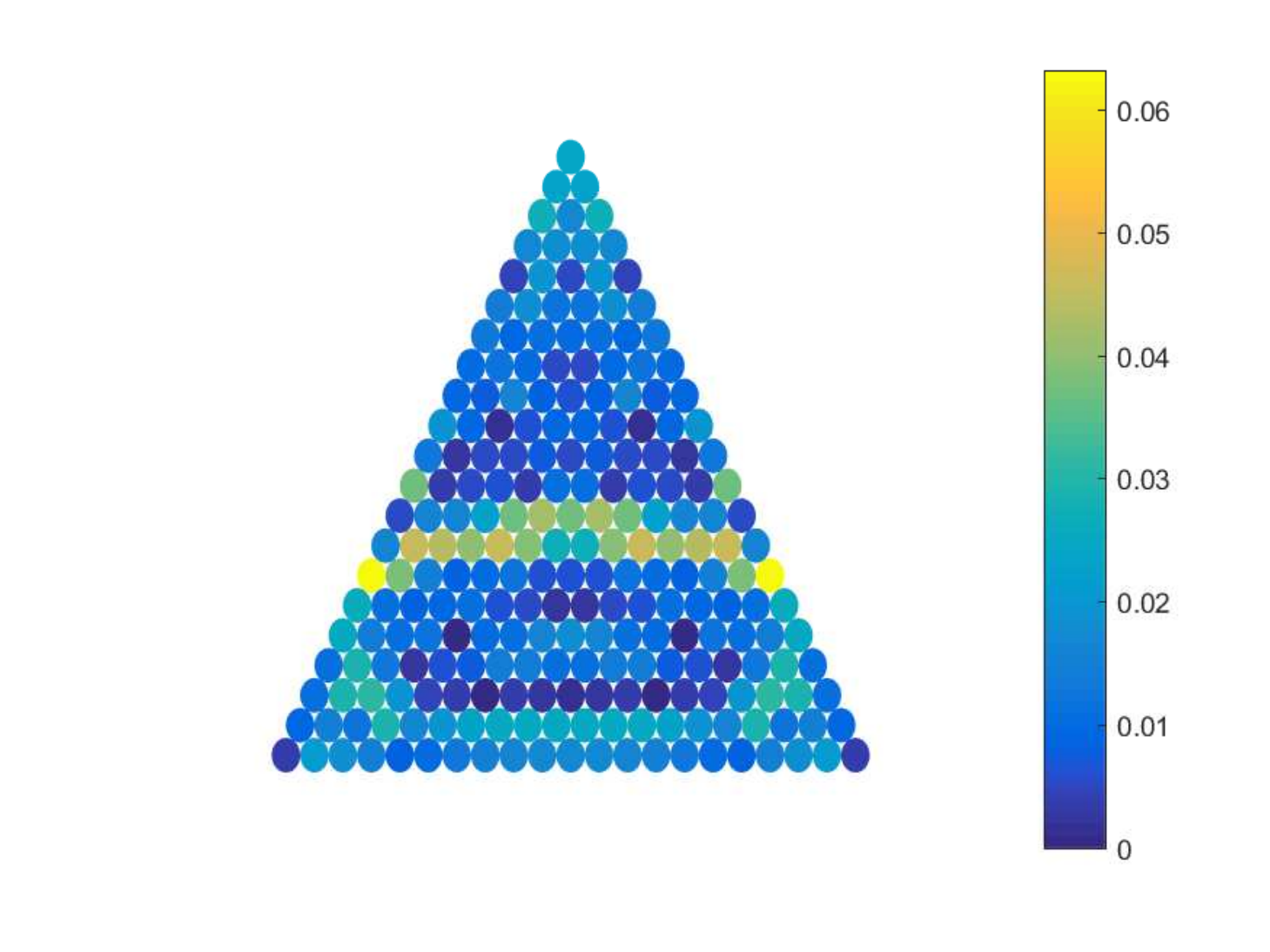}}
\subfigure[]{\includegraphics[width=0.32\linewidth]{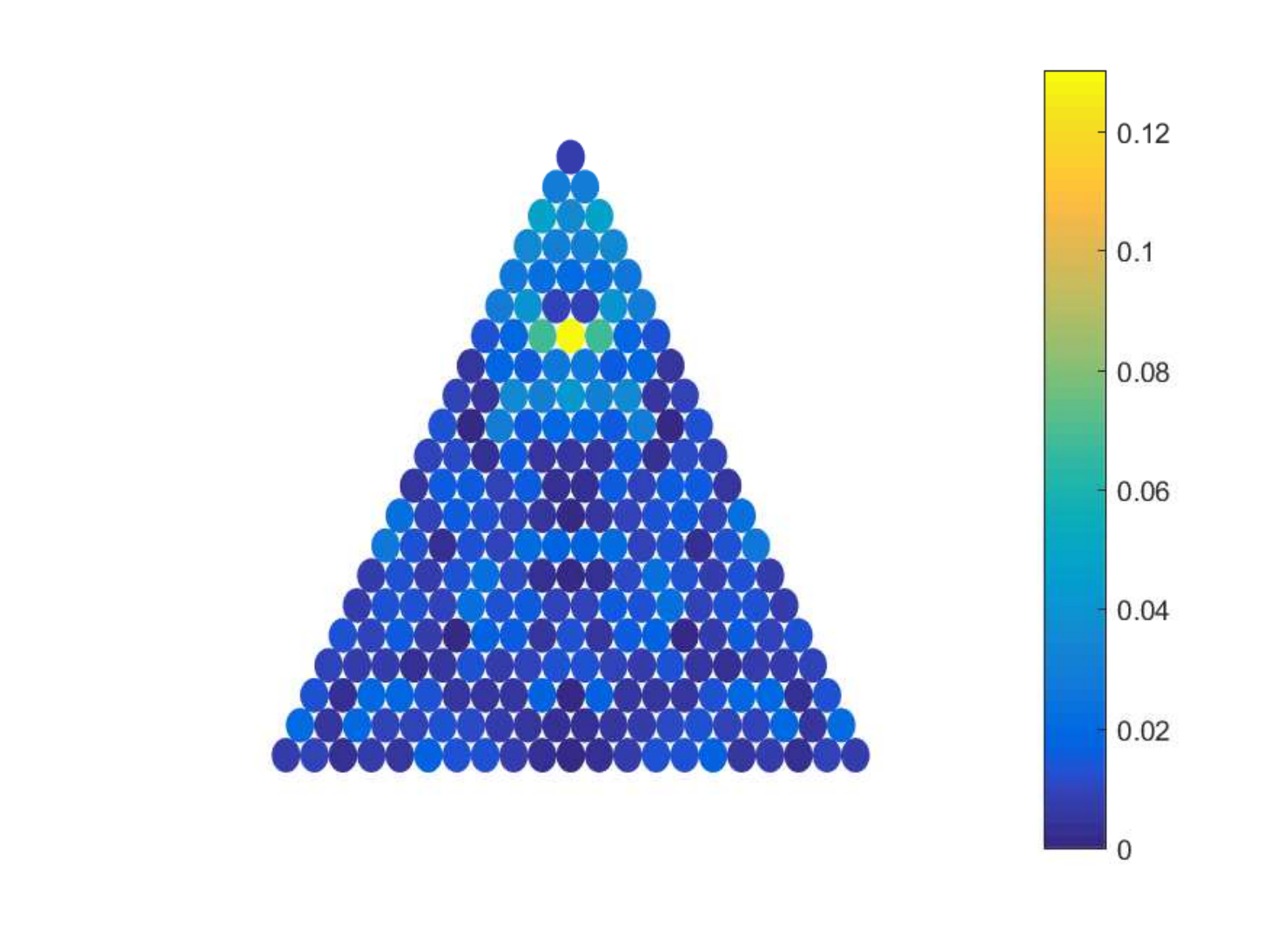}}
\vspace{-0.2in}
\caption{Velocity magnitudes at each sphere as a function of time for a triangular basin filled with a hexagonal packing of spheres. The packing has 21 spheres on each edge. A striker will hit the triangle vertically from the top vertex. The initial velocity of the striker is $0.4m/s$. Here (a) $t=0.500$ms, (b) $t=1.20$ms, (c) $t=2.10$ms. Note that (a) is before the main wave reflection, while (b)-(c) are after the main wave reflection.}
\label{WaveTriCorner}
\end{figure*}

\begin{figure*}
\centering
\subfigure[]{\includegraphics[width=0.32\linewidth]{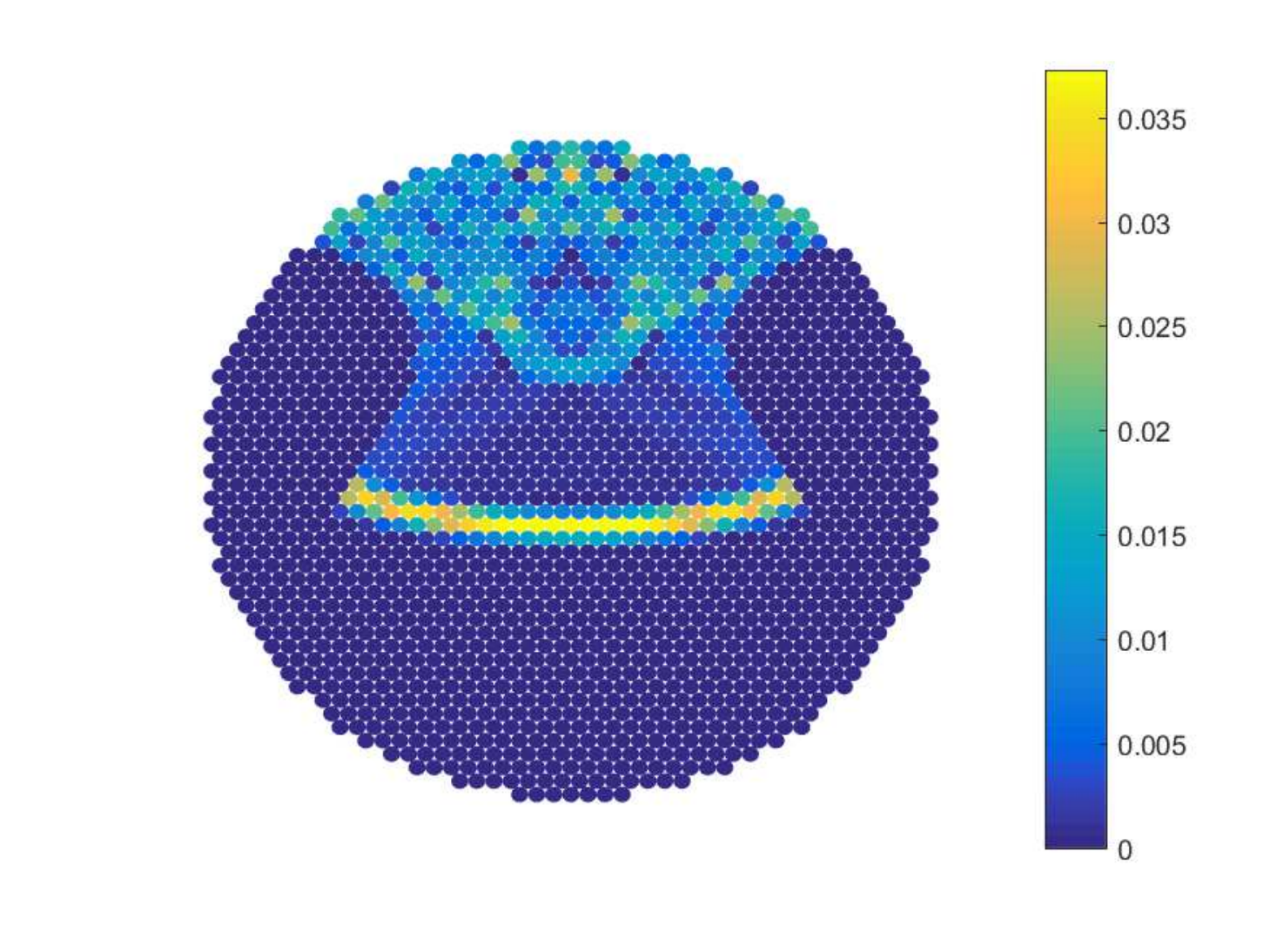}}
\subfigure[]{\includegraphics[width=0.32\linewidth]{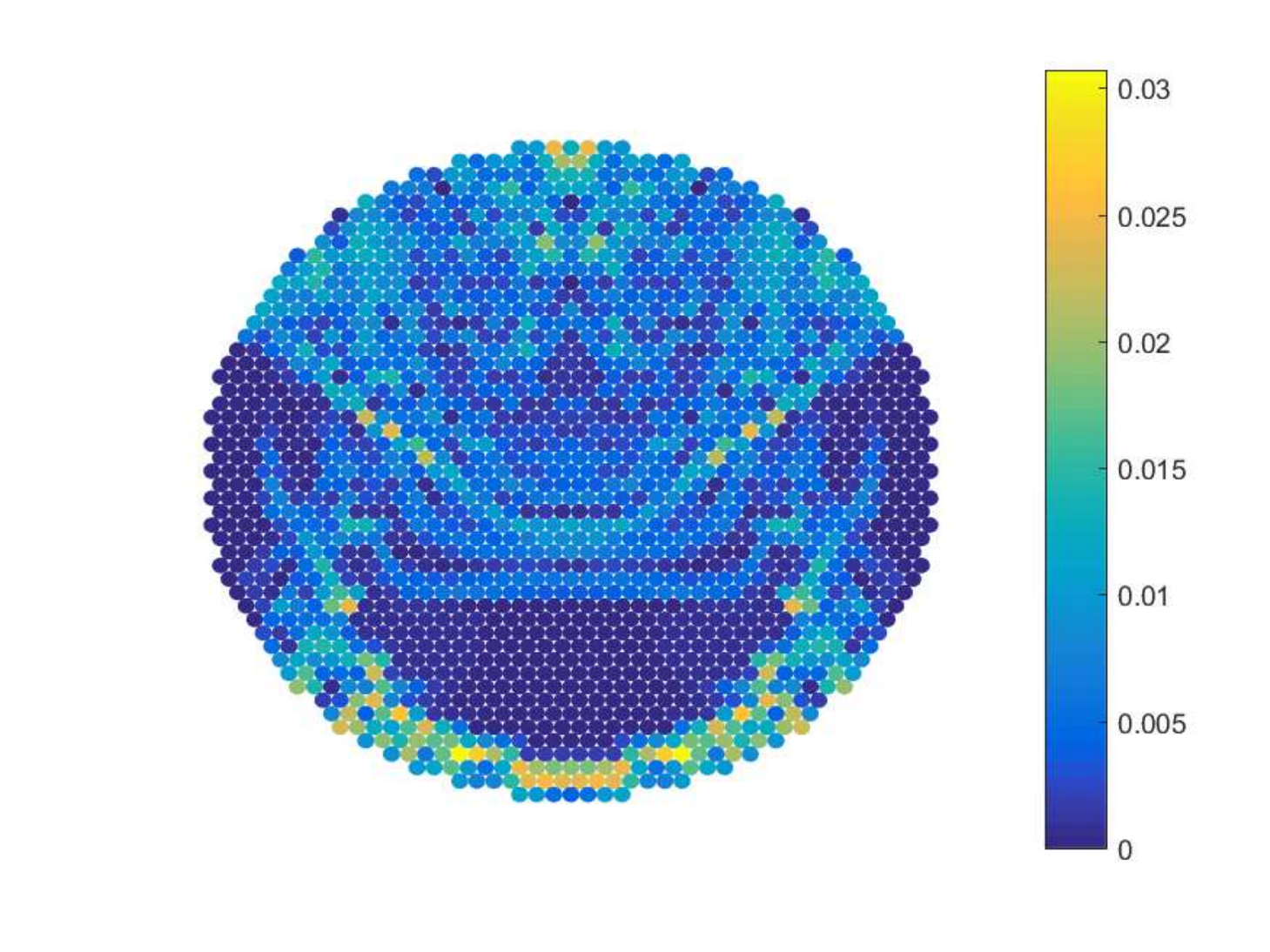}}
\subfigure[]{\includegraphics[width=0.32\linewidth]{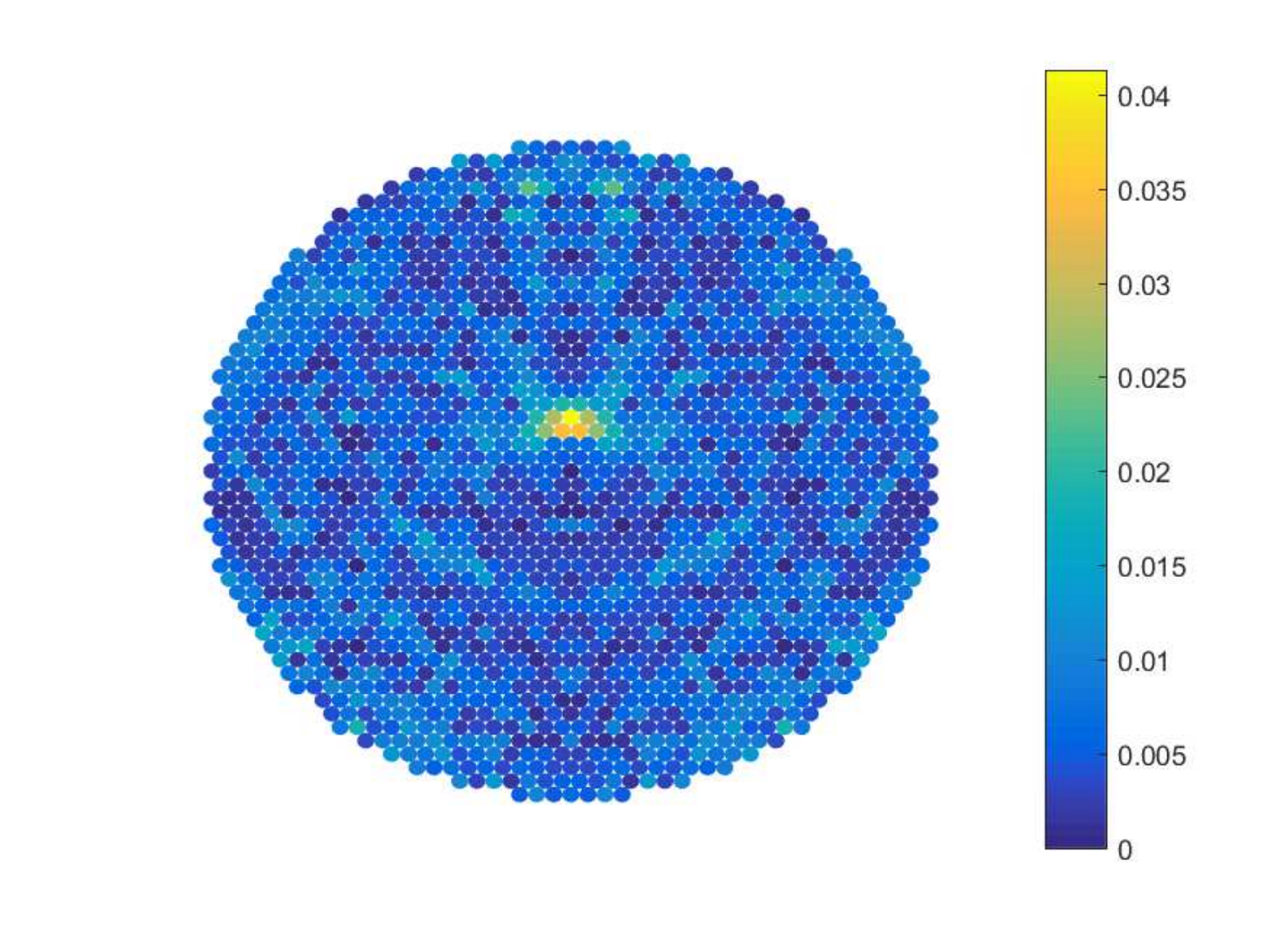}}
\vspace{-0.2in}
\caption{Velocity magnitudes at each sphere as a function of time for a circular basin filled with a hexagonal packing of spheres. The packing has 43 spheres on its longest row. A striker will hit at the centre of the first row. The initial velocity of the striker is $0.4m/s$. Here (a) $t=1.20$ms, (b) $t=2.20$ms, (c) $t=3.25$ms. Note that (a)-(b) are before the main wave reflection, while (c) is after the main wave reflection.}
\label{WaveCircleUpstriker}
\end{figure*}

In Fig. \ref{WaveTriPer}, we consider the case where a striker hits one edge perpendicularly. The shape of the wave is similar to the corresponding case in hexagonal geometry. Two wave fronts on the sides reach the boundaries and are reflected first, while the middle wave front (in the direction of the strike) will have its side reach the boundary, which the center is focused and increases in velocity as it is focused toward the opposing corner. This manner of focusing was not observed in the hexagonal basin.

Next, consider the case where a striker hits the system from a corner or vertex of the triangle, as shown in Fig. \ref{WaveTriCorner}. As the angle between the two boundaries is smaller in this case than in the hexagon case, greater interactions will take place between the wave front and the boundaries. As the wave front approaches the opposite boundary, waves emerge around the initially struck vertex due to reflections. The primary wave front will then reflected back, finally forming a highly localized wave (roughly the size of a single sphere) near the center of the basin. The reflected wave will reach the top vertex at $t=1.75$ms and remains highly localized, and contrast to the broadening of the waves observed in hexagonal basins. 

It is also possible to plot the wave speeds resulting from varying striking angle, but the results will be qualitatively similar to those observed in Fig. \ref{VaryingAngle}, and are omitted.  

Such a triangular basin configuration as we consider here was recently considered in \cite{lisyansky2015primary}. In that study, instead of striking the corner or vertex, the vertex of the triangular basin was placed into contact with a larger rectangular basin, and a wave initiated in the rectangular basin then impacted the single sphere at the vertex of the triangular basin. The results of \cite{lisyansky2015primary} look very much like our results in the configuration where we strike the triangular basin directly, and the initial strike velocity can likely be normalized in such a way so that the results of \cite{lisyansky2015primary} can be recovered directly from a strike without the need for the additional rectangular region.

\subsection{Circular basin}
Another basin geometry that can be considered is an approximation to a circular disc. Such a configuration will not form a perfect disc, but with a larger number of spheres, the boundaries will be smoother and the the shape will eventually look more like a disc. The positioning of the boundaries are defined as follows. A tilt boundary will be prescribed at the end of rows whose number of spheres is only different by one to the adjacent rows and the number of spheres is either monotonically increasing or decreasing among those rows. A vertical boundary will be prescribed at the ends, except no boundary will be assigned at the end of rows whose number of particles is smaller than the both adjacent rows. A horizontal boundary will be set for spheres which have only adjacent spheres either above or below them. For demonstration, we use the configuration with a maximum of 43 spheres in a row.

\begin{figure*}
\centering
\subfigure[]{\includegraphics[width=0.32\linewidth]{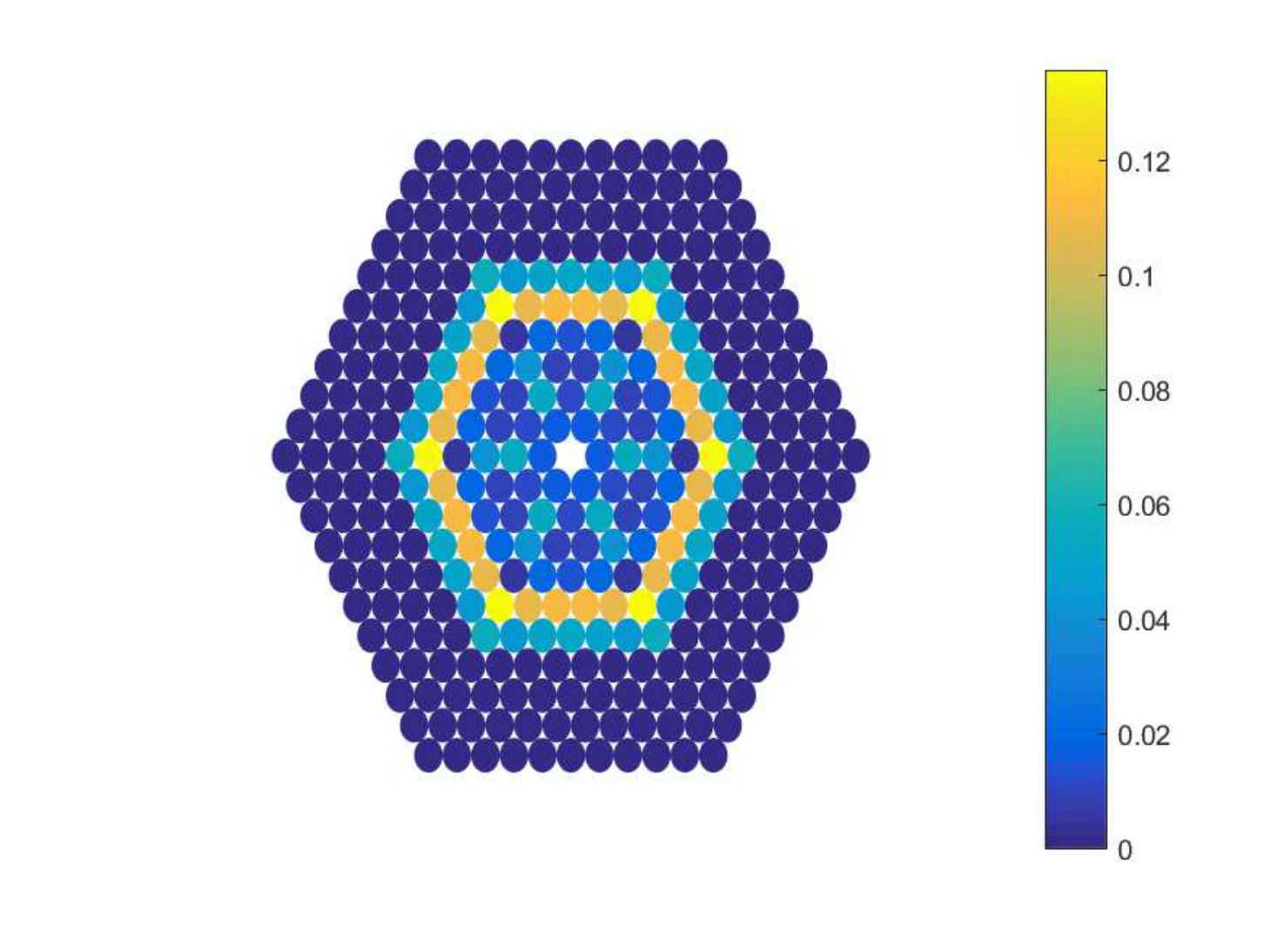}}
\subfigure[]{\includegraphics[width=0.32\linewidth]{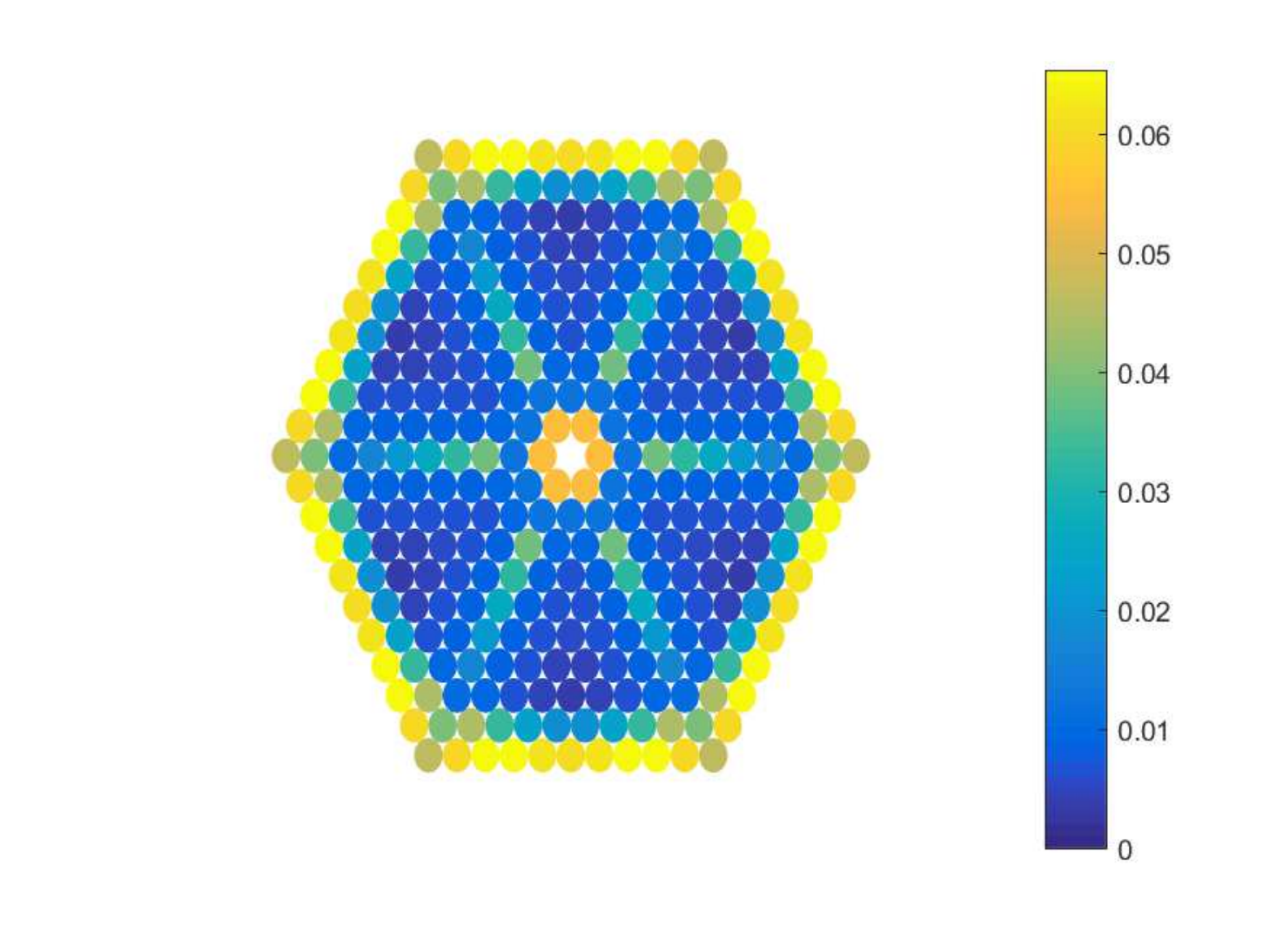}}
\subfigure[]{\includegraphics[width=0.32\linewidth]{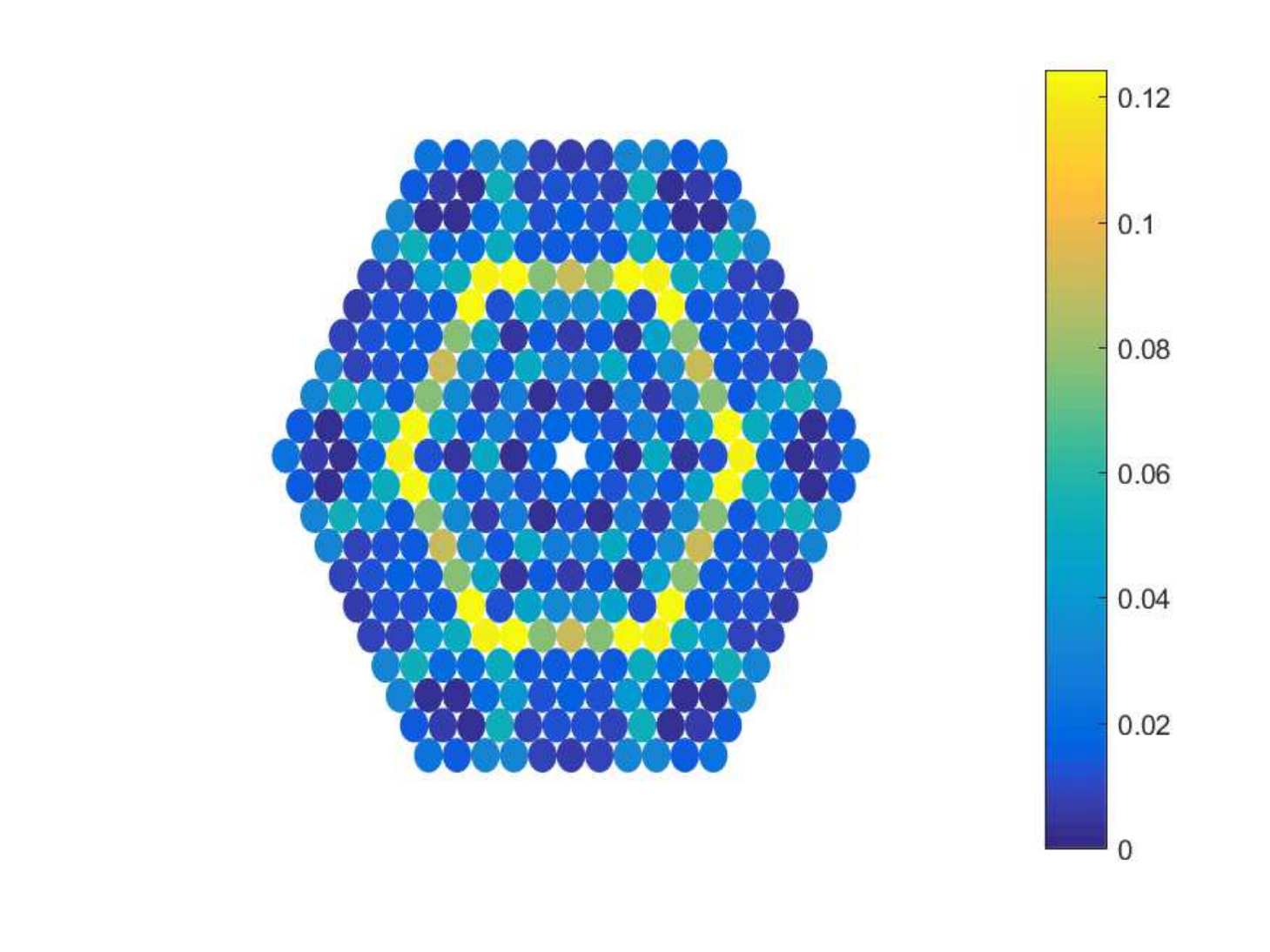}}
\subfigure[]{\includegraphics[width=0.32\linewidth]{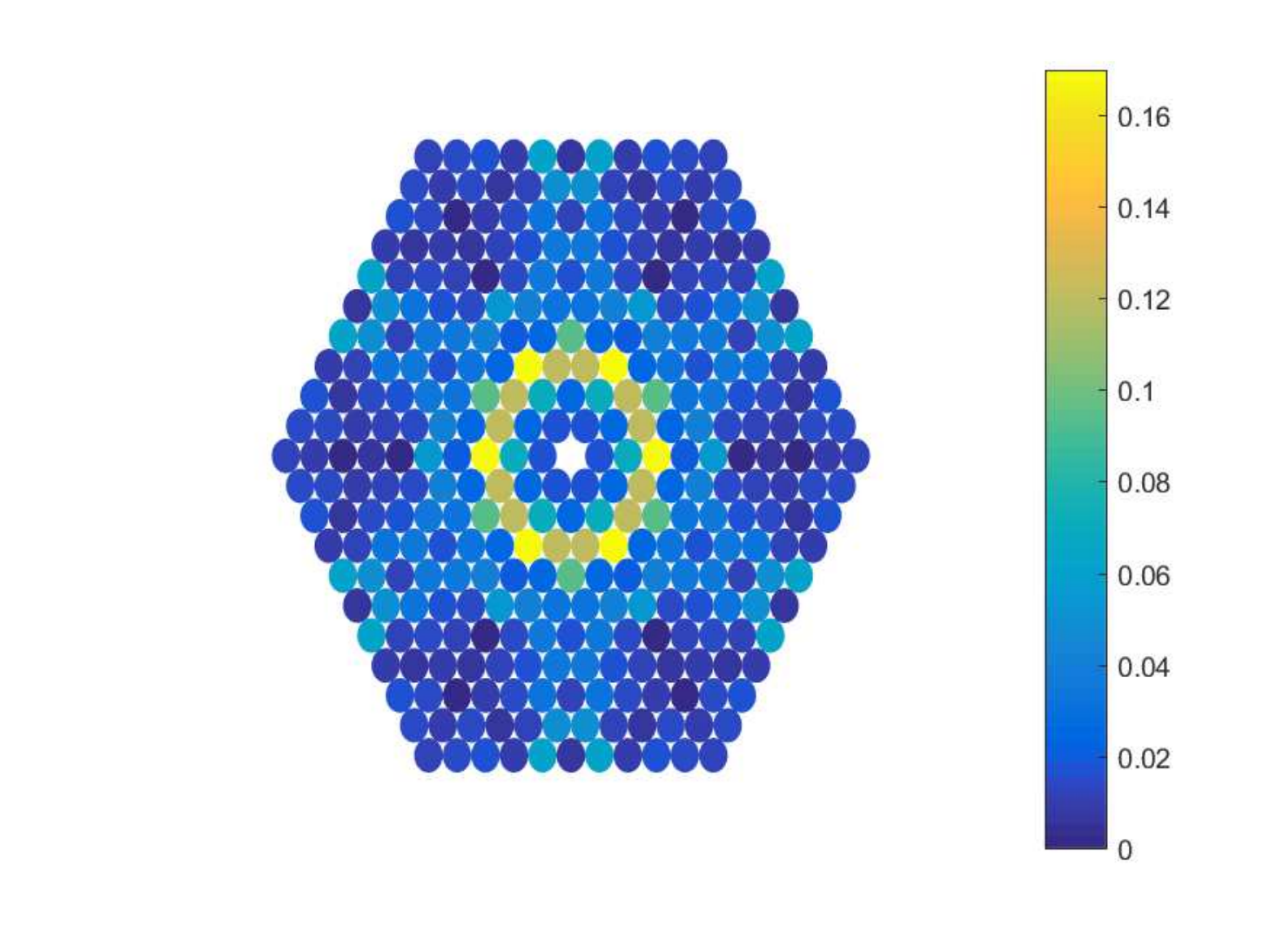}}
\subfigure[]{\includegraphics[width=0.32\linewidth]{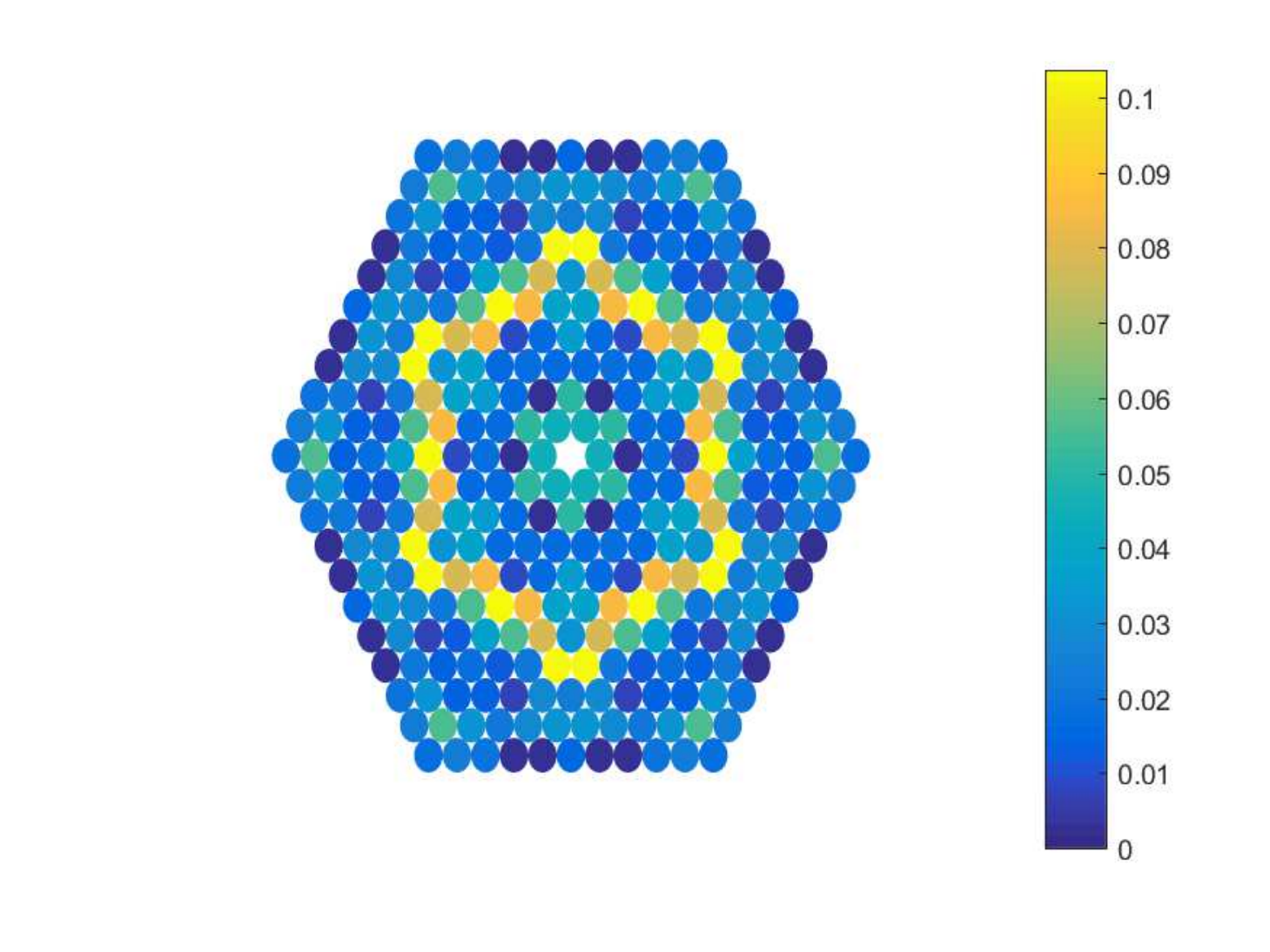}}
\subfigure[]{\includegraphics[width=0.32\linewidth]{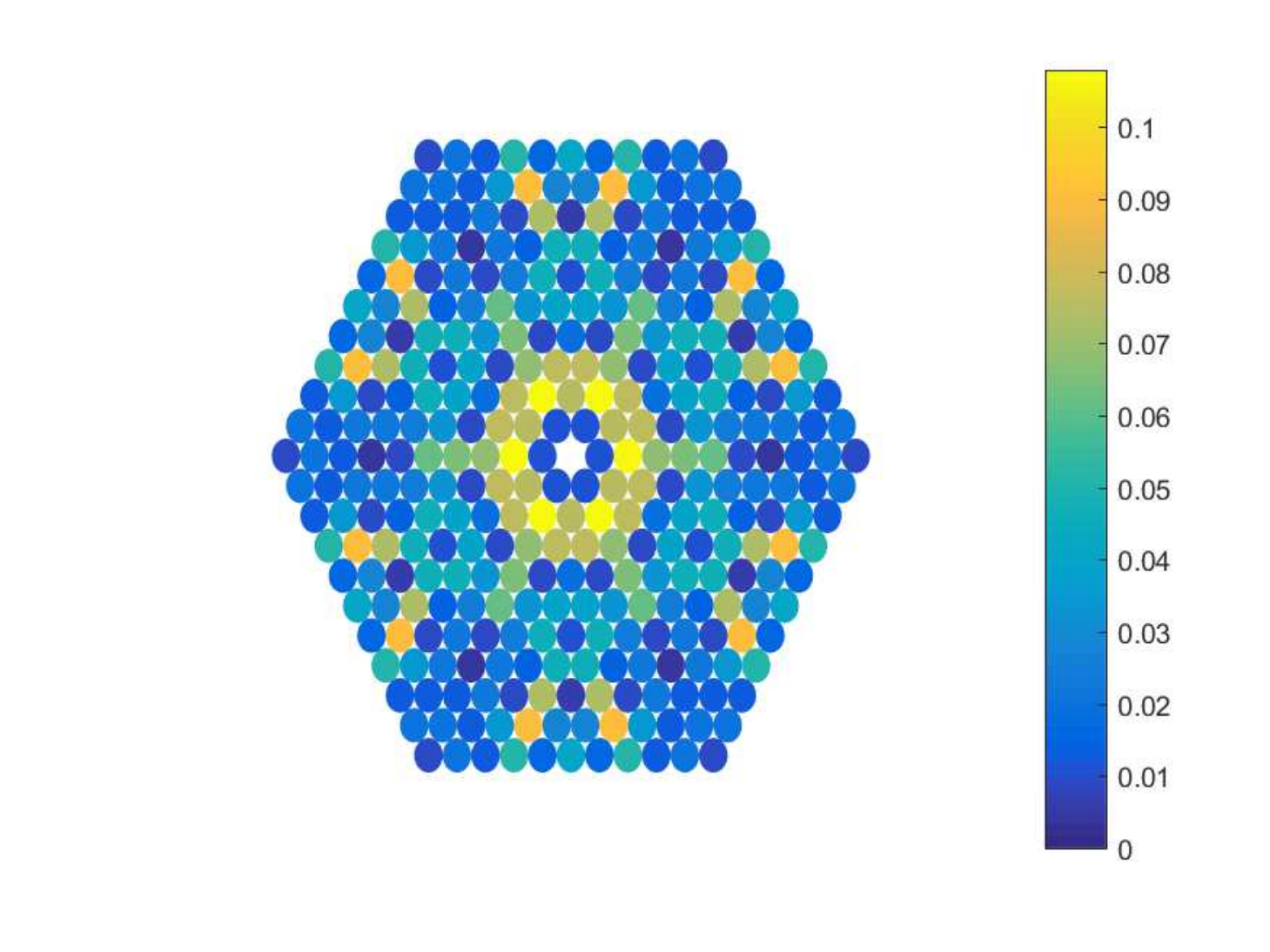}}
\vspace{-0.2in}
\caption{\label{hexstrike1} Velocity magnitudes at each sphere as a function of time for a hexagonal basin filled with a hexagonal packing of spheres. The packing has 11 spheres on each edge. The central sphere is replaced with six walls acting as strikers. The interior walls will be fixed after bouncing back to the initial position.  The initial velocity of each striker is $0.4m/s$. Here (a) $t=0.525$ms, (b) $t=1.00$ms, (c) $t=1.53$ms, (d) $t=1.88$ms, (e) $t=2.75$ms, (f) $t=4.00$ms. This gives an example of symmetric pattern formation. Note that (a)-(b) are before the main wave reflection, while (c)-(f) are after the main wave reflection.}
\end{figure*}

Though the sphere packing remains hexagonal, the complex settings of the boundaries shall influence the characteristics of the waves. As shown in Fig. \ref{WaveCircleUpstriker}, due to the reflected waves from the boundaries, a bottle neck structure has emerged, which is not seen in the hexagonal or triangular basins. There are also several layers of waves caused by reflection behind the leading wave front, which result from nonlinear boundaries around the striking position. After the wave front is reflected by the lower segments of the boundary, the wave front meets with the reflected waves, and a highly localized wave structure is formed and propagates upward in the location of the initial strike. The wave will eventually break and the system will transition into disorder.

The localized wave structure is more broad than that observed in the triangular basin, and has the appearance of a peak with small bands extending from the center, creating an X shape with the maximal velocity at the center of the X. This is somewhat akin to what is seen in certain solutions to shallow water equations, such as KPII \cite{kadomtsev1970stability,infeld2000nonlinear}, and these X patterns can be formed from the superposition of ``line-solitons" \cite{biondini2003family,chakravarty2008classification,chakravarty2013construction}. In the case of hexagonal basins, a strike at a corner (such as in Fig. \ref{Hex:StainlessSteel_OneCornerstriker}) can also generate a reflected wave with large amplitude, however such waves are far less localized. That said, those waves do still exhibit something of a skewed X-shaped pattern.

\section{Wave propagation due to interior strikers}
Instead of striking the system from the exterior at boundaries, we can consider a configuration where we strike the domain from within the interior of a basin. Interior intruders have previously been considered in granular media composed of square packings of spheres \cite{Sze12,Sze13}, where they have modified the propagation of a wave initially created due to a boundary striker. Here we consider the case where an interior sphere is replaced with six walls such that the contact directions between other spheres and the walls are the same as those spheres with the removed sphere. This results in a non-convex basin. We also assume that each of these six new walls has the same weight as the removed sphere. That is, the interior striker can be viewed as six individual strikers, each acting on the adjacent sphere. At $t=0$, the walls will move towards other spheres and this motion results in an interior strike. When the walls bounce back to the initial positions, they will be fixed at those positions.

\begin{figure*}
\centering
\subfigure[]{\includegraphics[width=0.32\linewidth]{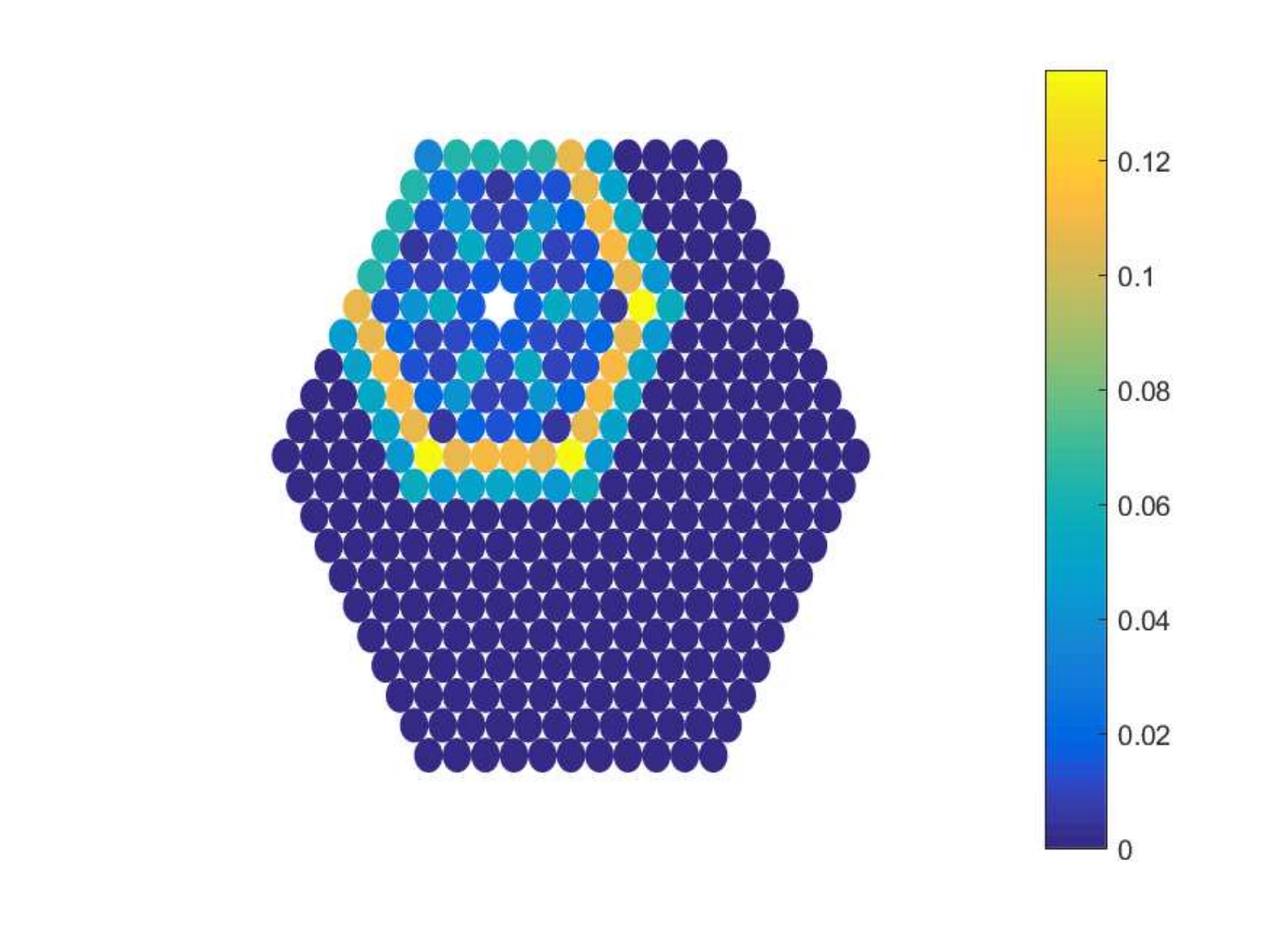}}
\subfigure[]{\includegraphics[width=0.32\linewidth]{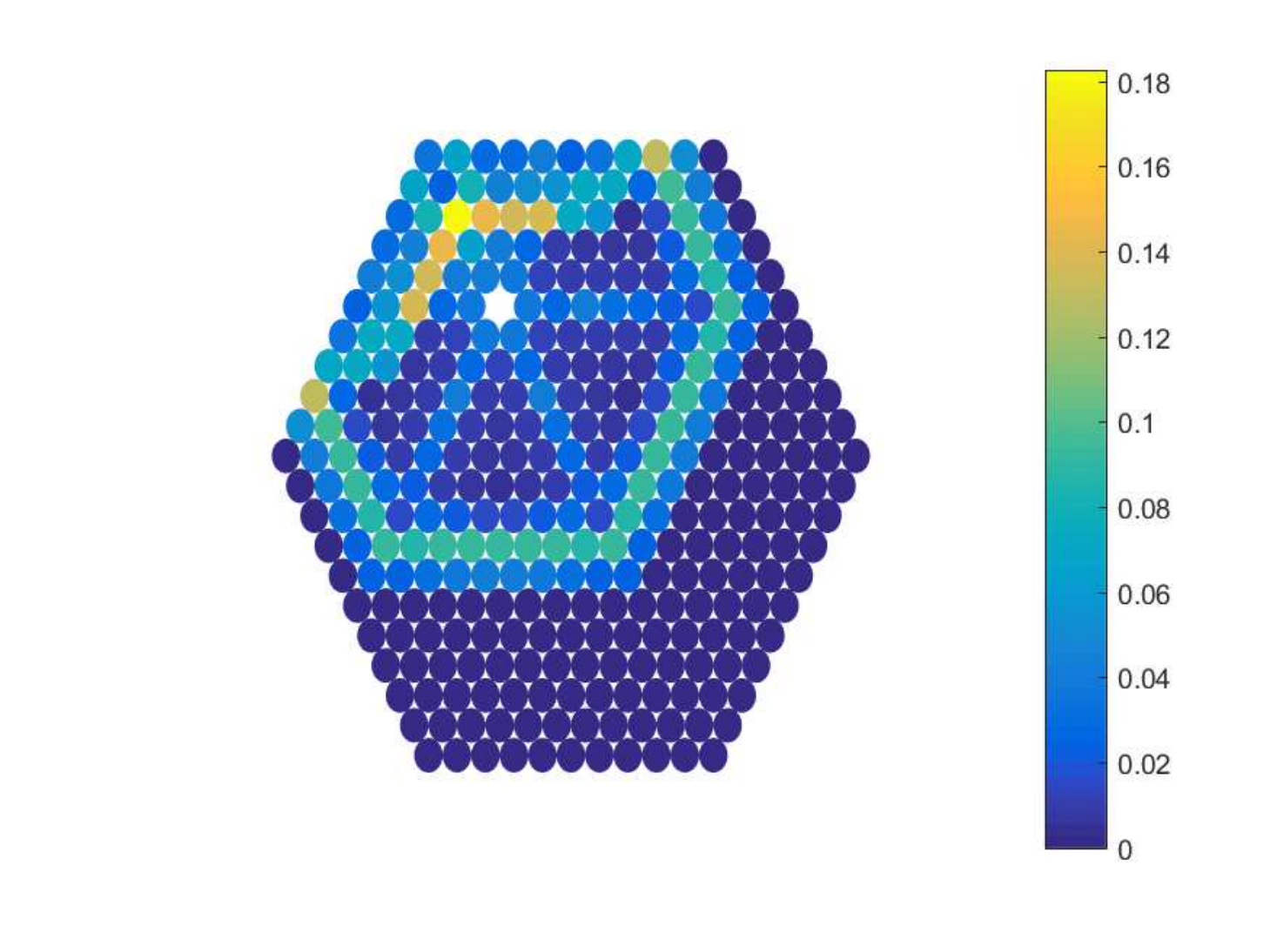}}
\subfigure[]{\includegraphics[width=0.32\linewidth]{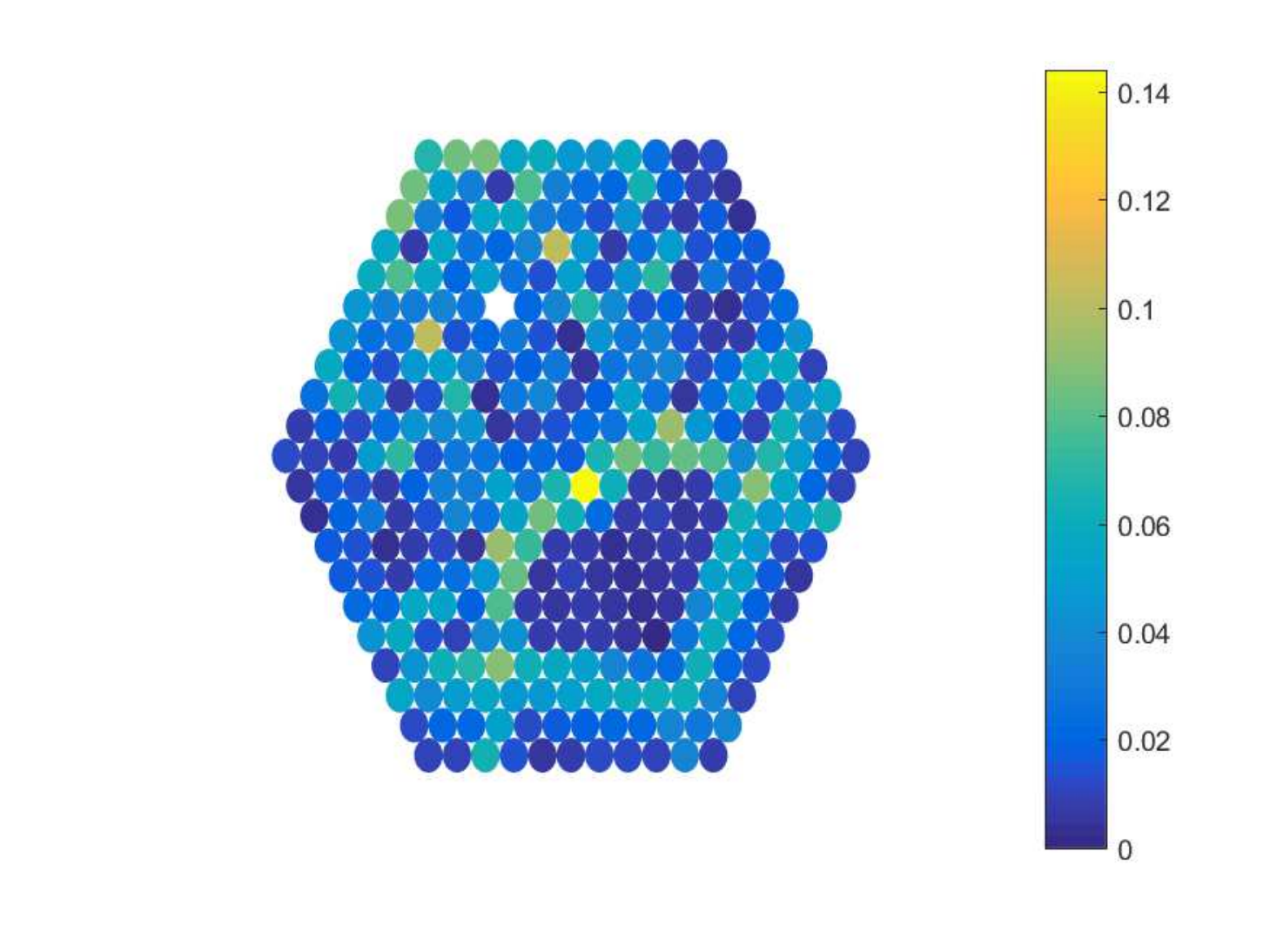}}
\vspace{-0.2in}
\caption{\label{hexstrike2} Velocity magnitudes at each sphere as a function of time for a hexagonal basin filled with a hexagonal packing of spheres. The packing has 11 spheres on each edge. A sphere off the centre is replaced with six walls acting as strikers. The interior walls will be fixed after bouncing back to the initial position.  The initial velocity of the striker is $0.4m/s$. Here (a) $t=0.525$ms, (b) $t=1.20$ms, (c) $t=1.88$ms. Note that (a)-(b) are before the main wave reflection, while (c) is after the main wave reflection.}
\end{figure*}

\begin{figure*}
\centering
\subfigure[]{\includegraphics[width=0.32\linewidth]{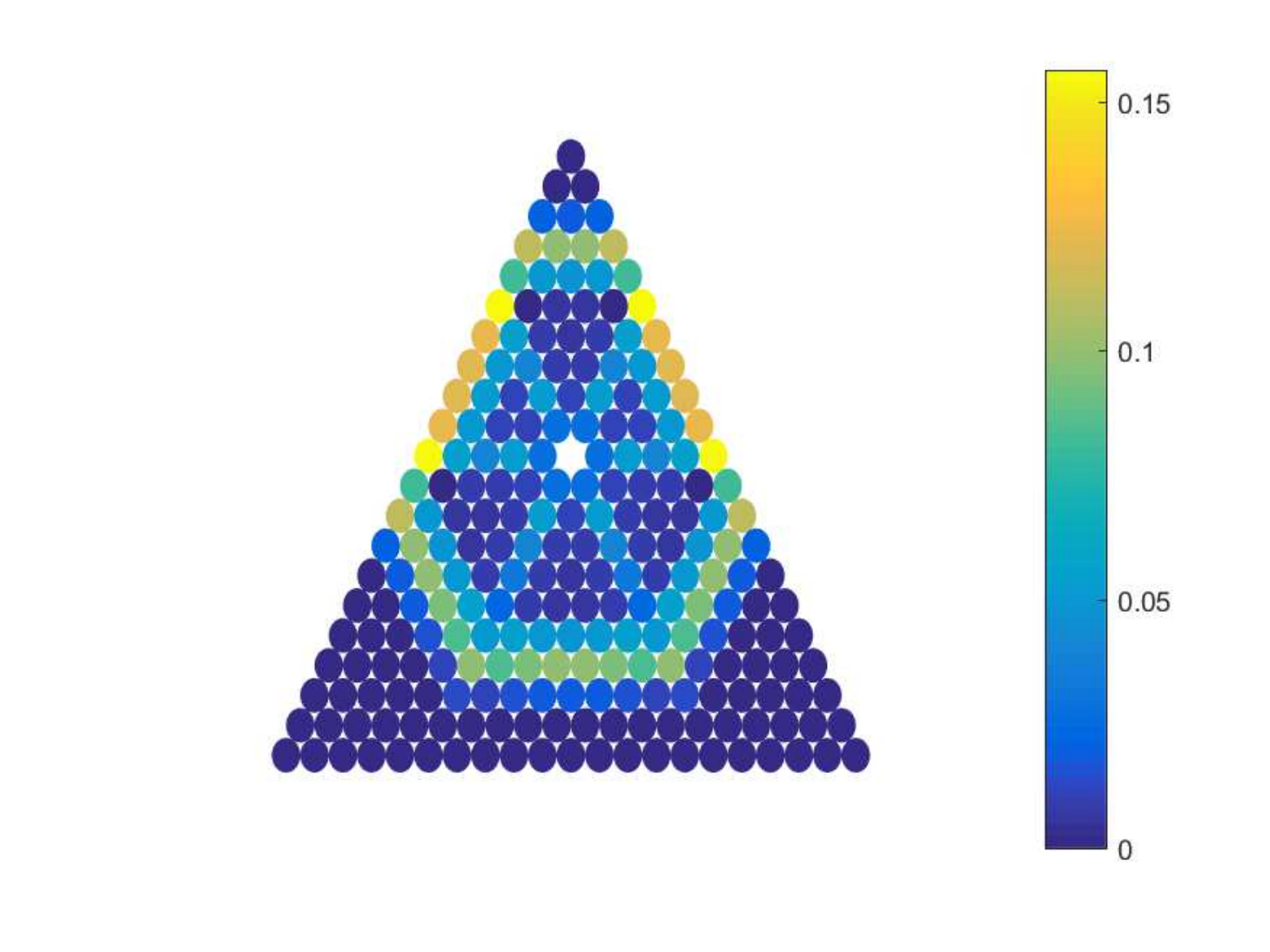}}
\subfigure[]{\includegraphics[width=0.32\linewidth]{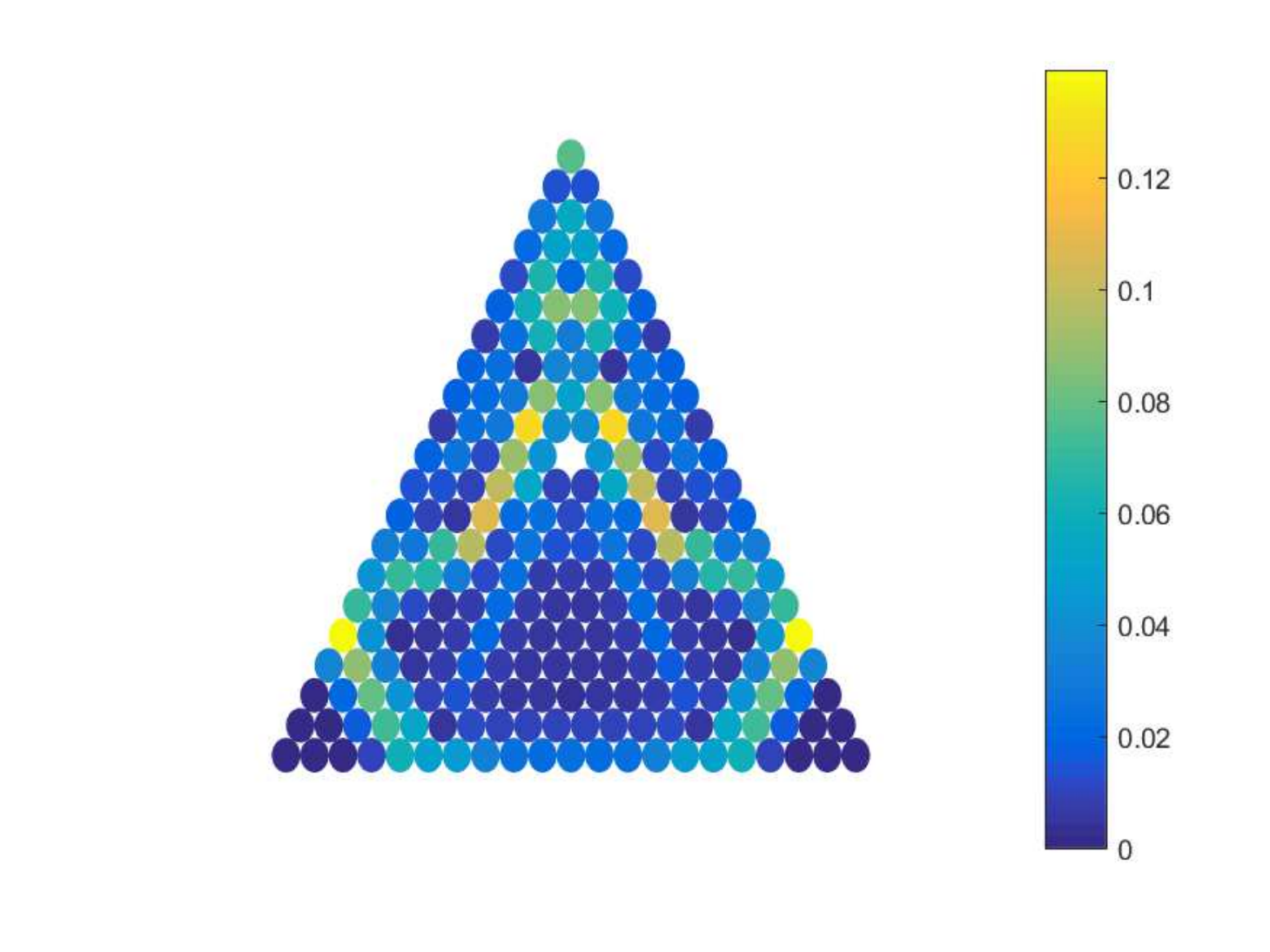}}
\subfigure[]{\includegraphics[width=0.32\linewidth]{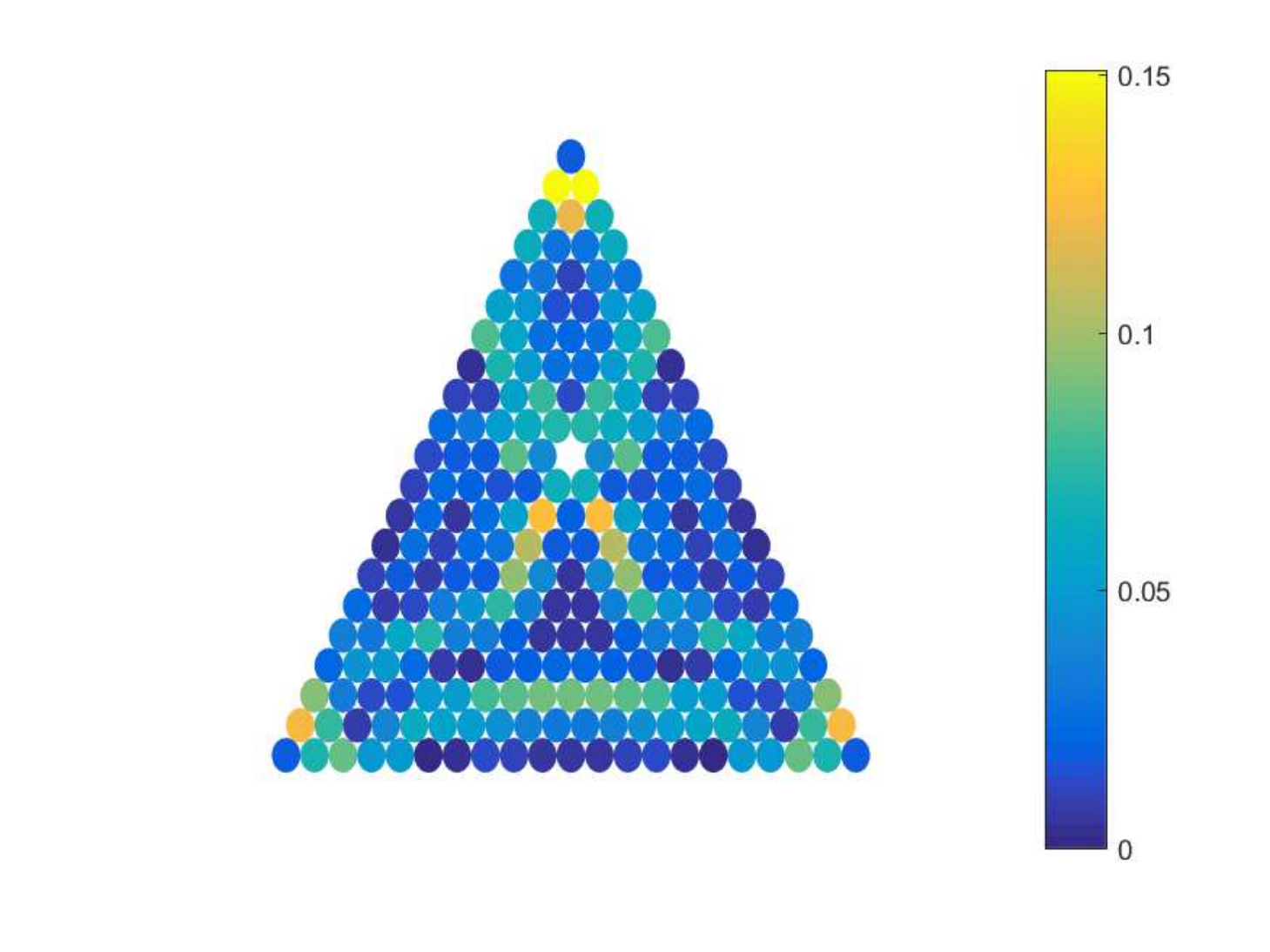}}
\vspace{-0.2in}
\caption{\label{tristrike} Velocity magnitudes at each sphere as a function of time for a triangular basin filled with a hexagonal packing of spheres. The packing has 21 spheres on each edge. The central sphere is removed and replaced with 6 walls. The interior walls will hit other particles, and then bounce back to the original position and stay at that position. The initial velocity of the striker is $0.4m/s$. Here (a) $t=0.250$ms, (b) $t=0.400$ms, (c) $t=0.500$ms. Note that (a)-(b) are before the main wave reflection, while (c) is after the main wave reflection.}
\end{figure*}

Consider first a hexagonal basin with the center sphere removed and replaced with a striker assembly, as described above. At small times after impact, a hexagonal wave is formed, as shown in Fig. \ref{hexstrike1}. The hexagonal wave hits the boundary at around $t=1$ms and is reflected back in the shape of a hexagon. After reflection at the exterior boundary, the hexagonal wave will propagate back and forth between the inner and outer boundary, maintaining a symmetric pattern. For large enough time, the localized hexagonal wave front is lost, although the global wave pattern still exhibits strong symmetry, with new patterns emerging.

Next we consider a hexagonal basin with a sphere removed and replaced with a striker in a way that is asymmetric relative to the symmetry of the basin, as shown in Fig. \ref{hexstrike2}. For small time, a hexagonal wave is generated, however after the first two wave fronts hit the nearest boundaries, some reflected waves will interfere with other wave fronts. The energy will start to concentrate at a small region and start to travel to the boundaries. At later times, the wave generated by the first reflection eventually passes the interior walls, and propagates toward the right-lower corner. This case naturally exhibits far less structure as time grows relative to the initially symmetric case shown in Fig. \ref{hexstrike1}.

The same idea can be applied to triangular basins. In Fig. \ref{tristrike}, an interior striker is placed at the center of the triangular basin. A hexagonal wave is again generated for small time. The three wave fronts in the upper half of the geometry will hit boundaries first and are reflected back. After collisions, the reflected waves from the upper half of the basin propagate into the lower half of the basin. The waves will eventually interact with the waves returning from the bottom, and there will be nodes of higher velocity present near boundaries due to these interactions.

\begin{figure*}
\centering
\subfigure[]{\includegraphics[width=0.32\linewidth]{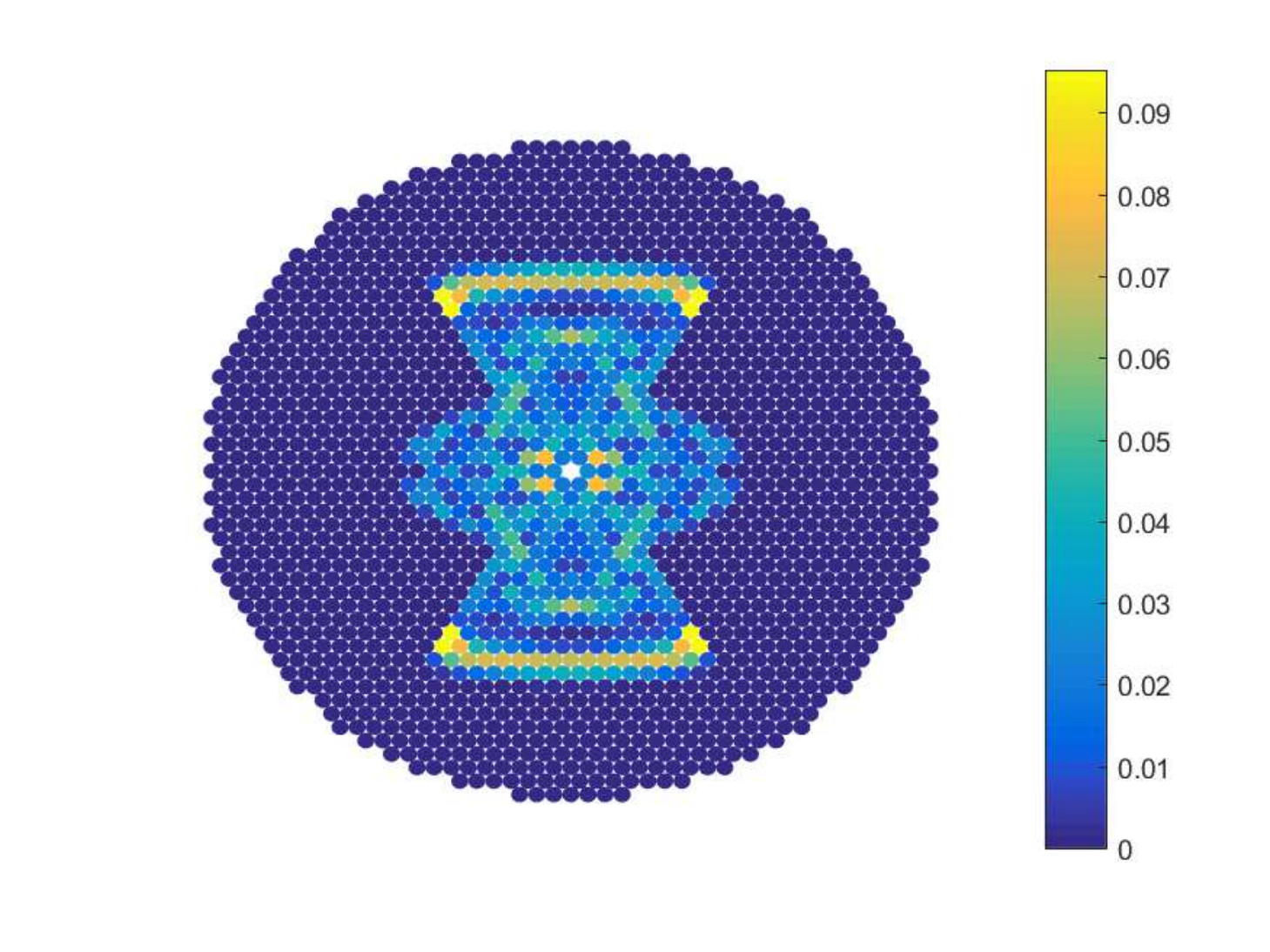}}
\subfigure[]{\includegraphics[width=0.32\linewidth]{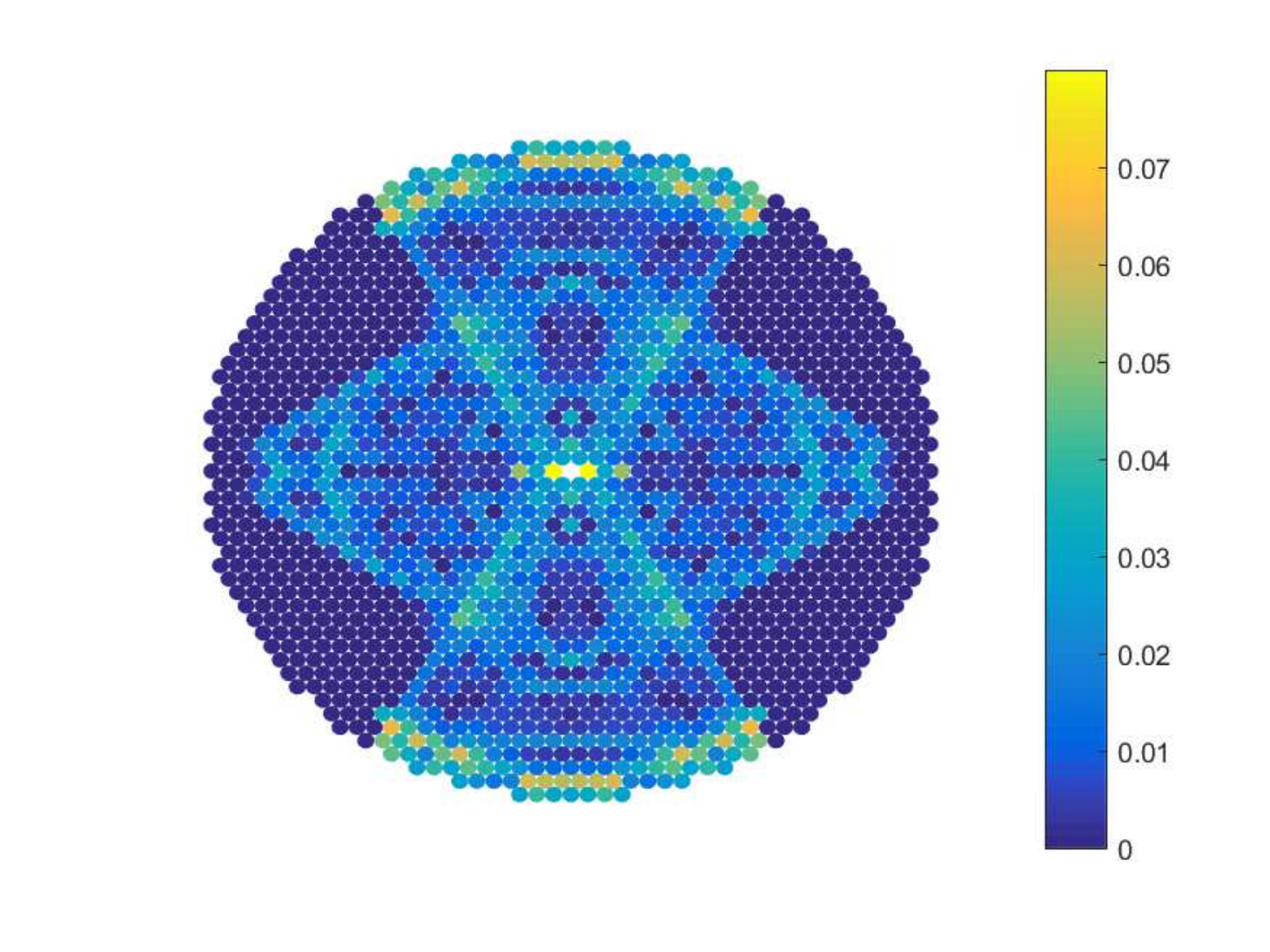}}
\subfigure[]{\includegraphics[width=0.32\linewidth]{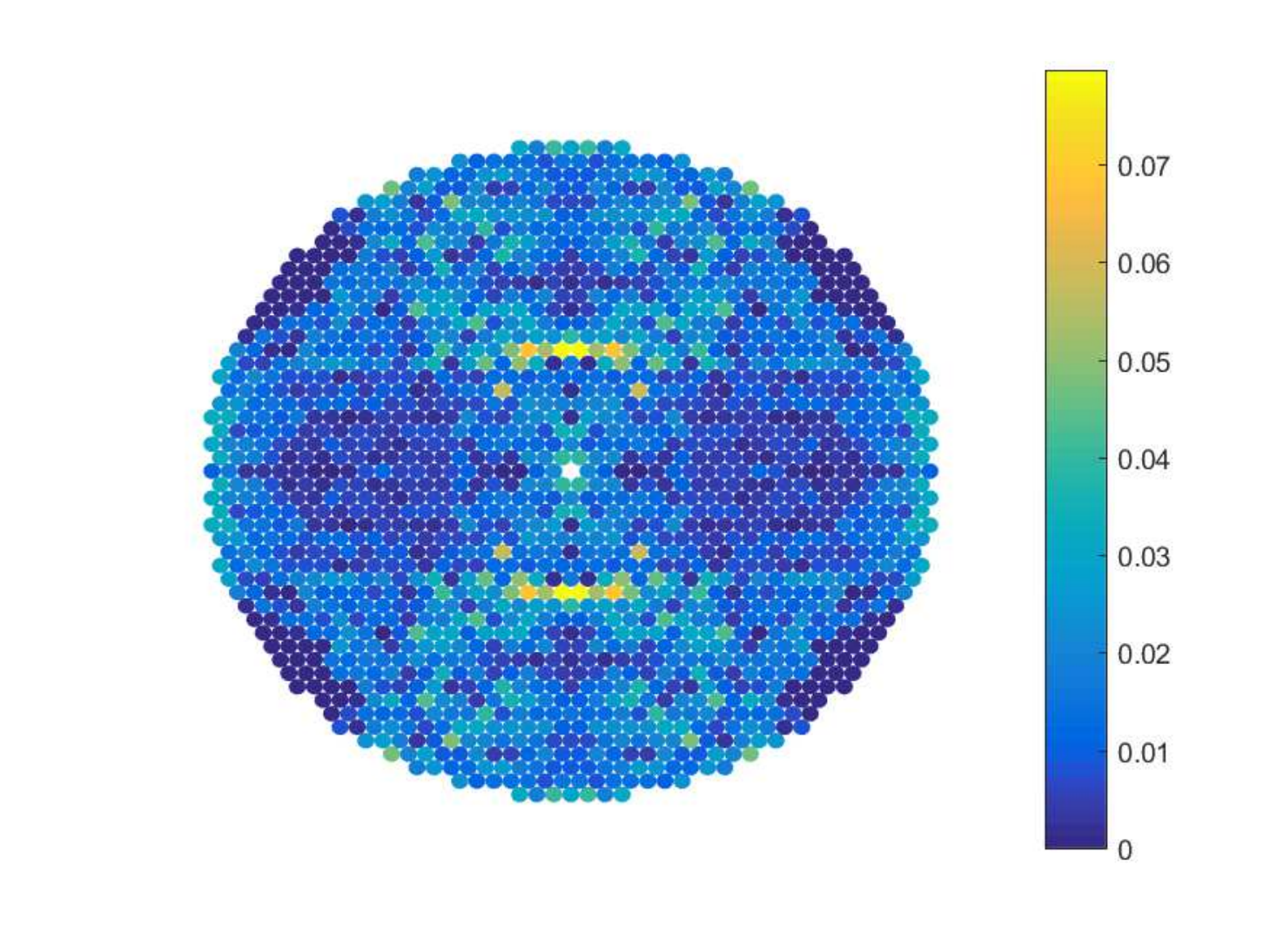}}
\vspace{-0.2in}
\caption{Velocity magnitudes at each sphere as a function of time for a circular basin filled with a hexagonal packing of spheres. The packing has 43 spheres on its longest row. The central sphere is removed and replaced with 6 walls. The interior walls will strike a sphere and bounce back to the original position, becoming fixed there. The initial velocity of the striker is $0.4m/s$. Here (a) $t=0.525$ms, (b) $t=1.00$ms, (c) $t=1.53$ms. Note that (a)-(b) are before the main wave reflection, while (c) is after the main wave reflection.}
\label{WaveCirCentre}
\end{figure*}

\begin{figure*}
\centering
\subfigure[]{\includegraphics[width=0.32\linewidth]{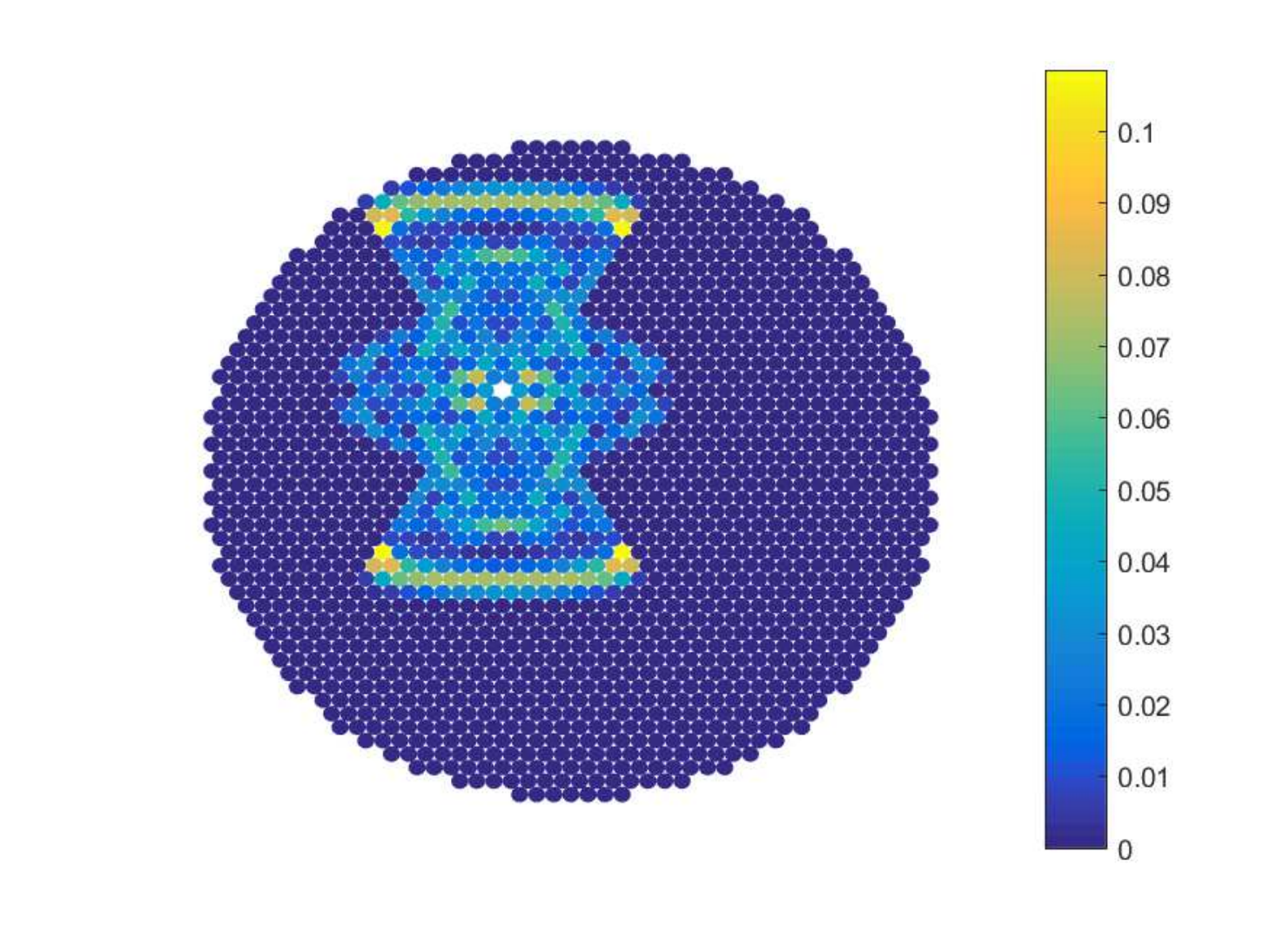}}
\subfigure[]{\includegraphics[width=0.32\linewidth]{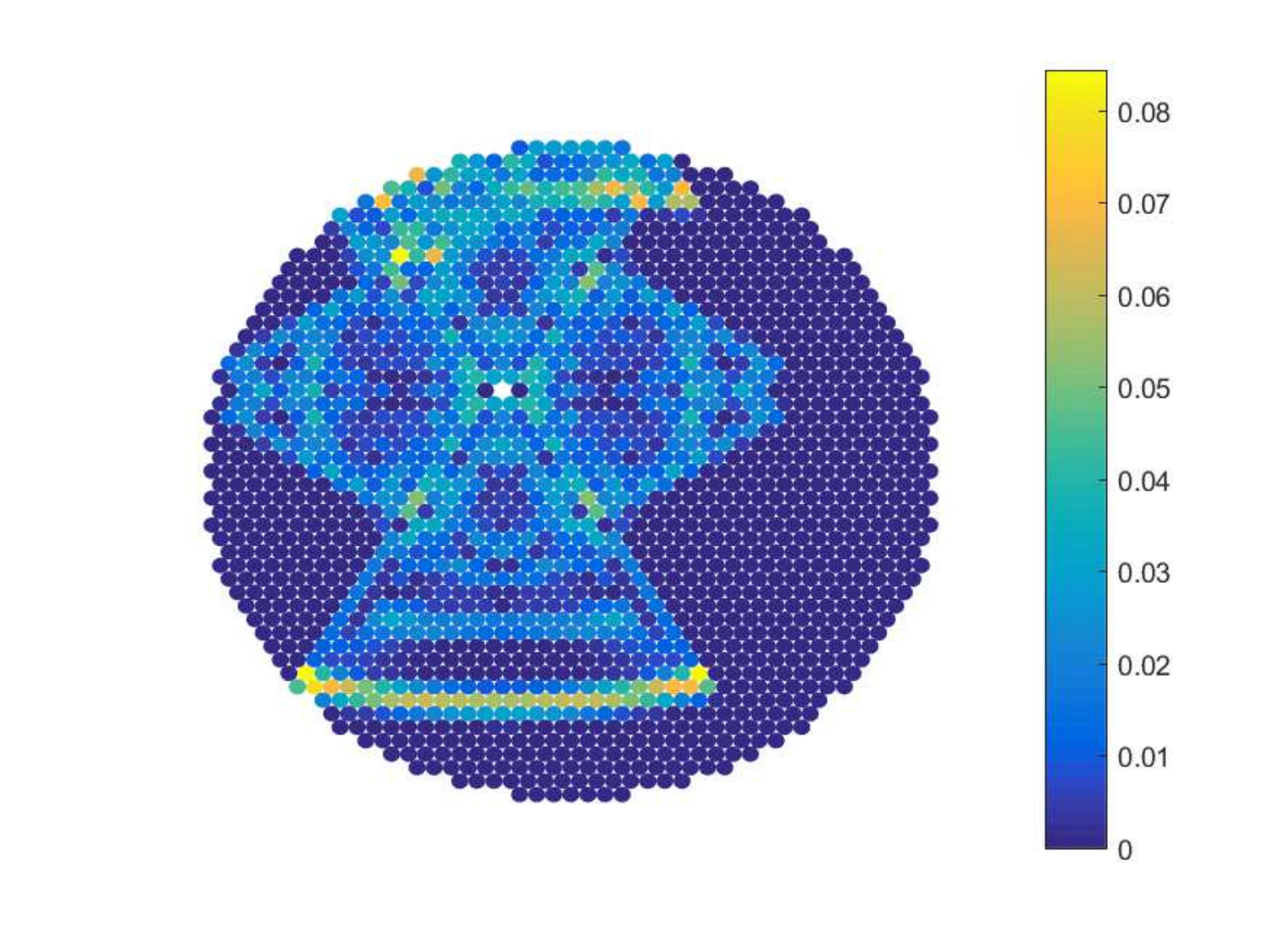}}
\subfigure[]{\includegraphics[width=0.32\linewidth]{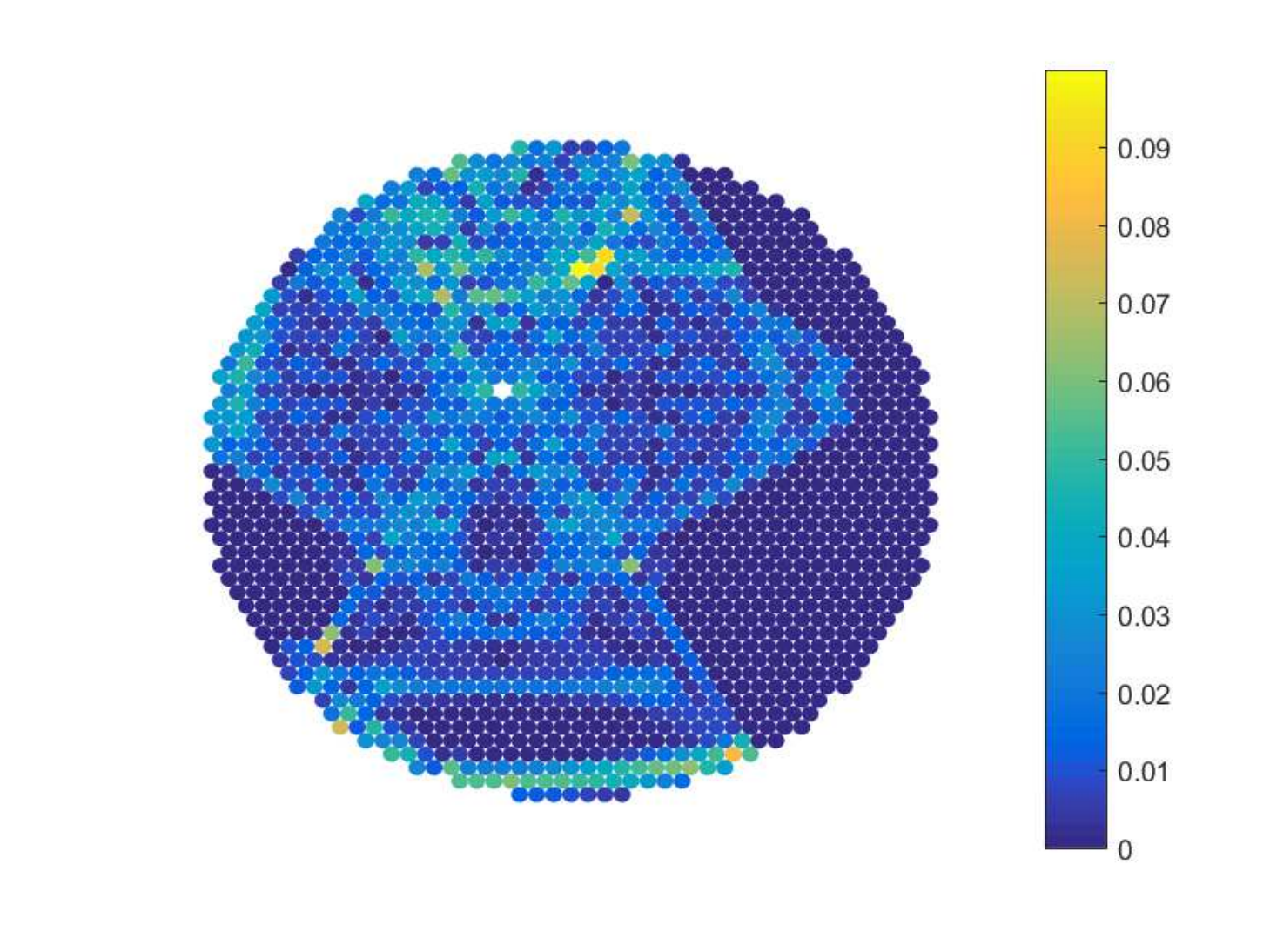}}
\vspace{-0.2in}
\caption{Velocity magnitudes at each sphere as a function of time for a circular basin filled with a hexagonal packing of spheres. The packing has 43 spheres on its longest edge. A sphere off the center is removed and replaced with 6 walls. The interior walls will strike a sphere and bounce back to the original position, becoming fixed there. The initial velocity of the striker is $0.4m/s$. Here (a) $t=0.525$ms, (b) $t=0.875$ms, (c) $t=1.10$ms. Note that (a) is before the main wave reflection, while (b) -(c) are after the main wave reflection.}
\label{WaveCirOffCentre}
\end{figure*}

The geometry and exterior boundaries of the circular basins will be the same as discussed in Section 3, only now we remove one interior sphere and replace it with a six striker assembly. In Fig. \ref{WaveCirCentre}, the results from a striker placed at the center of the circular basin are shown. Unlike in the previous basins considered, two leading wave fronts, moving toward the top and bottom of the domain, have appeared. These leading wave fronts are followed by a circular wave of lower velocity. The shape of the leading waves are similar to the bottle neck structure seen in Fig. \ref{WaveCircleUpstriker}, while the circular wave behind the wave front evolves into a diamond shape. At later times, wave reflection from boundaries results in a loss of order, although highly localized structures of velocity much higher than surrounding locations are observed. In Fig. \ref{WaveCirOffCentre} we give similar results for an asymmetric placement of an internal striker assembly. For small time the wave patterns are nearly identical, yet due to the asymmetry the larger time dynamics become more disordered. There are still small localized regions of higher velocity, although these are no longer symmetrically placed within the basin. 

Of course, we could remove more than one sphere from the center of a basin, to obtain more complicated structures. However, the general principle would be the same. In symmetric cases, the waves would bounce between the exterior and interior boundaries, while for asymmetric cases there will be more rapid transition to disorder. 

\section{Wave propagation across interfaces between two distinct granular crystals}\label{twomedia}
Thus far we have focused on homogeneous granular crystals. However, it is possible to consider heterogeneous structures composed of joining granular crystals composed of distinct materials. Of course, it is also possible to consider other configurations, but we shall restrict our attention to the case where to distinct homogeneous regions are joined together. 

\begin{figure*}
\centering
\subfigure[]{\includegraphics[width=0.32\linewidth]{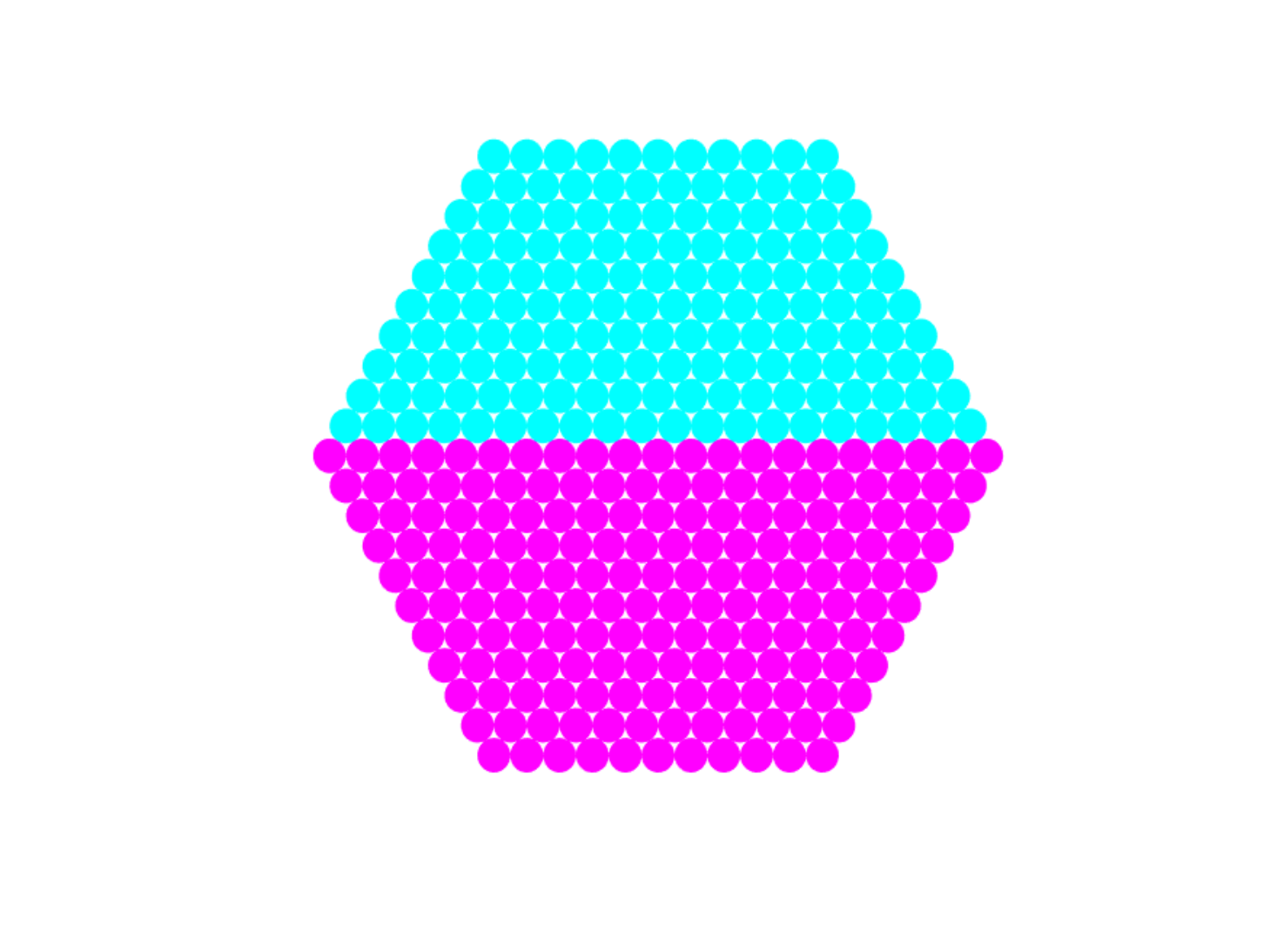}}
\subfigure[]{\includegraphics[width=0.32\linewidth]{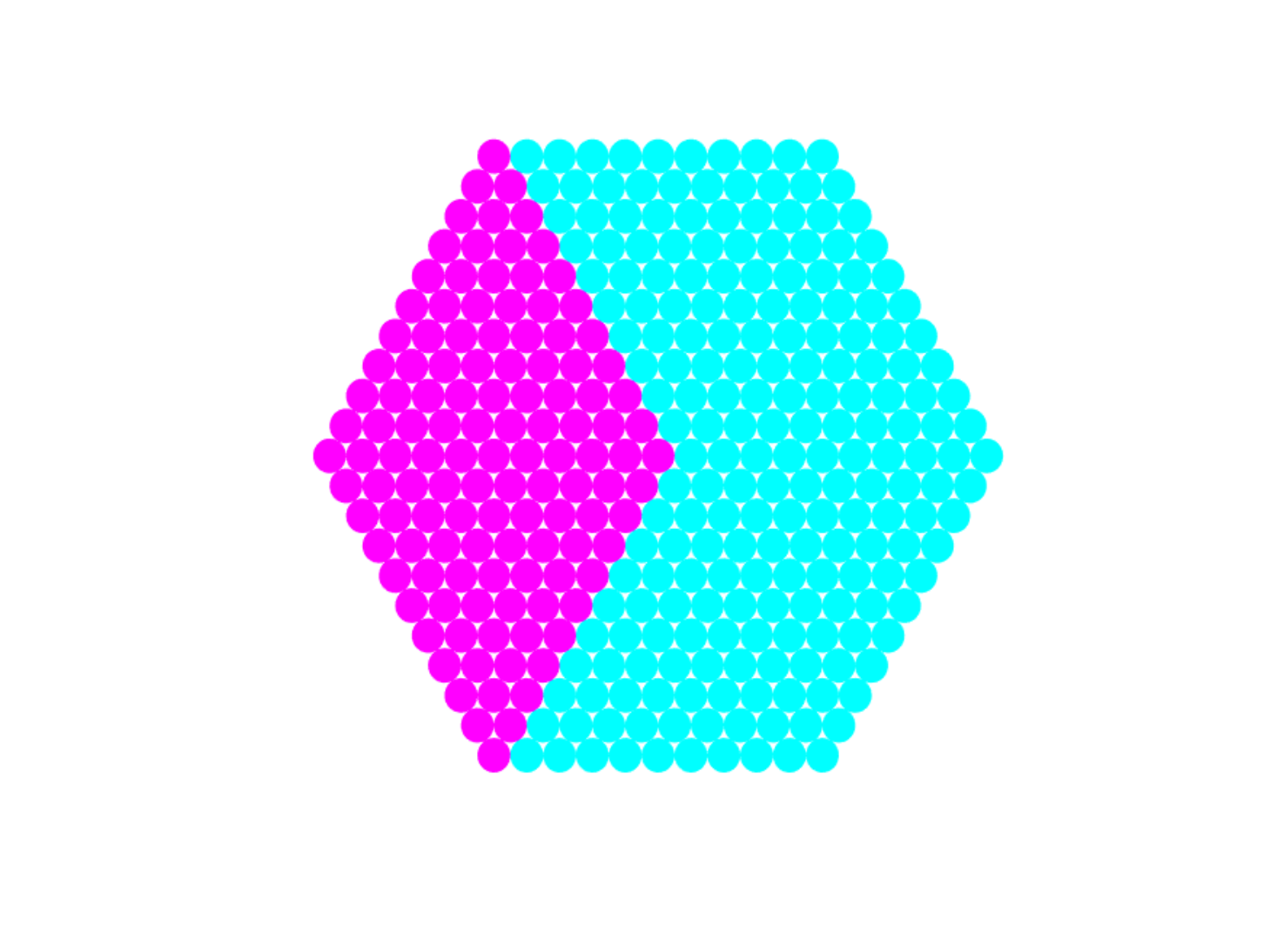}}
\subfigure[]{\includegraphics[width=0.32\linewidth]{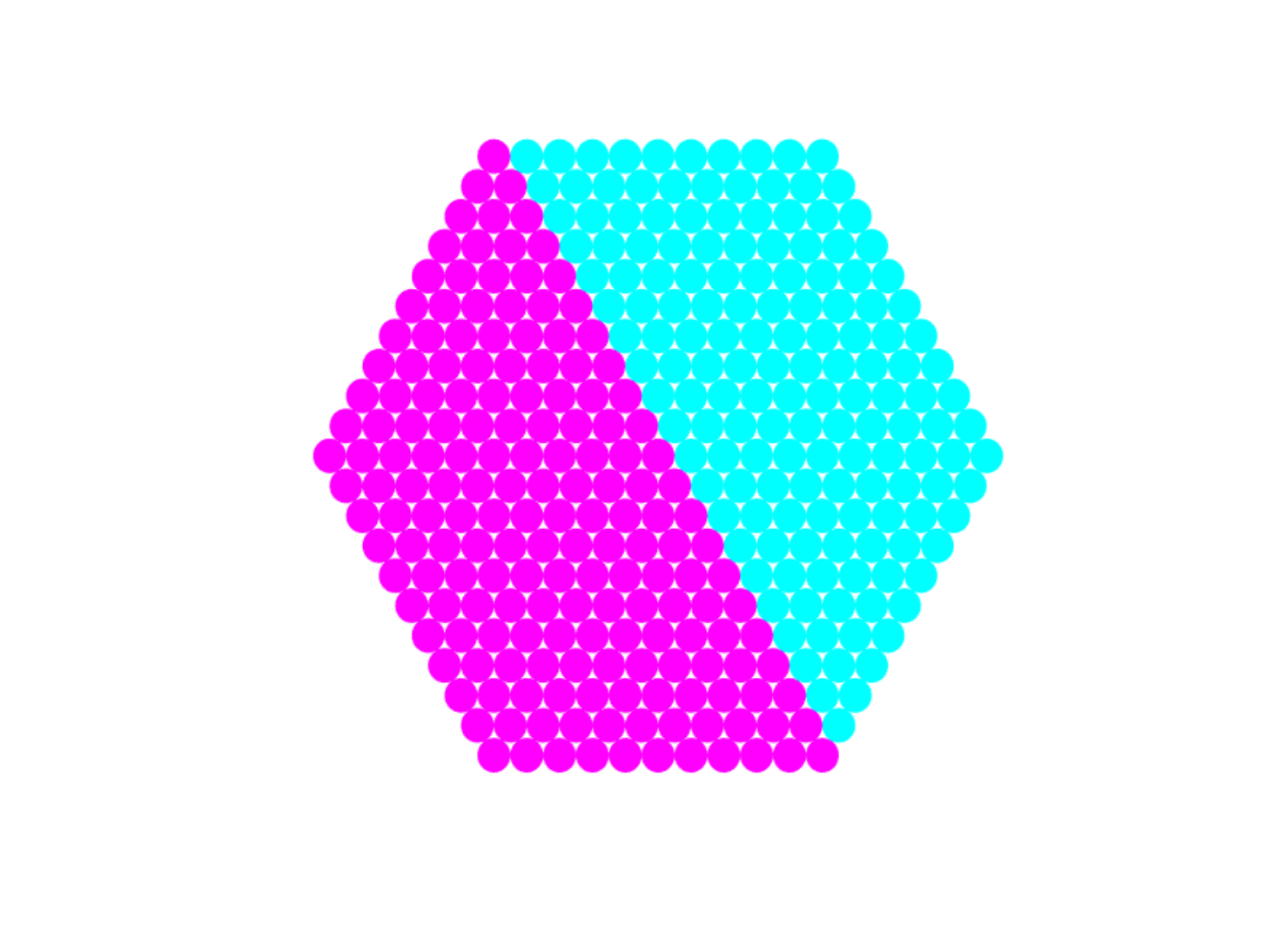}}
\vspace{-0.2in}
\caption{Geometries of various hexagonal packings within hexagonal basins composed of two distinct materials. Spheres marked as blue are made of stainless steel, and purple spheres are made of polycarbonate. See Table \ref{table} for properties of each material.}
\label{IntGeom}
\end{figure*}

\begin{figure*}
\centering
\subfigure[]{\includegraphics[width=0.32\linewidth]{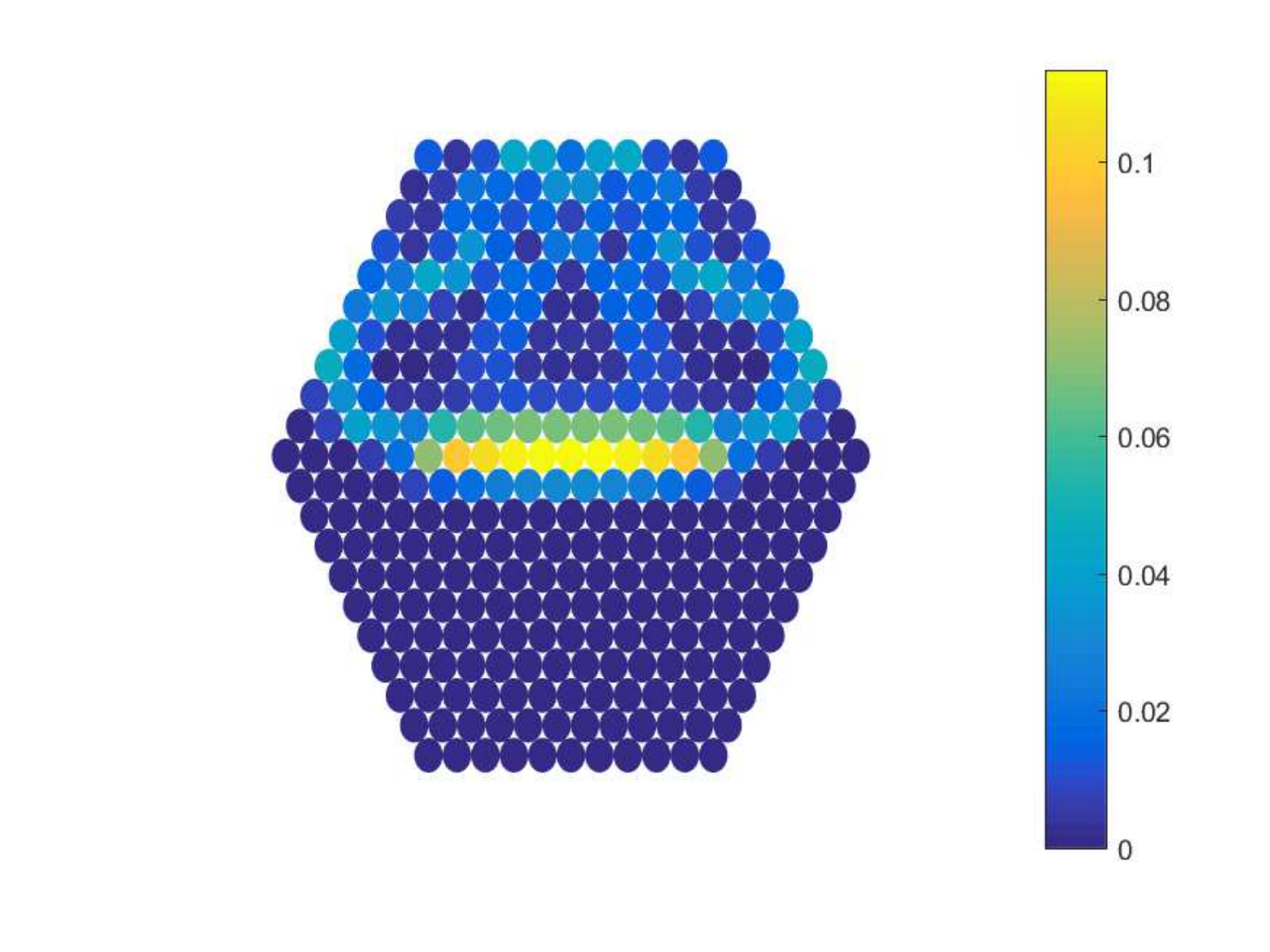}}
\subfigure[]{\includegraphics[width=0.32\linewidth]{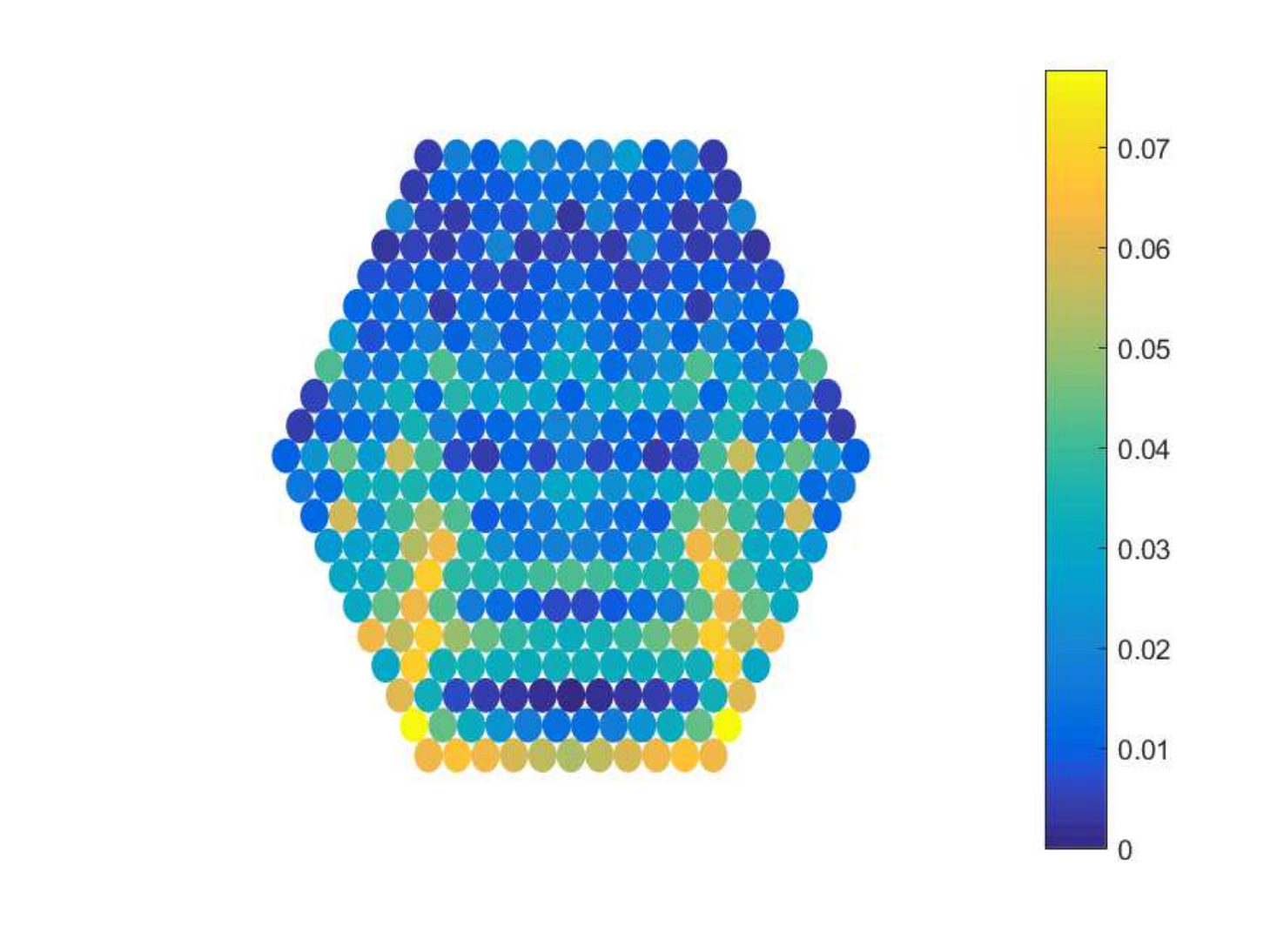}}
\subfigure[]{\includegraphics[width=0.32\linewidth]{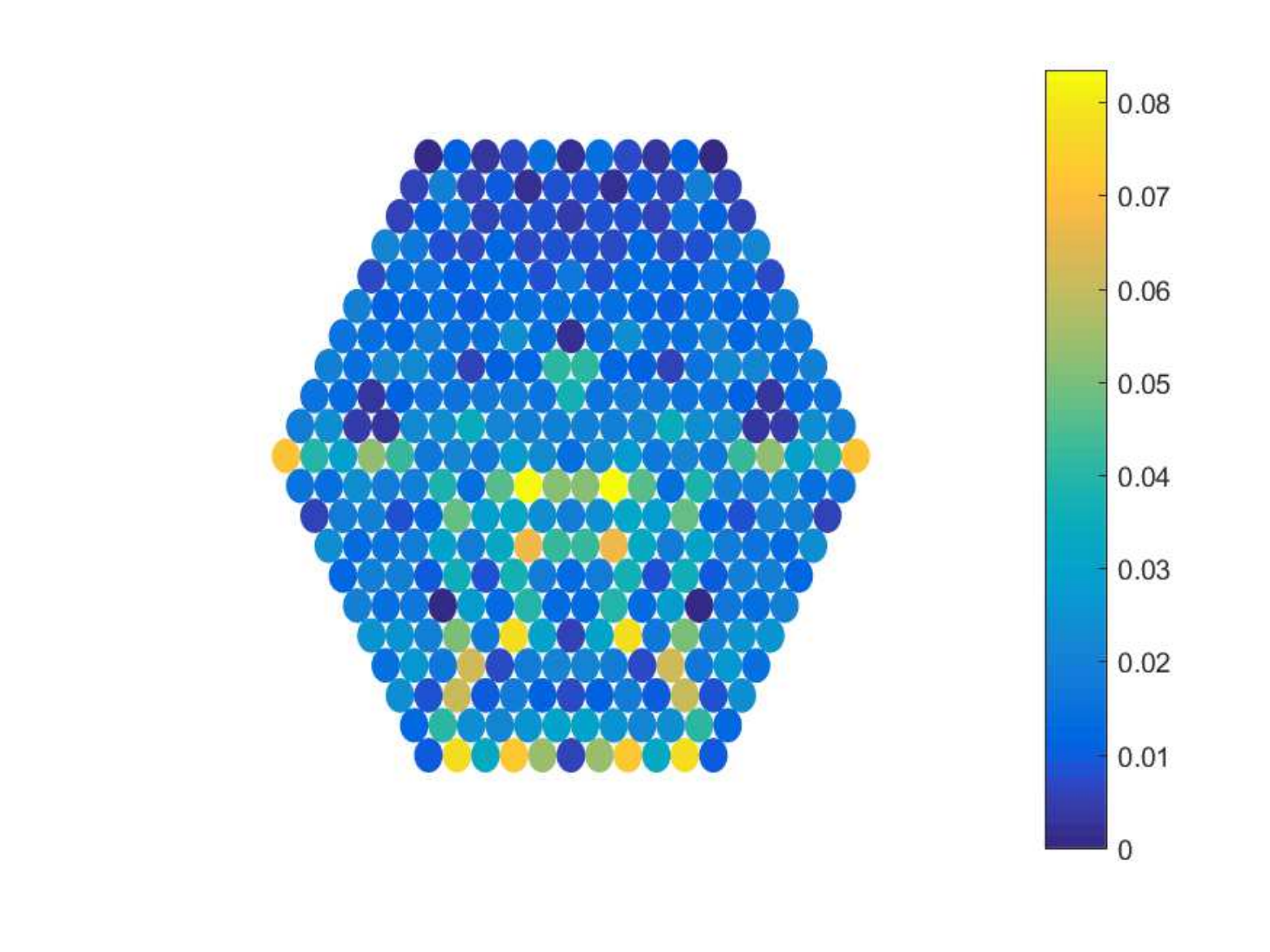}}
\vspace{-0.2in}
\caption{Velocity magnitudes at each sphere as a function of time for a hexagonal basin filled with hexagonal packings of spheres comprised of two distinct materials as in Fig \ref{IntGeom}(a). The packing has 11 spheres on each edge and is struck perpendicularly on one of the edge. The initial velocity of the striker is $0.4m/s$. Here (a) $t=0.525$ms, (b) $t=1.525$ms, (c) $t=2.75$ms. Note that (a)-(b) are before the main wave reflection, while (c) is after the main wave reflection.}
\label{WaveHexInter1}
\end{figure*}

The topic has been investigated in one dimensional granular crystals, and was previously shown that granular crystals with interfaces between spheres of different materials can be used to change the direction of waves under strong precompression\cite{Burgoyne}. The partial transmission and reflection of the incoming wave have been observed for a solitary wave propagating across an interface from a medium with light grains to one with heavier grains \cite{VER01,VER02,sen}. Similar effects to those at interfaces between granular crystals with different bead masses have been observed for mass impurities. Such impurities are created by changing the mass of a single bead in an otherwise monodisperse granular crystal \cite{HAS01,sen,HON02}. For a wave traveling in the opposite direction, the incoming wave is fully transmitted and disintegrates forming a multipulse structure \cite{NES01,VER02}. A second multipulse structure is formed with a delay to the first one. This delay is due to a gap opening and closing between beads at the interface. 

When the solitary wave moves from a lighter to a denser medium, most of its energy gets divided into a reflected and a transmitted pulse, whose respective amplitudes can be estimated by using the conservation of linear momentum and energy
\cite{NES02}. By contrast, when the incident solitary wave moves from a denser to a lighter medium, a train of (multiple) solitary waves is generated in the lighter medium \cite{NES02}.

Few works have considered such configurations in two spatial dimensions. However, \cite{tichler2013transmission} have considered two-dimensional rectangular basins with hexagonal packings of spheres where an interface is created between regions of spheres of different masses. By treating the solitary wave as a quasiparticle with an effective mass, they construct a model based on energy and linear momentum conservation to predict the amplitudes of the transmitted solitary waves generated when an incident solitary-wave front, parallel to the interface, moves from a denser to a lighter granular hexagonal lattice. They extend this model to oblique interfaces, where the angle of refraction and reflection of a solitary wave follows (below a critical value), an analogue of Snell's law in which the solitary-wave speed replaces the speed of sound.

We consider examples where one sub-basin is composed of spheres made of stainless steel and one sub-basin composed of spheres made of polycarbonate (see Table \ref{table} for properties of each). Diagrams of the basins considered are shown in Fig. \ref{IntGeom}.

Fig. \ref{WaveHexInter1} demonstrates wave propagation in a configuration where the upper half of the basin consists of stainless steel spheres, while the lower half is formed of polycarbonate spheres. A strike at $t=0$ produces a wave which propagates perpendicularly toward the interface of the two materials. For small time, the wave looks similar to that shown in Fig. \ref{WaveHexUpOnestriker}. However, after crossing the interface, multiple layers of waves travelling after the leading wave front, likely caused by reflection at the interface. We see that after a long time, the spheres with higher velocities are mainly polycarbonate. The reason behind this is that the polycarbonate spheres are much lighter than the stainless steel ones, and with a roughly equipartition of energy, the spheres made of polycarbonate will have higher velocities after impact.

\begin{figure*}
\centering
\subfigure[]{\includegraphics[width=0.32\linewidth]{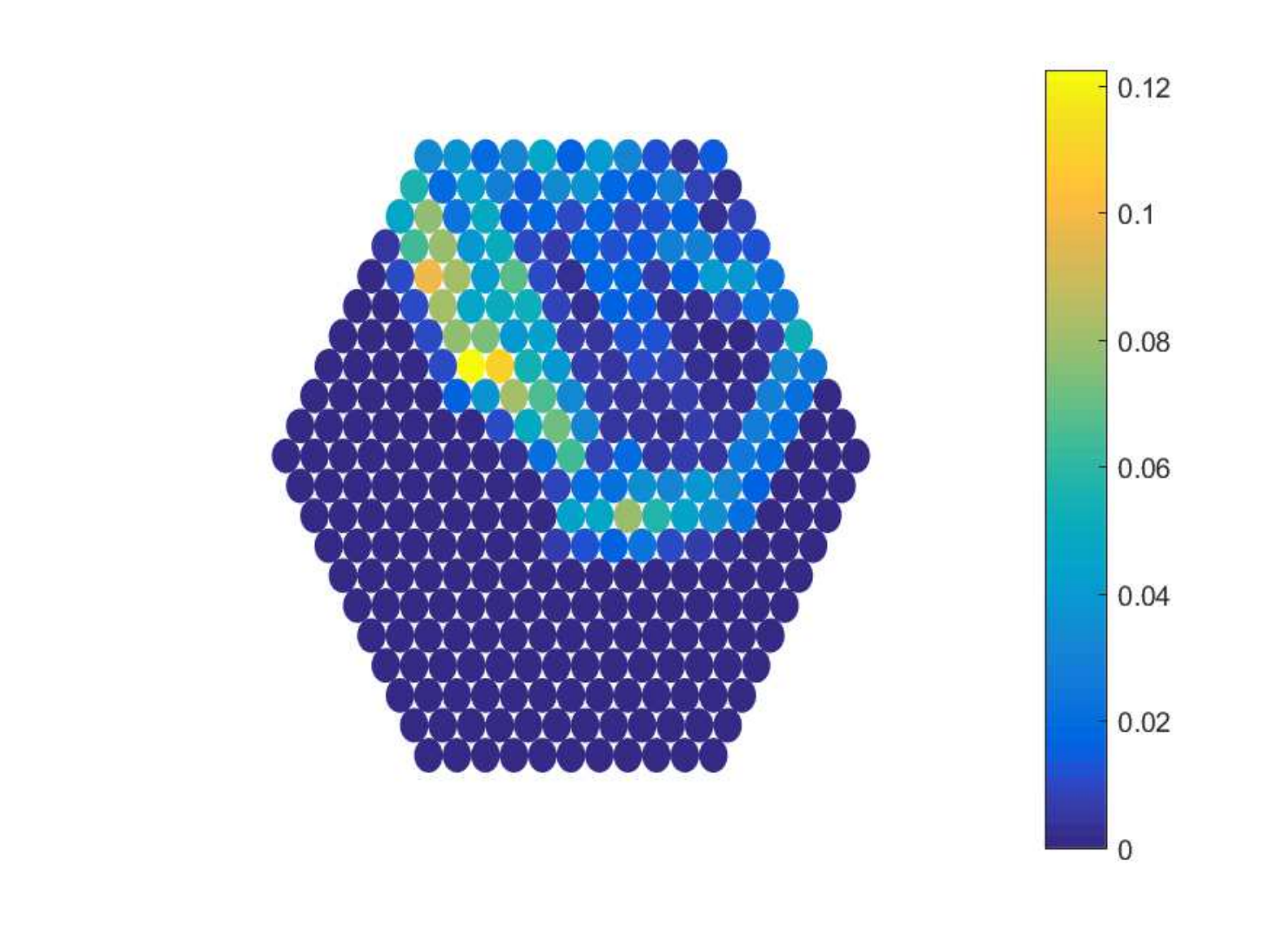}}
\subfigure[]{\includegraphics[width=0.32\linewidth]{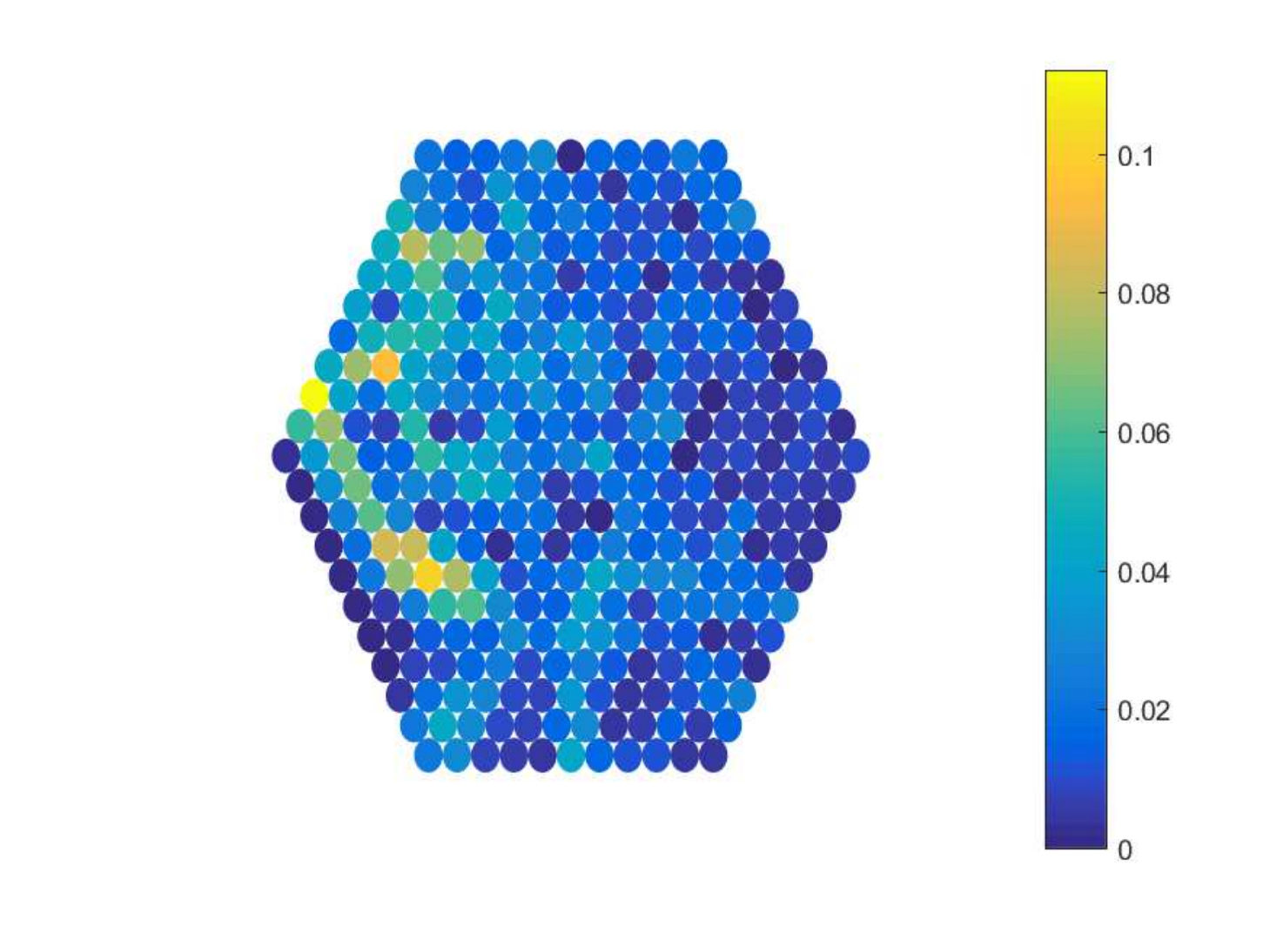}}
\subfigure[]{\includegraphics[width=0.32\linewidth]{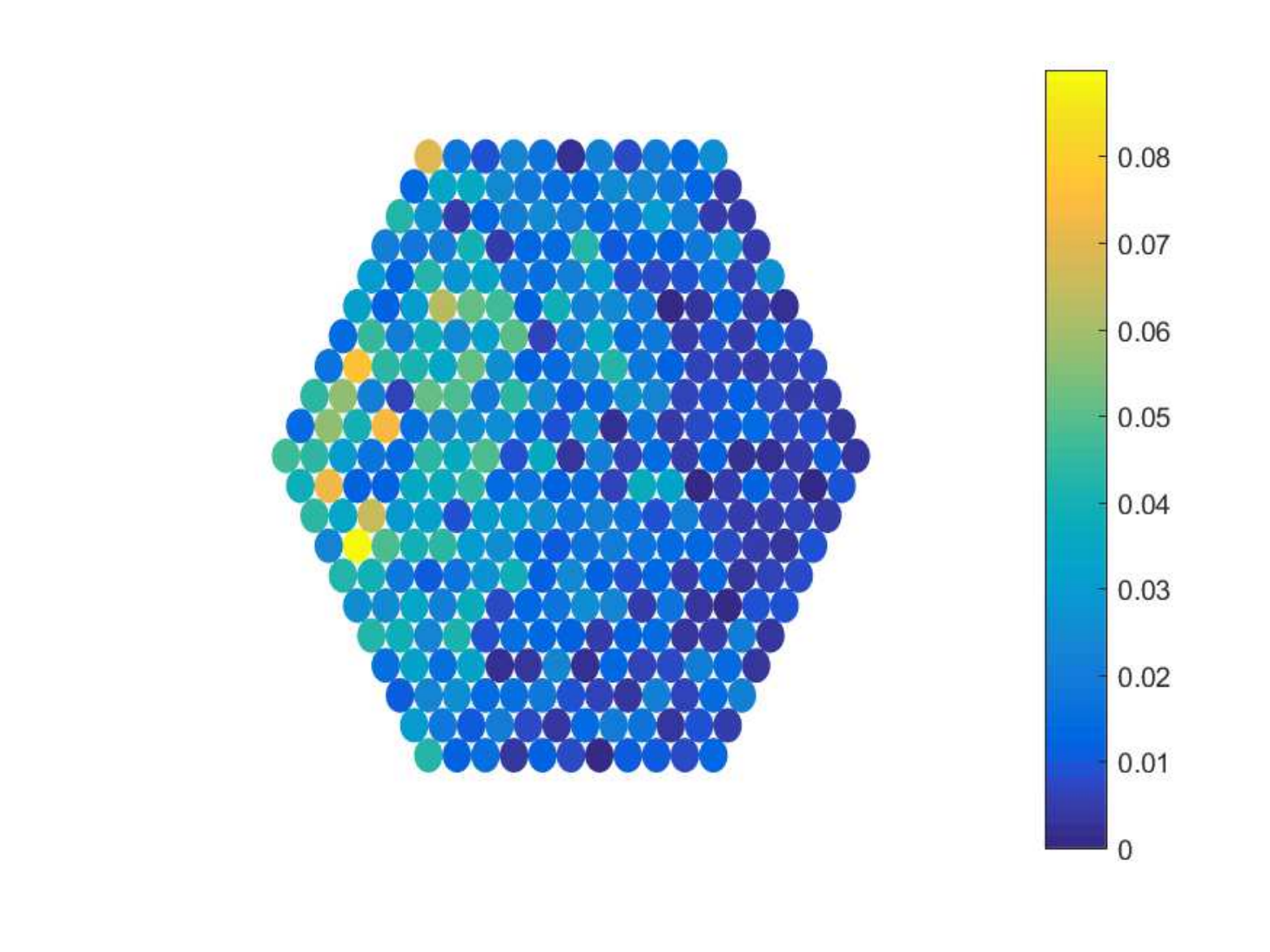}}
\vspace{-0.2in}
\caption{Velocity magnitudes at each sphere as a function of time for a hexagonal basin filled with hexagonal packings of spheres comprised of two distinct materials as in Fig \ref{IntGeom}(b). The packing has 11 spheres on each edge. An interface is created. The initial velocity of the striker is $0.4m/s$. Here (a) $t=0.525$ms, (b) $t=1.20$ms, (c) $t=4.00$ms. Note that (a)-(b) are before the main wave reflection, while (c) is after the main wave reflection.}
\label{WaveHexInter3}
\end{figure*}

\begin{figure*}
\centering
\subfigure[]{\includegraphics[width=0.32\linewidth]{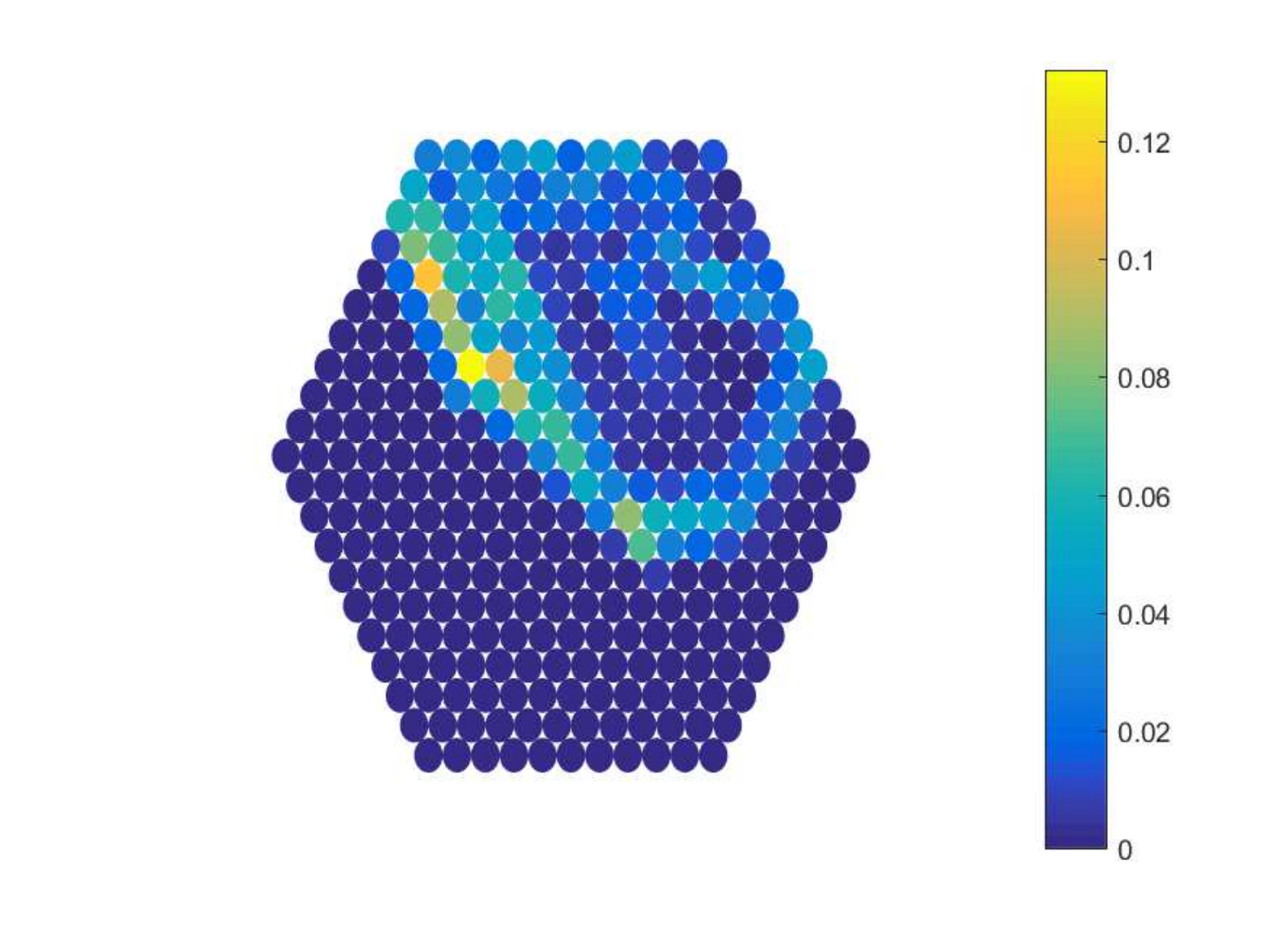}}
\subfigure[]{\includegraphics[width=0.32\linewidth]{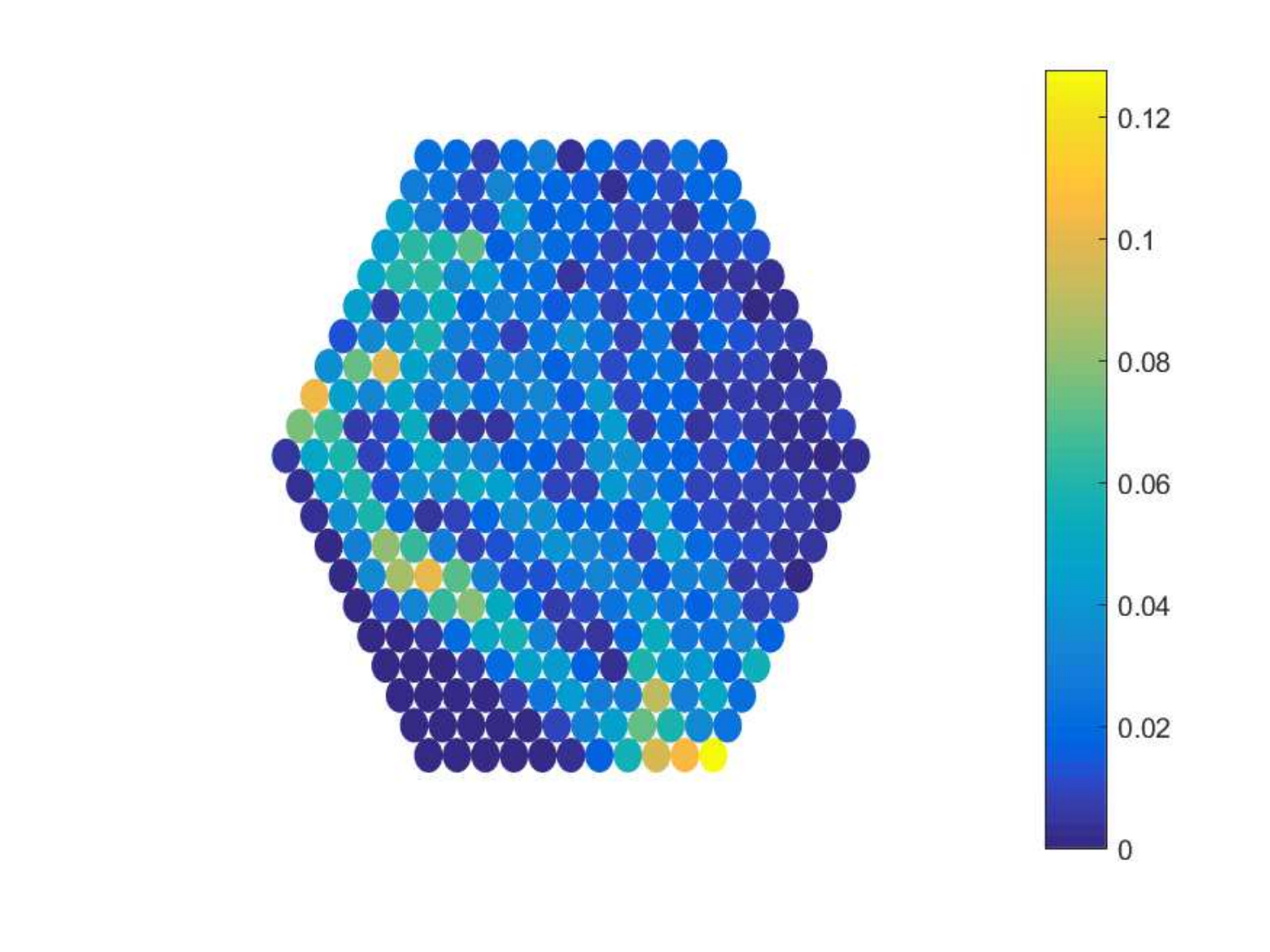}}
\subfigure[]{\includegraphics[width=0.32\linewidth]{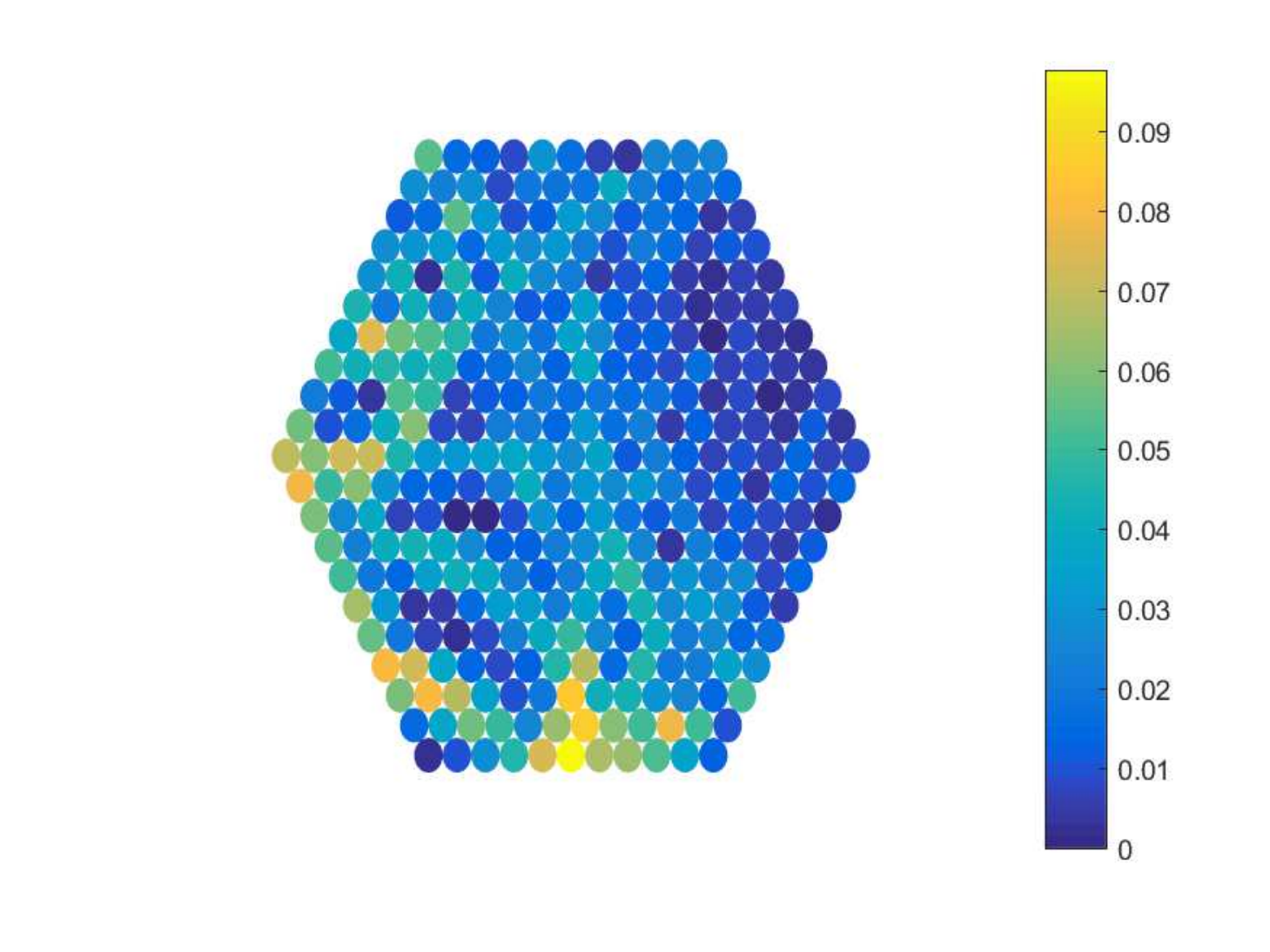}}
\vspace{-0.2in}
\caption{Velocity magnitudes at each sphere as a function of time for a hexagonal basin filled with hexagonal packings of spheres comprised of two distinct materials as in Fig \ref{IntGeom}(c). The packing has 11 spheres on each edge and is struck perpendicularly on one of the edge. The initial velocity of the striker is $0.4m/s$. Here (a) $t=0.525$ms, (b) $t=1.20$ms, (c) $t=1.53$ms. Note that (a)-(b) are before the main wave reflection, while (c) is after the main wave reflection.}
\label{WaveHexInter4}
\end{figure*}

In Fig. \ref{WaveHexInter3}, the arrangement of beads is no longer symmetric in the horizontal direction. As time increases, the interface has slowed the wave propagation and a reflected wave is observed. For similar reasoning to the previous numerical experiment, the spheres in the polycarbonate region will have a higher velocity after a sufficiently long time. In Fig. \ref{WaveHexInter4}, the interface between the two materials is at an angle to the incident wave front. Multiple reflected waves are again observed at the interface. A wave that concentrates at the bottom of the geometry is captured within the polycarbonate region, which can be a result of the collision between the slowed wave front within the polycarbonate region and reflected waves entering from the stainless steel region.

These results demonstrate that interfaces between different material can create reflected waves, as previously seen in the one-dimensional case and in the rectangular basin results of \cite{tichler2013transmission}. Furthermore, higher velocity waves may become ``trapped" in one of the materials, depending upon the relative properties of each material and the basin geometry, with this phenomenon being observed for different geometric configurations. This was seemingly not observed in \cite{tichler2013transmission}, as that study considered large lattices in order to study the local behavior near the interface, rather than the global behavior of smaller granular crystals, for which boundary location and reflected waves will place a stronger role in the dynamics at small timescales.

\section{Discussion}
We have considered wave propagation in two-dimensional granular crystals composed of hexagonally packed spheres. Different configurations of basin geometries, including hexagonal, triangular, and circular regions, have been considered. Various configurations of external boundary strikers are considered, as are interior strikers. We also consider granular media composed of two granular crystals of distinct media joined together, in order to study wave propagation across the internal interface of such media. The results are presented through plots of numerical simulations are fixed time steps.

The results for boundary strikes on a hexagonal basin were examined first. A semi-hexagonal-shaped solitary wave will emerge upon strike (as previously observed in \cite{Leo14}), while there will be significant waves behind the wave front as a result of the complex interaction between spheres and reflections from the boundaries. The waves will collide with each other and the system will eventually fall into disorder. For the case of two strikers on different edges, the wave front which is the closest to the other striker will carry less energy as a result of suppression from waves generated by that striker. 

In the case where the system is struck on a corner, the structure of the wave will be in the shape of an arrow. Reflected waves from different edges will meet each other and form a wave that concentrates in a small region. Changing the direction of the striker will still generate a hexagonal-shaped wave front initially, although the symmetry in velocity is destroyed. The direction of strikes will not change the wave speed significantly.

For a triangular basin, a hexagonal-shaped wave will also be generated when the basin is struck on an edge, while the wave front will look like a line when the basin is struck at a corner. In this basin geometry, the angle the striker makes with the edge can be shown to have some influence over the wave speed, although the wave pattern remains similar.

For a strike at the boundary of a circular basin, the complex structure of the boundaries (due to the discrete approximation of a circle) will result in a significant change to the wave propagation compared with what is seen in the hexagonal or triangular basins. A bottle-neck structure is generated after a boundary strike, and multiple waves are generated at the region near a strike. These waves will be reflected to the center, and the energy will be concentrated there after some time.

We next considered cases where a striker is placed in the interior of a basin. Hexagonal-shaped waves appear in hexagonal and triangular basins, while a similar bottle-neck structure to that mentioned above will emerge when the circular packing contains an interior striker. The shapes of waves observed when an interior striker is used at the center of a basin will look like a mirror reflection of the wave structures obtained in the boundary striker case. However, if the interior striker is placed away from the center of the basin, then the wave pattern will naturally exhibit a lack of symmetry, and the system will tend for fall into disorder more rapidly.

The influence of having a granular media composed of two distinct granular crystals, each homogeneous, was considered next. These structures will feature an interface between the two materials, which the wave generated by a striker must pass through. For this case we focused on the hexagonal basin, although similar results were observed for numerical simulations using other basins. A lighter material was used in one of the regions, and the waves are seen to be slowed and reflected when crossing the interface to enter this region. These results are in agreement with previous experiments and simulations using one-dimensional granular chains, and the behavior near an interface is in agreement with prior results of \cite{tichler2013transmission} for two-dimensional granular crystals. In addition to those results, we observed trapping of waves within specific regions, depending on their respective mass ratios. The wave trapping we observe was previously observed under a rather different two-dimensional geometric configuration which resulted in passive wave arrest \cite{hasan2016nonlinear}. 

On a more theoretical note, the utility of one-dimensional granular chains for demonstrating interesting nonlinear wave structures has been discussed previously \cite{chong2014damped}, and our findings suggest that two-dimensional hexagonally packed granular crystals may similarly provide an additional application where interesting nonlinear wave structures can be found. We find that certain basin geometries and striker configurations result in the emergence of X and Y type patters as individual waves interact. These waves are akin to what one sees in two-dimensional generalizations of the KdV equation \cite{korteweg1895xli}, namely KP equations \cite{kadomtsev1970stability,infeld2000nonlinear}, and these X and Y type patterns can be formed from the superposition of ``line-solitons" (see \cite{biondini2003family,chakravarty2008classification,chakravarty2013construction} for general constructions of such solutions). Such patterns have also been observed in the real world in shallow water at beaches \cite{ablowitz2012nonlinear} and hence are not only of mathematical interest. The present application to granular media demonstrates that such wave patterns as occurring in line-soliton interactions may be ubiquitous within a variety of two-dimensional phenomena, rather than particular to shallow water. Indeed, our application shows that such waves may be possible for atomized systems of relatively few particles. 

To more properly compare our results to PDE solutions of shallow water wave models in order to best classify X or Y type waves, one should develop an appropriate continuum approximation to the model we consider. Such models already exist in the one-dimensional case (and in some limits, one can recover equations akin to the KdV equation), and can fairly easily be extended in a natural way to the two-dimensional case when spherical packings are used (as they dynamics are quasi-one-dimensional), however the development is more involved in the case of two-dimensional hexagonal packings. Nevertheless, this will be one area of future work. Compactons have been shown to exist in the one-dimensional granular chain \cite{Nesterenko1983,mackay1999solitary,sen,james2012periodic}, and exist in two-dimensional square packings, as well, as such configurations are essentially quasi-one-dimensional systems \cite{Leo13}. However, in our simulations (and those of \cite{Leo14}) we observe a smoothing of wave fronts in the two-dimensional hexagonal packed basins, due to the spreading of energy among all neighboring spheres. This suggests that more smooth solutions may be possible for the PDE analogue of the discrete two-dimensional model with hexagonally packed spheres. A more systematic study comparing a relevant PDE model for hexagonal packings to PDE models of shallow water equations, such as the KP equations, may cast light on wave structures and pattern formation observed at transient timescales, and would be one interesting topic of future work.

Future work could also involve considering nonuniform spheres in various basins. It may also be worthwhile to consider models where transversal motion of the constituent spheres are included; significant rotational effects were observed in a thin 1-2-1 hexagonally packed granular channel were considered in \cite{yang2015nonlinear}. Another direction could be to consider various basin geometries placed on a curved base in 3D, resulting in a quasi-two-dimensional model on which a two-dimensional granular crystal sits on a curved surface. This would be the higher-dimensional analogue of the quasi-one-dimensional model of a curved granular chain studied in \cite{Yang,Yang2}. One could also change the physics included in the model: For instance, recent results include the vibration of wetted hexagonally packed granular crystals, with the formation of liquid bridges modifying the dynamics \cite{baur2017dynamics}. Wave propagation in a hexagonal monolayer of spheres on a substrate has also recently been studied \cite{vega2017contact}.

In the case of a granular material constructed by joining two granular crystals composed of distinct materials, the interface dynamics we observe do not give results which are unexpected from the one-dimensional theory. There will be reflection at an interface between two materials, while the transmitted wave will undergo speed and amplitude changes in the new material. However, one could consider more complicated interfaces; instead of linear interfaces, one could create granular media in which interfaces are non-linear curves. As an example, a circular disc of one material might be enclosed within an annular disc of another material. Such configurations may prove useful in applications such as guided impact mitigation \cite{Burgoyne}.

Returning to possible application of waves in granular media, studies on actuating devices considered employing the highly nonlinear waves emergent from a 1D granular chain for Non Destructive Evaluation and Structural Health Monitoring (NDE/SHM) \cite{khatri2009coupling}. That paper considers the role which defects may play in modifying the effectiveness of the approach. In 2D hexagonal configurations of granular media, waves may propagate in part around certain defects, perhaps lending a degree of robustness to the approach. 

There is a propensity for waves to scatter in finite 2D hexagonally packed media. This scattering may be practically useful in devices such as sound scramblers, which were previously studied in 1D chains \cite{daraio2005strongly,NES01}, as the added 2D hexagonal geometry may be more useful in dispersing a wave in finite time, in some configurations. This will depend upon the overall geometry of the basin, of course, as undesirable designs may actually amplify the waves, as observed for triangular or circular basins in Figs. \ref{WaveTriCorner} and \ref{WaveCircleUpstriker}, respectively. Choosing a basin which scatters waves in an asymmetric manner may lead to spatio-temporal irregularities in the wave patterns more rapidly than would more symmetric domains. Even in symmetric domains, however, the amplitude of the waves (as seen in our simulations) tends to decrease as the number of successive reflections increases, with energy being spread out in a more uniform manner over the domain for large time. In the case of granular media composed of multiple materials, such as those studied in Sec. \ref{twomedia}, the scattering can be enhanced due to the interface between the two media. Perhaps a more interesting possibility would be of a three-layer media, with a middle region which traps most of the wave energy. The idea would be the same as that considered in Sec. \ref{twomedia}, with the middle region playing the role of the purple region in Figure \ref{IntGeom}, which was shown to often trap wave energy (see Figs. \ref{WaveHexInter1}-\ref{WaveHexInter4}). Such a configuration could prove useful for acoustic insulation or impact mitigation.

On the other hand, certain geometric configurations of hexagonally packed particles can amplify waves spatially in ways which would not be possible in 1D or quasi-1D chains, hence one could extend results on the generation and control of sound bullets \cite{spadoni2010generation} (where 1D and 2D square packings were shown) using configurations such as those employed to produce the highly spatially localized waves shown in Figs. \ref{WaveTriCorner} and \ref{WaveCircleUpstriker}. Similarly, such amplification or confinement of localized waves within hexagonal packings might also provide an alternative route toward the design of acoustic switches and rectification devices \cite{boechler2011bifurcation} which are able to exploit this localized structure.


\begin{thebibliography}{10}
\providecommand{\url}[1]{{#1}}
\providecommand{\urlprefix}{URL }
\expandafter\ifx\csname urlstyle\endcsname\relax
  \providecommand{\doi}[1]{DOI \discretionary{}{}{}#1}\else
  \providecommand{\doi}{DOI \discretionary{}{}{}\begingroup
  \urlstyle{rm}\Url}\fi

\bibitem{Burgoyne}
H.A. Burgoyne, J.A. Newman, W.C. Jackson, C.~Dariao, Guided impact mitigation
  in 2d and 3d granular crystals, Procedia Engineering \textbf{103}, 52 (2015)

\bibitem{NES02}
V.F. Nesterenko, \emph{Dynamics of heterogeneous materials} (Springer, New
  York, 2001)

\bibitem{sen}
S.~Sen, J.~Hong, J.~Bang, E.~Avalos, R.~Doney, Solitary waves in the granular
  chain, Physics Reports \textbf{462}(2), 21 (2008)

\bibitem{M1}
M.A. Porter, P.G. Kevrekidis, C.~Daraio, Granular crystals: Nonlinear dynamics
  meets materials engineering, Physics Today \textbf{68}, 44 (2015)

\bibitem{M2}
C.~Chong, M.A. Porter, P.~Kevrekidis, C.~Daraio, Nonlinear coherent structures
  in granular crystals, Journal of Physics: Condensed Matter \textbf{29}(41),
  413003 (2017)

\bibitem{JOH01}
K.L. Johnson, \emph{Contact mechanics} (Cambridge University Press, Cambridge,
  1985)

\bibitem{Leo14}
A.~Leonard, C.~Chong, P.~Kevrekidis, C.~Dariao, Traveling waves in 2d hexagonal
  granular crystal lattices, Granular Matter \textbf{16(4)}, 531 (2014)

\bibitem{VER01}
L.~Vergara, Scattering of solitary waves from interfaces in granular media,
  Phys. Rev. Lett. \textbf{95}, 108002 (2005)

\bibitem{VER02}
L.~Vergara, Delayed scattering of solitary waves from interfaces in a granular
  container, Phys. Rev. E \textbf{73}, 066623 (2006)

\bibitem{kekic2018wave}
A.~Keki{\'c}, R.A. Van~Gorder, Wave propagation across interfaces induced by
  different interaction exponents in ordered and disordered hertz-like granular
  chains, Physica D: Nonlinear Phenomena  (2018).
\newblock \doi{https://doi.org/10.1016/j.physd.2018.07.007}

\bibitem{Leo13}
A.~Leonard, F.~Franternali, C.~Dariao, Directional wave propagation in a highly
  nonlinear square packing of spheres, Experimental Mechanics \textbf{53(3)},
  327 (2013)

\bibitem{Sze12}
I.~Szelengowicz, P.~Kevrekidis, C.~Dariao, Wave propagation in square granular
  crystals with sphereical interstitial intruders, Physical Review E
  \textbf{86(6)}, 061306 (2012)

\bibitem{Sze13}
I.~Szelengowicz, M.~Hasan, Y.~Starosvetsky, A.~Vakakis, C.~Dariao, Energy
  equipartition in two-dimensional granular systems with spherical intruders,
  Physical Review E \textbf{87(3)}, 032204 (2013)

\bibitem{Leo12}
A.~Leonard, C.~Dariao, A.~Awasthi, P.~Geubelle, Effects of weak disorder on
  stress-wave anisotropy in centered square nonlinear granular crystals,
  Physical Review E \textbf{86(3)}, 031305 (2012)

\bibitem{Hasan}
M.A. Hasanm, A.F. Vakakis, D.M. McFarland, Nonlinear localization, passive wave
  arrest and traveling breathers in two-dimensional granular networks with
  discontinuous lateral boundary conditions, Wave Motion \textbf{60}, 196
  (2016)

\bibitem{Goncu}
F.~Goncu, S.~Luding, K.~Bertoldi, Exploiting pattern transformation to tune
  phononic band gaps in a two dimensional granular crystal, The Journal of the
  Acoustical Society of America \textbf{131(6)}, EL475 (2012)

\bibitem{Wattis}
J.A. Wattis, L.M. James, Discrete breathers in honeycomb fermi–pasta–ulam
  lattices, Journal of Physics A: Mathematical and Theoretical \textbf{47}(34),
  345101 (2014)

\bibitem{Yang}
J.~Yang, S.~Dunatunga, C.~Daraio, Amplitude-dependent attenuation of
  compressive waves in curved granular crystals constrained by elastic guides,
  Acta Mechanica \textbf{223(3)}, 549 (2012)

\bibitem{Yang2}
J.~Yang, C.~Dariao, Frequency-and amplitude-dependent transmission of stress
  waves in curved one-dimensional granular crystals composed of diatomic
  particles, Experimental Mechanics \textbf{53(3)}, 469 (2013)

\bibitem{Yi}
C.~Yi, Y.~Liu, Q.~Mu, Y.~Qi, Force transmission in three-dimensional
  hexagonal-close-packed granular arrays with point defect submitted to a point
  load, Granular Matter \textbf{9}(3-4), 195 (2007)

\bibitem{manjunath2014plane}
M.~Manjunath, A.P. Awasthi, P.H. Geubelle, Plane wave propagation in 2d and 3d
  monodisperse periodic granular media, Granular Matter \textbf{16}(1), 141
  (2014)

\bibitem{Chong}
C.~Chong, P.G. Kevrekidis, M.J. Ablowitz, Y.P. Ma, Conical wave propagation and
  diffraction in two-dimensional hexagonally packed granular lattices, Physical
  Review E \textbf{93(3)}, 012909 (2016)

\bibitem{wallen2017shear}
S.P. Wallen, N.~Boechler, Shear to longitudinal mode conversion via second
  harmonic generation in a two-dimensional microscale granular crystal, Wave
  Motion \textbf{68}, 22 (2017)

\bibitem{xu2016stress}
J.~Xu, B.~Zheng, Stress wave propagation in two-dimensional buckyball lattice,
  Scientific Reports \textbf{6}, 37692 (2016)

\bibitem{waymel2018propagation}
R.F. Waymel, E.~Wang, A.~Awasthi, P.H. Geubelle, J.~Lambros, Propagation and
  dissipation of elasto-plastic stress waves in two dimensional ordered
  granular media, Journal of the Mechanics and Physics of Solids \textbf{120},
  117 (2018)

\bibitem{li2018two}
L.l. Li, X.q. Yang, W.~Zhang, Two interactional solitary waves propagating in
  two-dimensional hexagonal packing granular system, Granular Matter
  \textbf{20}(3), 49 (2018)

\bibitem{hanaor}
D.A. Hanaor, Y.~Gan, I.~Einav, Contact mechanics of fractal surfaces by spline
  assisted discretisation, International Journal of Solids and Structures
  \textbf{59}, 121 (2015)

\bibitem{lisyansky2015primary}
A.~Lisyansky, D.~Meimukhin, Y.~Starosvetsky, Primary wave transmission in the
  hexagonally packed, damped granular crystal with a spatially varying cross
  section, Communications in Nonlinear Science and Numerical Simulation
  \textbf{27}(1), 193 (2015)

\bibitem{kadomtsev1970stability}
B.B. Kadomtsev, V.I. Petviashvili, in \emph{Sov. Phys. Dokl}, vol.~15 (1970),
  vol.~15, pp. 539--541

\bibitem{infeld2000nonlinear}
E.~Infeld, G.~Rowlands, \emph{Nonlinear waves, solitons and chaos} (Cambridge
  university press, 2000)

\bibitem{biondini2003family}
G.~Biondini, Y.~Kodama, On a family of solutions of the kadomtsev--petviashvili
  equation which also satisfy the toda lattice hierarchy, Journal of Physics A:
  Mathematical and General \textbf{36}(42), 10519 (2003)

\bibitem{chakravarty2008classification}
S.~Chakravarty, Y.~Kodama, Classification of the line-soliton solutions of
  kpii, Journal of Physics A: Mathematical and Theoretical \textbf{41}(27),
  275209 (2008)

\bibitem{chakravarty2013construction}
S.~Chakravarty, Y.~Kodama, Construction of kp solitons from wave patterns,
  Journal of Physics A: Mathematical and Theoretical \textbf{47}(2), 025201
  (2013)

\bibitem{HAS01}
E.~Hasco{\"e}t, H.J. Herrmann, Shocks in non-loaded bead chains with
  impurities, The European Physical Journal B - Condensed Matter and Complex
  Systems \textbf{14}(1), 183 (2000)

\bibitem{HON02}
J.~Hong, A.~Xu, Nondestructive identification of impurities in granular medium,
  Applied Physics Letters \textbf{81}(25), 4868 (2002)

\bibitem{NES01}
V.F. Nesterenko, C.~Daraio, E.B. Herbold, S.~Jin, Anomalous wave reflection at
  the interface of two strongly nonlinear granular media, Physical Review
  Letters \textbf{95}, 158702 (2005)

\bibitem{tichler2013transmission}
A.M. Tichler, L.R. Gomez, N.~Upadhyaya, X.~Campman, V.F. Nesterenko,
  V.~Vitelli, Transmission and reflection of strongly nonlinear solitary waves
  at granular interfaces, Physical review letters \textbf{111}(4), 048001
  (2013)

\bibitem{hasan2016nonlinear}
M.A. Hasan, A.F. Vakakis, D.M. McFarland, Nonlinear localization, passive wave
  arrest and traveling breathers in two-dimensional granular networks with
  discontinuous lateral boundary conditions, Wave Motion \textbf{60}, 196
  (2016)

\bibitem{chong2014damped}
C.~Chong, F.~Li, J.~Yang, M.O. Williams, I.G. Kevrekidis, P.G. Kevrekidis,
  C.~Daraio, Damped-driven granular chains: An ideal playground for dark
  breathers and multibreathers, Physical Review E \textbf{89}(3), 032924 (2014)

\bibitem{korteweg1895xli}
D.J. Korteweg, G.~{De Vries}, On the change of form of long waves advancing in
  a rectangular canal, and on a new type of long stationary waves, The London,
  Edinburgh, and Dublin Philosophical Magazine and Journal of Science
  \textbf{39}(240), 422 (1895)

\bibitem{ablowitz2012nonlinear}
M.J. Ablowitz, D.E. Baldwin, Nonlinear shallow ocean-wave soliton interactions
  on flat beaches, Physical Review E \textbf{86}(3), 036305 (2012)

\bibitem{Nesterenko1983}
V.~Nesterenko, Propagation of nonlinear compression pulses in granular media,
  Journal of Applied Mechanics and Technical Physics \textbf{24}(5), 733 (1983)

\bibitem{mackay1999solitary}
R.S. MacKay, Solitary waves in a chain of beads under hertz contact, Physics
  Letters A \textbf{251}(3), 191 (1999)

\bibitem{james2012periodic}
G.~James, Periodic travelling waves and compactons in granular chains, Journal
  of nonlinear science \textbf{22}(5), 813 (2012)

\bibitem{yang2015nonlinear}
J.~Yang, M.~Sutton, Nonlinear wave propagation in a hexagonally packed granular
  channel under rotational dynamics, International Journal of Solids and
  Structures \textbf{77}, 65 (2015)

\bibitem{baur2017dynamics}
M.~Baur, K.~Huang, Dynamics of wet granular hexagons, Physical Review E
  \textbf{95}(3), 030901 (2017)

\bibitem{vega2017contact}
A.~Vega-Flick, R.~Duncan, S.~Wallen, N.~Boechler, C.~Stelling, M.~Retsch,
  J.~Alvarado-Gil, K.~Nelson, A.~Maznev, Contact-based and spheroidal
  vibrational modes of a hexagonal monolayer of microspheres on a substrate,
  Wave Motion \textbf{76}, 122 (2018)

\bibitem{khatri2009coupling}
D.~Khatri, C.~Daraio, P.~Rizzo, in \emph{Sensors and Smart Structures
  Technologies for Civil, Mechanical, and Aerospace Systems 2009}, vol. 7292
  (International Society for Optics and Photonics, 2009), vol. 7292, p. 72920P

\bibitem{daraio2005strongly}
C.~Daraio, V.~Nesterenko, E.~Herbold, S.~Jin, Strongly nonlinear waves in a
  chain of teflon beads, Physical Review E \textbf{72}(1), 016603 (2005)

\bibitem{spadoni2010generation}
A.~Spadoni, C.~Daraio, Generation and control of sound bullets with a nonlinear
  acoustic lens, Proceedings of the National Academy of Sciences
  \textbf{107}(16), 7230 (2010)

\bibitem{boechler2011bifurcation}
N.~Boechler, G.~Theocharis, C.~Daraio, Bifurcation-based acoustic switching and
  rectification, Nature materials \textbf{10}(9), 665 (2011)

\end{thebibliography}
\end{document}